\documentclass[11pt, a4paper]{article}
\pdfoutput=1

\usepackage[height=8.85in,width=6.75in]{geometry}

\usepackage{graphicx,rotating}     
\usepackage[bookmarksopen,colorlinks=true,linkcolor=steelblue,
citecolor=darkred,urlcolor=darkred,linktoc=all]{hyperref}

\usepackage{amsmath,amssymb}
\usepackage{mathtools}
\usepackage{slashed}
\usepackage{bm}
\usepackage{bbm}
\usepackage{cite}
\usepackage{comment}
\usepackage[utf8]{inputenc}
\usepackage{accents}
\usepackage[footnotesize]{caption}
\usepackage{cancel}
\usepackage{physics}
\usepackage[export]{adjustbox}

\usepackage[lf]{Baskervaldx} 
\usepackage[bigdelims,vvarbb]{newtxmath} 
\usepackage[cal=boondoxo]{mathalfa} 

\usepackage{stmaryrd}
\usepackage{trimclip}
\usepackage[framemethod=default]{mdframed}
\newmdenv[skipabove=7pt,
skipbelow=7pt,
rightline=false,
leftline=false,
topline=false,
bottomline=false,
backgroundcolor=gray!10,
linecolor=gray,
innerleftmargin=5pt,
innerrightmargin=5pt,
innertopmargin=5pt,
innerbottommargin=5pt,
leftmargin=0cm,
rightmargin=0cm,
linewidth=4pt]{eBox}
\newmdenv[skipabove=7pt,
skipbelow=7pt,
rightline=true,
leftline=true,
topline=true,
bottomline=true,
backgroundcolor=white,
linecolor=gray,
innerleftmargin=5pt,
innerrightmargin=5pt,
innertopmargin=5pt,
innerbottommargin=5pt,
leftmargin=0cm,
rightmargin=0cm,
linewidth=1pt]{eBox2}

\makeatletter
\@addtoreset{equation}{section}

\makeatother

\usepackage{xcolor}
\definecolor{darkred}{rgb}{0.7, 0., 0.}
\definecolor{orangered}{rgb}{1,0.27,0.}
\definecolor{steelblue}{rgb}{0.275,0.51, 0.706}
\definecolor{forestgreen}{rgb}{0.13,0.55,0.13}

\usepackage{tikz}
\usepackage{tikz-feynman}
\tikzfeynmanset{compat=1.0.0}
\pgfdeclarelayer{bg}    
\pgfsetlayers{bg,main}  
\usetikzlibrary{decorations}

\pgfdeclaredecoration{dashsoliddouble}{initial}{
  \state{initial}[width=\pgfdecoratedinputsegmentlength]{
    \pgfmathsetlengthmacro\lw{.3pt+.5\pgflinewidth}
    \begin{pgfscope}
      \pgfpathmoveto{\pgfpoint{0pt}{\lw}}%
      \pgfpathlineto{\pgfpoint{\pgfdecoratedinputsegmentlength}{\lw}}%
      \pgfmathtruncatemacro\dashnum{%
        round((\pgfdecoratedinputsegmentlength-3pt)/6pt)
      }
      \pgfmathsetmacro\dashscale{%
        \pgfdecoratedinputsegmentlength/(\dashnum*6pt + 3pt)
      }
      \pgfmathsetlengthmacro\dashunit{3pt*\dashscale}
      \pgfsetdash{{\dashunit}{\dashunit}}{0pt}
      \pgfusepath{stroke}
      \pgfsetdash{}{0pt}
      \pgfpathmoveto{\pgfpoint{0pt}{-\lw}}%
      \pgfpathlineto{\pgfpoint{\pgfdecoratedinputsegmentlength}{-\lw}}%
      \pgfusepath{stroke}
    \end{pgfscope}
  }
}

\begin{document}

\hypersetup{pageanchor=false}
\begin{titlepage}

\begin{center}

\hfill UMN-TH-4331/24 \\
\hfill FTPI-MINN-24-18 \\
\hfill KEK-TH-2653

\vskip 0.5in

{\Huge \bfseries 
Cutting rule for in-in correlators 
\\[5mm] and cosmological collider
} \\
\vskip .8in

{\Large Yohei Ema,$^{1,2}$ Kyohei Mukaida$^{3,4}$}

\vskip .3in
\begin{tabular}{ll}
$^{1}$ & \!\!\!\!\!\emph{School of Physics and Astronomy, University of Minnesota, Minneapolis, MN 55455, USA}\\
$^{2}$ & \!\!\!\!\!\emph{William I. Fine Theoretical Physics Institute, School of Physics and Astronomy,}\\[-.15em]
& \!\!\!\!\!\emph{University of Minnesota, Minneapolis, MN 55455, USA}\\
$^{3}$ & \!\!\!\!\!\emph{Theory Center, IPNS, KEK, 1-1 Oho, Tsukuba, Ibaraki 305-0801, Japan}\\
$^{4}$ & \!\!\!\!\!\emph{Graduate University for Advanced Studies (Sokendai), }\\[-.15em]
& \!\!\!\!\!\emph{1-1 Oho, Tsukuba, Ibaraki 305-0801, Japan}
\end{tabular}

\end{center}
\vskip .6in

\begin{abstract}
\noindent
We derive a cutting rule for equal-time in-in correlators including cosmological correlators
based on Keldysh $r/a$ basis, which decomposes diagrams into fully retarded functions 
and cut-propagators consisting of Wightman functions.
Our derivation relies only on basic assumptions such as 
unitarity, locality, and the causal structure of the in-in formalism,
and therefore holds for theories with arbitrary particle contents and local interactions at any loop order.
As an application, we show that non-local cosmological collider signals arise solely from cut-propagators
under the assumption of microcausality.
Since the cut-propagators do not contain (anti-)time-ordering theta functions, 
the conformal time integrals are factorized, simplifying practical calculations. 

\end{abstract}

\end{titlepage}

\tableofcontents
\renewcommand{\thepage}{\arabic{page}}
\renewcommand{\thefootnote}{$\natural$\arabic{footnote}}
\setcounter{footnote}{0}
\hypersetup{pageanchor=true}

\section{Introduction}
\label{sec:introduction}

Cosmic inflation is an essential piece of modern cosmology~\cite{Guth:1980zm,Sato:1981qmu,
Linde:1981mu,Albrecht:1982wi}.
The most important implication of inflation is arguably that
it generates large-scale cosmological perturbations.
Statistical patterns of the perturbations, measured through different cosmological observables
such as cosmic microwave background (CMB) and large-scale structure (LSS) of the universe,
provide important clues of physics in the very early universe including the inflationary energy scale, 
an inflaton potential, and possible inflaton couplings to other particles
(see \textit{e.g.,} \cite{Baumann:2018muz,Achucarro:2022qrl} for recent reviews).
In particular, since the inflationary energy scale can be extremely high, it offers an exciting opportunity
to probe heavy particles never producible by any near-future terrestrial collider experiments,
dubbed as cosmological collider (CC) physics
(see~\cite{Chen:2009we,Chen:2009zp,Baumann:2011nk,Assassi:2012zq,
Sefusatti:2012ye,Norena:2012yi,Chen:2012ge,Noumi:2012vr,Cespedes:2013rda,Gong:2013sma,Kehagias:2015jha,
Liu:2015tza,Arkani-Hamed:2015bza} for earlier works on the topic).

Primordial cosmological perturbations are calculated in the in-in formalism,
where cosmological correlators are evaluated at future infinity, corresponding to the end of inflation,
with the initial conditions fixed (usually taken as the Bunch--Davies vacuum)~\cite{Maldacena:2002vr,Weinberg:2005vy,
Adshead:2009cb,Chen:2017ryl}.
In the context of CC physics, a cutting rule of the in-in formalism attracted recent attention
as it indicates a factorization of the conformal time integrals,
significantly simplifying practical calculations.\footnote{
	See \textit{e.g.,} \cite{Weldon:1983jn,Caron-Huot:2007zhp,Ghiglieri:2020dpq} for the cutting rules in the in-in (or real-time) formalism in 
	the context of thermal field theory.
}
Indeed, a cutting rule of cosmological in-in correlators that efficiently extracts CC signals
has been proposed for tree-level processes in~\cite{Tong:2021wai} and extended to multi-loops 
(for non-local CC signals) in~\cite{Qin:2023bjk,Qin:2023nhv}.\footnote{
	A cutting rule on the wavefunction of the universe, instead of the in-in correlators, is discussed 
	in~\cite{Goodhew:2020hob,Jazayeri:2021fvk,Melville:2021lst,Goodhew:2021oqg,AguiSalcedo:2023nds,
	Stefanyszyn:2023qov,Ghosh:2024aqd},
	relying only on minimal assumptions of quantum field theory.
	See also~\cite{Donath:2024utn} 
	for a recent discussion on a cutting rule on cosmological correlators from the in-out formalism,
	and~\cite{Werth:2024mjg} for relations among cosmological correlators 
	derived from the cosmological largest time equation.
	See the end of Sec.~\ref{subsec:cutting_proof} and Sec.~\ref{subsec:microcausality} 
	for a comparison.
}
Despite being successful in extracting CC signals,
however, the cutting rules discussed in these references 
focus on tree-level processes, or rely on explicit forms of propagators.
This is in contrast to the in-out (or $S$-matrix) formalism, 
where the cutting rule is widely known under the name of Cutkosky rule~\cite{Cutkosky:1960sp},
and can be shown by Veltman's largest time equation~\cite{Veltman:1963th,Veltman:1994wz} 
independently of theory details such as particle contents and interactions,
and without any additional assumptions/approximations.
Therefore, it would be fair to say that the cutting rule for cosmological in-in correlators is still immature 
compared to its in-out counterpart.

Our main goal in this paper is to fill this gap.
We derive a cutting rule for equal-time in-in correlators.
Our derivation relies only on the minimal assumptions such as unitarity, locality, and the causal structure of the in-in formalism,
and hence our cutting rule holds for arbitrary theories and at any loop order, 
in analogy with the Cutkosky rule of the in-out formalism. 
The equal-time and commuting nature of the cosmological correlators, 
together with Keldysh $r/a$ basis for the in-in fields,
allows us to write down a particularly simple form of the cutting rule.
For instance, it is expressed diagrammatically as
\begin{align}
	\includegraphics[valign=c]{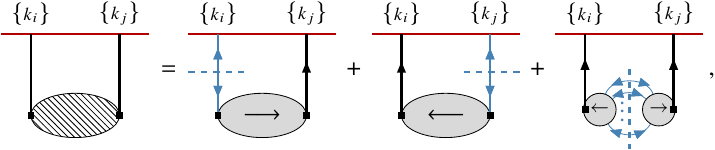}
\end{align}
for two-point bulk correlators, and 
\begin{align}
	\includegraphics[valign=c]{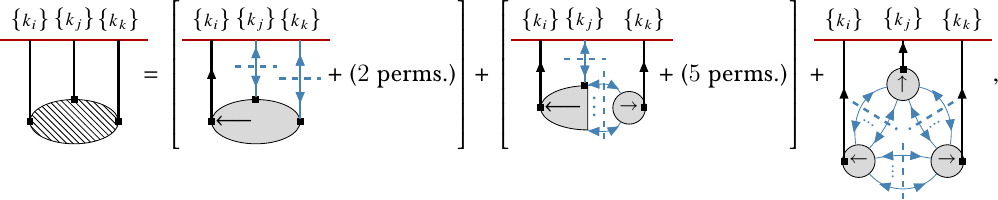}
\end{align}
for three-point bulk correlators, where the black thick lines represent the bulk-to-boundary propagators collectively, 
the arrows represent the causal flow, and the blobs with the arrows represent fully retarded functions.
This rule has a straightforward generalization to $n$-point bulk correlators; see Sec.~\ref{sec:cutting} for details.
With the assumption of microcausality,  
we show that CC signals of non-local type arise solely from cut-propagators,
the cyan solid lines with the dashed cuts in the above diagrams, in the cutting rule.
Since the bulk-to-boundary propagator does not contain non-local CC signals, 
only diagrams with cuts on bulk correlators, or ``bulk-cut'' diagrams, generate non-local CC signals,
which factorizes the conformal time integrals and therefore simplifies the calculations.

This paper is organized as follows. In Sec.~\ref{sec:cutting}, after reviewing the in-in formalism 
and the $r/a$ basis suitable for our purpose, 
we derive a cutting rule for equal-time in-in correlators.
We prove it by duplicating the in-in contours, 
which is analogous to the ``circled'' vertices in the largest time equation representing the hermitian conjugate.
We then explicitly check our cutting rule with several examples, starting from the standard in-in Feynman rules.
Our discussion in Sec.~\ref{sec:cutting} is intended to be as general as possible,
and applies to arbitrary theories even beyond the context of cosmological correlators.
In Sec.~\ref{sec:application}, we narrow down our focus and study CC signals.
In particular, we show that the microcausality restricts the origin of the non-local CC signals as cut-propagators.
We apply the cutting rule to several examples and explicitly check that it indeed extracts the non-local CC signals efficiently.
Finally, Sec.~\ref{sec:summary} is devoted to a summary and possible future applications of our cutting rule.

\section{Cutting rule for in-in correlators}
\label{sec:cutting}

In this section, we derive a cutting rule for equal-time in-in correlators.
Our cutting rule is best described in the $r/a$ basis, defined by a linear transformation of fields
in the standard ``$1/2$'' (often denoted as ``$+/-$'' in cosmological contexts) in-in contours.
Therefore, in Sec.~\ref{subsec:Keldysh_basis}, we review its basic properties with an emphasis on its causal structure.
Sec.~\ref{subsec:cutting_proof} provides the most important result in this paper; we prove a cutting rule
for in-in correlators based solely on unitarity, locality, and the causal structure of the in-in formalism,
which is particularly manifest in the $r/a$ basis.
Our proof in Sec.~\ref{subsec:cutting_proof} relies on a technique analogous to 
the largest time equation for the in-out formalism~\cite{Veltman:1963th,Veltman:1994wz}, 
and may look somewhat abstract for those unfamiliar with the formal derivation of the Cutkosky rule.
Therefore, to convince even the most skeptical readers (if any), we provide several examples
and confirm the validity of the cutting rule by starting from the standard in-in Feynman rules in Sec.~\ref{subsec:cutting_examples}.

In this section, we keep our discussion as general as possible.
Indeed, our cutting rule applies to theories with arbitrary particle contents and (local) interactions,
and at any loop orders, even beyond the context of cosmological correlators.
In the next section, 
as an application, we show that our cutting rule efficiently extracts non-local CC signals.

\subsection{In-in formalism and Keldysh $r/a$ basis}
\label{subsec:Keldysh_basis}
In the study of quantum field theory at zero temperature, we are often interested in the $S$-matrix.
There the correlators of the time-ordered operator product play central roles owing to the LSZ reduction formula.
The correlators of the time-ordered operator product are computed in the in-out formalism, where the expectation value is taken between the far past/future vacuum states.

Instead, when we are interested in the time evolution of certain operator products as an initial value problem, the in-in formalism provides a transparent way of computing the correlators under a given initial state~\cite{Schwinger:1960qe,Bakshi:1962dv,Bakshi:1963bn,Keldysh:1964ud}.
In the following, we briefly review the basics of in-in formalism (see also \textit{e.g.,} Refs.~\cite{Laine:2016hma,Ghiglieri:2020dpq}).
Then we introduce the $r/a$ basis convenient to see the causal structure, and discuss its basic properties.
Finally we introduce the generalization of the retarded self-energy to the $n$-point vertex, namely the fully retarded function.

\subsubsection*{Schwinger--Keldysh contour in a nutshell}
\begin{figure}[t]
	\centering
    \includegraphics[width=0.8\linewidth]{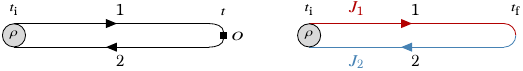}
	\caption{\small \emph{Left:} the back-and-forth contour of $1/2$, \textit{i.e.,} the Schwinger--Keldysh contour, corresponding to the expectation value of $O(t)$ given in Eq.~\eqref{eq:expO_skcontour}. The state $\rho$ is inserted at $t = t_\text{i}$ and an operator $O$ is inserted at $t$.
    \emph{Right:} the back-and-forth contour of $1/2$ corresponding to the generating functional given in Eq.~\eqref{eq:Z_def_12_PI}.
    Contrary to the left case, the time evolution on the contour $1$ is executed in the presence of $J_1$ while that on the contour $2$ is done in the presence of $J_2$.
    We use different colors for the $1/2$ time evolution to clarify this point.
	}
	\label{fig:skcontour}
\end{figure}

Let $\rho$ be a state, which does not evolve in time as we take the Heisenberg picture.
Suppose that we are interested in the expectation value of an operator $O (t)$, given by
\begin{align}
    \left\langle O (t) \right\rangle 
    &\coloneqq \trace \qty[ \rho O(t) ] \nonumber \\
    &=
    \int \dd \varphi_{1i} \dd \varphi_{2i} \dd \varphi_{1\text{f}} \dd \varphi_{2\text{f}}\,
    \rho [\varphi_{1\text{i}},\varphi_{2\text{i}};t_\text{i}]
    \braket*{\varphi_{2\text{i}},t_\text{i}}{\varphi_{2 \text{f}},t}
    O[ \varphi_{2\text{f}}, \varphi_{1\text{f}}; t]
    \braket*{\varphi_{1\text{f}},t}{\varphi_{1 \text{i}},t_\text{i}}
    \,,
    \label{eq:O_exp}
\end{align}
with the relevant matrix elements being
\begin{align}
    \rho [\varphi_{1\text{i}},\varphi_{2\text{i}};t_\text{i}] \coloneqq \bra{\varphi_{1\text{i}},t_\text{i}}\rho \ket{\varphi_{2\text{i}},t_\text{i}}\,, \qquad
    O[ \varphi_{2\text{f}}, \varphi_{1\text{f}}; t] \coloneqq \bra{\varphi_{2\text{f}},t} O(t) \ket{\varphi_{1\text{f}},t}\,.
\end{align}
Note that the Hermiticity of $\rho$ implies $\rho [\varphi_{1\text{i}},\varphi_{2\text{i}};t_\text{i}] = \rho^\ast [\varphi_{2\text{i}},\varphi_{1\text{i}};t_\text{i}]$.
Here an eigenstate for a field operator $\ket{\varphi, t}$ is introduced, which fulfills
\begin{align}
    \varphi (t, {\bm x}) \ket{\varphi', t} = \varphi' ({\bm x}) \ket{\varphi',t}\,.
\end{align}
The unitary time evolution from $t_\text{i}$ to $t$ is denoted by $\mathcal{U} (t,t_\text{i})$ obeying the Schr\"odinger equation
\begin{align}
    i \partial_t \mathcal{U} (t,t_\text{i}) = H \mathcal{U} (t,t_\text{i})\,, \qquad \mathcal{U} (t_\text{i},t_\text{i}) = \mathbb{1}\,.
\end{align}
Recalling that the eigenstates should evolve ``backwards'' in time, \textit{i.e.},
$
    \ket{\varphi', t} = \mathcal{U}^\dag (t,t_\text{i}) \ket{\varphi', t_\text{i}}
$,
one may express the probability amplitude between $\ket{\varphi_\text{i},t_\text{i}}$ and $\ket{\varphi_\text{f},t}$ by means of the path integral
\begin{align}
    \braket*{\varphi_{1\text{f}},t}{\varphi_{1 \text{i}},t_\text{i}}
    &=
    \bra{\varphi_{1\text{f}}, t_\text{i}} \mathcal{U} (t, t_\text{i}) \ket{\varphi_{1\text{i}}, t_\text{i}}
    =
    \int^{\varphi_1(t) = \varphi_{1\text{f}}}_{\varphi_1(t_\text{i}) = \varphi_{1\text{i}}} \mathcal D \varphi_1\, e^{i S[\varphi_1]}\,, \\
    \braket*{\varphi_{2\text{i}},t_\text{i}}{\varphi_{2 \text{f}},t}
    &=
    \bra{\varphi_{2\text{i}}, t_\text{i}} \mathcal{U}^\dag (t, t_\text{i}) \ket{\varphi_{2\text{f}}, t_\text{i}}
    =
    \int^{\varphi_2(t) = \varphi_{2\text{f}}}_{\varphi_2(t_\text{i}) = \varphi_{2\text{i}}} \mathcal D \varphi_2\, e^{-i S[\varphi_2]}
    \,.
\end{align}
Inserting these expressions, we rewrite the expectation value as follows
\begin{align}
    \left\langle O (t) \right\rangle 
    = \int \dd \varphi_{1i} \dd \varphi_{2i} \dd \varphi_{1\text{f}} \dd \varphi_{2\text{f}}\,
    \rho [\varphi_{1\text{i}},\varphi_{2\text{i}};t_\text{i}] O[ \varphi_{2\text{f}}, \varphi_{1\text{f}}; t]
    \int^{\varphi_1(t) = \varphi_{1\text{f}}}_{\varphi_1(t_\text{i}) = \varphi_{1\text{i}}} \mathcal D \varphi_1
    \int^{\varphi_2(t) = \varphi_{2\text{f}}}_{\varphi_2(t_\text{i}) = \varphi_{2\text{i}}} \mathcal D \varphi_2
    e^{i S[\varphi_1] - i S[\varphi_2]}\,.
    \label{eq:expO_skcontour}
\end{align}
This expression clarifies that the expectation value of a certain operator $O$ with respect to a given state $\rho$ can be understood as the back-and-forth path integral denoted by $1$ and $2$ contours respectively, known as the Schwinger--Keldysh contour, with insertions of operators $\rho$ at $t_\text{i}$ and $O$ at $t$ (see Fig.~\ref{fig:skcontour}).

In general, a correlator of multiple operators can be of our interest.
To this end, we introduce a generating functional on the Schwinger--Keldysh contour by doubling the source terms in $1/2$ basis
\begin{align}
    e^{i W[J_1, J_2]}
    &\coloneqq
    \trace \qty[ \rho\,  \qty{\tilde{T} e^{ - i \int_{t_\text{i}}^{t_\text{f}} \dd^4 x \varphi (x) J_2 (x) }} \qty{T e^{i \int_{t_\text{i}}^{t_\text{f}} \dd^4 x \varphi (x) J_1 (x) }} ] \label{eq:Z_def_12} \\
    &=
    \int \dd \varphi_{1i} \dd \varphi_{2i} \dd \varphi_\text{f}
    \, \rho [\varphi_{1\text{i}},\varphi_{2\text{i}};t_\text{i}]
    \int^{\varphi_1(t) = \varphi_{\text{f}}}_{\varphi_1(t_\text{i}) = \varphi_{1\text{i}}} \mathcal D \varphi_1
    \int^{\varphi_2(t) = \varphi_{\text{f}}}_{\varphi_2(t_\text{i}) = \varphi_{2\text{i}}} \mathcal D \varphi_2\,
    e^{i \qty(S[\varphi_1] - S[\varphi_2] 
    + J_1 \cdot \varphi_1 - J_2 \cdot \varphi_2 ) }\,, 
    \label{eq:Z_def_12_PI}
\end{align}
where, in the second line, we have used a shorthanded notation
\begin{align}
    J \cdot \varphi \coloneqq \int^{t_\text{f}}_{t_\text{i}} \dd^4 x\, J (x) \varphi (x)\,.
\end{align}
The time-ordered and anti-time-ordered products are indicated by $T$ and $\tilde T$ respectively.
The future end point of the Schwinger--Keldysh contour $t_\text{f}$ is taken to be sufficiently late and the external sources fulfill $J_1 = J_2$ at $t = t_\text{f}$.
See also Fig.~\ref{fig:skcontour}.
The generating functional on the Schwinger--Keldysh contour must meet the following requirements from the unitarity~\cite{Glorioso:2016gsa}
\begin{eBox}
    \textit{Unitarity constraints on the generating functional in the $1/2$ basis}
    \begin{alignat}{2}
        \label{eq:Z_norm}
        &\text{(Normalization)} & \qquad & 
        W [J,J] = 0\,, \\
        \label{eq:Z_rflc}
        &\text{(Reflectivity)} & \qquad & 
        W [J_1, J_2] = - W^\ast [J_2, J_1]\,, \\
        \label{eq:Z_pos}
        &\text{(Positivity)} & \qquad &
        \Im W[J_1,J_2] \geq 0\,.
    \end{alignat}
\end{eBox}
\paragraph{\textit{Proof.---}}
Notice that the time-ordered product of the external source in Eq.~\eqref{eq:Z_def_12} can be understood as the unitary time evolution under $S[\varphi] + J \cdot \varphi$.
Hence, the generating functional can be rewritten as
\begin{align}
    \label{eq:Z_def_12_unitary}
    e^{i W[J_1, J_2]} = \trace \qty[ \rho \mathcal{U}_{J_2}^\dag (t_\text{f}, t_\text{i}) \mathcal{U}_{J_1} (t_\text{f},t_\text{i}) ]\,,
\end{align}
with the unitary time evolution under the influence of $J$ being $\mathcal{U}_{J}$.
The first requirement immediately follows
\begin{align}
    e^{i W [J,J]} = \trace \qty[ \rho \mathcal{U}_{J}^\dag (t_\text{f}, t_\text{i}) \mathcal{U}_{J} (t_\text{f},t_\text{i}) ] = \trace \qty[\rho] = 1\,.
\end{align}
Also the second one is readily obtained from
\begin{align}
    e^{- i W^\ast [J_2, J_1]} = 
    \qty(e^{i W[J_2, J_1]})^\ast 
    = \trace \qty[\mathcal{U}_{J_2}^\dag (t_\text{f},t_\text{i}) \mathcal{U}_{J_1} (t_\text{f}, t_\text{i}) \rho ]
    = e^{i W [J_1, J_2]}\,.
\end{align}
The third one is a bit tricky.
In general, the state can be decomposed into
\begin{align}
    \rho = \sum_n p_n \ket{n} \bra{n}\,,
\end{align}
with $p_n$ being non-negative coefficients adding up to unity, \textit{i.e.,} $\sum_n p_n = 1$.
This implies
\begin{align}
    \abs{ e^{i W[J_1, J_2]} } 
    &=
    \abs{ \sum_n p_n \bra{n} \mathcal{U}_{J_2}^\dag (t_\text{f},t_\text{i}) \mathcal{U}_{J_1} (t_\text{f},t_\text{i}) \ket{n} }
    \leq
    \sum_n p_n \abs{ \bra{n} \mathcal{U}_{J_2}^\dag (t_\text{f},t_\text{i}) \mathcal{U}_{J_1} (t_\text{f},t_\text{i}) \ket{n} } \leq 1\,.
\end{align}
We have utilized the Cauchy--Schwarz inequality
\begin{align}
    \abs{ \bra{n} \mathcal{U}_{J_2}^\dag (t_\text{f},t_\text{i}) \mathcal{U}_{J_1} (t_\text{f},t_\text{i}) \ket{n} } \leq
    \norm{\mathcal{U}_{J_2} (t_\text{f},t_\text{i}) \ket{n}}\,
    \norm{\mathcal{U}_{J_1} (t_\text{f},t_\text{i}) \ket{n} } = 1\,.
\end{align}
This completes our proof.

\vspace{5mm}

The propagators in the $1/2$ basis are obtained by differentiating the generating functional with respect to the external source $J_{1,2}$:
\begin{align}
    \label{eq:prop_12_def}
    G_{ij} (x,y) &\coloneqq \left.\frac{\delta^2 i W}{\delta i J^i (x)\, \delta i J^j (y)}\right|_{J_1 = J_2 = 0} \\[.5em]
    &=\begin{pmatrix}
        \left\langle T \varphi(x) \varphi(y) \right\rangle & \qty(-)^{\abs{\varphi}} \left\langle \varphi(y) \varphi(x) \right\rangle\\
        \left\langle \varphi(x) \varphi(y) \right\rangle &  \left\langle \tilde{T} \varphi(x) \varphi(y) \right\rangle
    \end{pmatrix}
    \eqqcolon
    \begin{pmatrix}
        G_\text{F} (x,y) & G_< (x,y)\\
        G_> (x,y) & G_{\bar{\text{F}}} (x,y)
    \end{pmatrix}\,,
\end{align}
where we have introduced the external sources with a superscript, which are defined by $J^1 = J_1$, $J^2 = - J_2$ for notational convenience.
The Feynman propagator is $G_{\text{F}}$, the Wightman positive/negative functions are $G_{>/<}$, and the anti-Feynman propagator is $G_{\bar{\text{F}}}$.
The statistical property of an operator is denoted by $\abs{\varphi} = 0,1$ for boson and fermion respectively.
By definition, the Feynman and anti-Feynman propagators can be expressed by the positive and negative Wightman functions
\begin{align}
    \label{eq:GF}
    G_\text{F} (x,y) &= \theta (x_0-y_0) G_> (x,y) + \theta (y_0-x_0) G_< (x,y) \,, \\
    \label{eq:GbarF}
    G_{\bar{\text{F}}} (x,y) &= \theta (y_0-x_0) G_> (x,y) + \theta (x_0-y_0) G_< (x,y) \,,
\end{align}
with the unit step function being $\theta (x)$.

\subsubsection*{$r/a$ basis}

In the previous subsection, we have introduced the path-integral in the $1/2$ basis.
However, in many practical calculations, the $r/a$ basis is more convenient, which is defined by
\begin{align}
    \varphi_r \coloneqq \frac{1}{2} \qty( \varphi_1 + \varphi_2 )\,, \qquad
    \varphi_a \coloneqq \varphi_1 - \varphi_2\,.
\end{align}
This basis reduces the number of propagators and thereby diagrams by utilizing the special feature of the path-integral exponent, and moreover clarifies the causal flow in the Feynman diagrams as we will see shortly.
The external sources are also redefined in the $r/a$ basis as
\begin{align}
    J_r \coloneqq \frac{1}{2} \qty( J_1 + J_2 )\,,
    \qquad
    J_a \coloneqq J_1 - J_2\,.
\end{align}

The exponent of the path-integral in the $r/a$ basis reads
\begin{align}
    S[\varphi_1] + J_1 \cdot \varphi_1 - S[\varphi_2] - J_2 \cdot \varphi_2
    = 2 \sum_{n=1} \frac{\delta^{2n-1} S [\varphi_r]}{\delta \varphi_r^{2n-1}} \frac{1}{\qty(2 n -1)!} \qty( \frac{\varphi_a}{2} )^{2n-1}
    + J_a \cdot \varphi_r + J_r \cdot \varphi_a\,.
\end{align}
One readily sees that every vertex in the $r/a$ basis must have an odd number of $a$-fields, and a pair of $a$-fields involves an additional factor of $1/4$.
For instance, in the case of $\lambda\varphi^4$ theory, we have
\begin{align}
    \frac{\lambda}{4!} \qty( \varphi_1^4 - \varphi_2^4 )
    =
    \frac{\lambda}{3!} \, \varphi_r^3 \varphi_a 
    +
    \frac{\lambda}{4\cdot 3!}\, \varphi_r \varphi_a^3\,.
\end{align}

The generating functional can be rewritten in the $r/a$ basis by means of $J_r$ and $J_a$:
\begin{align}
    W[J_r,J_a] \coloneqq 
    \left.W[ J_1, J_2] \right|_{J_{1/2}= J_r \pm \frac{J_a}{2}}\,.
\end{align}
The unitarity constraints on the generating functional can be rewritten immediately as follows
\begin{eBox}
    \textit{Unitarity constraints on the generating functional in the $r/a$ basis}
    \begin{alignat}{2}
        \label{eq:Z_norm_ra}
        &\text{(Normalization)} & \qquad & 
        W [J_r,J_a = 0] = 0\,, \\
        \label{eq:Z_rflc_ra}
        &\text{(Reflectivity)} & \qquad & 
        W [J_r, J_a] = - W^\ast [J_r, -J_a]\,, \\
        \label{eq:Z_pos_ra}
        &\text{(Positivity)} & \qquad &
        \Im W[J_r,J_a] \geq 0\,.
    \end{alignat}
\end{eBox}
The $n$-point connected Green function in the $r/a$ basis is generated by differentiating with respect to $J_{r/a}$
\begin{align}
    G_{I_1 \cdots I_n} (x_1, \cdots, x_n) 
    \coloneqq
    \left.
    \frac{\delta^n i W}{\delta i J^{I_1}(x_1) \cdots \delta i J^{I_n}(x_n)}
    \right|_{J_r = J_a = 0}
    \,,
\end{align}
with the superscripts being $I_{m} = r,a$ for $m = 1, \cdots, n$.
We have introduced the external source with a superscript for notational clarity, which is defined by $J^r = J_a$ and $J^a = J_r$.
A non-trivial consequence of the unitarity constraint \eqref{eq:Z_norm_ra} on the $n$-point Green functions is the vanishing $a$-fields correlators
\begin{align}
    G_{a \cdots a} (x_1, \cdots , x_n) 
    &=
    \left.
        \frac{\delta^n i W[J_r,J_a]}{\delta i J_r (x_1) \cdots \delta i J_r (x_n)}
    \right|_{J_r = J_a = 0}
    =
    \left.
        \frac{\delta^n i W[J_r, 0]}{\delta i J_r (x_1) \cdots \delta i J_r (x_n)}
    \right|_{J_r = 0} \nonumber\\
    &= 0\,.
    \label{eq:vanishing_Gaaa}
\end{align}
In the last equality, we have used Eq.~\eqref{eq:Z_norm_ra}.

The propagators in the $r/a$ basis can be expressed as follows
\begin{align}
    \label{eq:prop_ra_def}
    G_{IJ} (x,y) &\coloneqq \left.\frac{\delta^2 i W}{\delta i J^I (x)\, \delta i J^J (y)}\right|_{J_r = J_a = 0}
    =\begin{pmatrix}
        \left\langle \varphi_r(x) \varphi_r(y) \right\rangle & \left\langle \varphi_r(y) \varphi_a(x) \right\rangle\\
        \left\langle \varphi_a(x) \varphi_r(y) \right\rangle &  \left\langle \varphi_a(x) \varphi_a(y) \right\rangle
    \end{pmatrix} \\[.5em]
    &=
    \begin{pmatrix}
        G_{rr} (x,y) & G_{ra} (x,y)\\
        G_{ar} (x,y) & 0
    \end{pmatrix}\,,
\end{align}
where the lower right component, $G_{aa}$, vanishes identically because of Eq.~\eqref{eq:vanishing_Gaaa}.
The remaining three propagators can be expressed as follows.
\begin{eBox}
\textit{Propagators in the $r/a$ basis}
\begin{alignat}{2}
    &\text{(Statistical propagator)} & \qquad 
    G_{rr} (x,y) 
    &= \frac{1}{2} \qty[ G_> (x,y) + G_< (x,y) ] \,, \\[.2em]
    &\text{(Retarded propagator)} & \qquad 
    G_{ra} (x,y) 
    &=
    \theta (x_0 - y_0) \qty[ G_> (x,y) - G_< (x,y) ]\,,
    \\[.5em]
    &\text{(Advanced propagator)} & \qquad 
    G_{ar} (x,y)
    &=
    - \theta (y_0 - x_0) \qty[ G_> (x,y) - G_< (x,y) ]\,.
\end{alignat}
\end{eBox}
\paragraph*{\textit{Proof.---}}
One may readily show this by the Leibniz rule:
\begin{align}
    \frac{\delta}{\delta J^a} =
    \frac{\delta}{\delta J_r} = \frac{\delta}{\delta J_1} + \frac{\delta}{\delta J_2}
    = \frac{\delta}{\delta J^1} - \frac{\delta}{\delta J^2}\,, \qquad
    \frac{\delta}{\delta J^r}
    =
    \frac{\delta}{\delta J_a} = \frac{1}{2}\frac{\delta}{\delta J_1} - \frac{1}{2}\frac{\delta}{\delta J_2}
    = \frac{1}{2}\frac{\delta}{\delta J^1} + \frac{1}{2}\frac{\delta}{\delta J^2}\,.
    \label{eq:leibniz}
\end{align}
From Eqs.~\eqref{eq:prop_12_def}, \eqref{eq:GF}, \eqref{eq:GbarF}, \eqref{eq:prop_ra_def}, and \eqref{eq:leibniz}, we obtain
\begin{align}
    G_{rr}(x,y) &= \frac{1}{4} \qty[ G_\text{F}(x,y) + G_<(x,y) + G_>(x,y) + G_{\bar{\text{F}}}(x,y) ] = \frac{1}{2} \qty[ G_>(x,y) + G_< (x,y)]\,, \\
    \label{eq:Gra_ret}
    G_{ra}(x,y) &= \frac{1}{2} \qty[ G_\text{F}(x,y) - G_<(x,y) + G_>(x,y) - G_{\bar{\text{F}}}(x,y) ] = \theta (x_0 - y_0) \qty[ G_>(x,y) - G_< (x,y)]\,, \\
    G_{ar}(x,y) &= \frac{1}{2} \qty[ G_\text{F}(x,y) + G_<(x,y) - G_>(x,y) - G_{\bar{\text{F}}}(x,y) ] = - \theta (y_0 - x_0) \qty[ G_>(x,y) - G_< (x,y)]\,.
\end{align}

\vspace{5mm}

\noindent
The upper-left component, $G_{rr}$, is referred to as the statistical propagator.
The upper-right/lower-left component, $G_{ra/ar}$, is nonzero only if $x$ resides within the future/past light-cone of $y$ under the assumption of microcausality, $[\varphi(x), \varphi(y)] = 0$ for $(x-y)^2 < 0$.
Hence, $G_{ra/ar}$ is the retarded/advanced propagator respectively.
To make the causal structure manifest, we use the following graphical notation for the Feynman diagram:
\begin{align}
    G_{rr} (x,y) &=
    \begin{tikzpicture}[baseline=(c)]
        \begin{feynman}[inline = (base.c),every blob={/tikz/fill=gray!30,/tikz/inner sep=2pt}]
            \vertex [label=\({\scriptstyle \varphi_r (x)}\)](f1);
            \vertex [right = 1.2 of f1,label=\({\scriptstyle \varphi_r(y)}\)] (v1);
            \vertex [below = 0.1 of f1] (c);
            \draw (f1) -- node[midway]{$\parallel$} (v1);
            \diagram*{
            (f1) -- [anti majorana] (v1),
            };
        \end{feynman}
    \end{tikzpicture}\,, \qquad
    G_{ra} (x,y) =
    \begin{tikzpicture}[baseline=(c)]
        \begin{feynman}[inline = (base.c),every blob={/tikz/fill=gray!30,/tikz/inner sep=2pt}]
            \vertex [label=\({\scriptstyle \varphi_r(x)}\)](f1);
            \vertex [right = 1.2 of f1,label=\({\scriptstyle \varphi_a(y)}\)] (v1);
            \vertex [below = 0.1 of f1] (c);
            \diagram*{
            (f1) -- [anti fermion] (v1),
            };
        \end{feynman}
    \end{tikzpicture}\,, \qquad
    G_{ar}(x,y) =
    \begin{tikzpicture}[baseline=(c)]
        \begin{feynman}[inline = (base.c),every blob={/tikz/fill=gray!30,/tikz/inner sep=2pt}]
            \vertex [label=\({\scriptstyle \varphi_a(x)}\)](f1);
            \vertex [right = 1.2 of f1,label=\({\scriptstyle \varphi_r (y)}\)] (v1);
            \vertex [below = 0.1 of f1] (c);
            \diagram*{
            (f1) -- [fermion] (v1),
            };
        \end{feynman}
    \end{tikzpicture}\,,
    \label{eq:prop_ra}
\end{align}
where the arrow indicates the direction of causation, namely the causal flow.
Basically, the causation flows from $a$-fields to $r$-fields.
The causal flow can be created by $G_{rr}$ but never annihilated by the propagators due to $G_{aa} = 0$.
With this notation, each vertex has an odd number of outgoing causal arrows. 
Moreover, a vertex gains a factor of $1/4$ by increasing the number of the outgoing arrows by two.

Finally, we will provide some constraints originating from the causality.
The first one is intuitively obvious, which just states that the causation cannot flow backward in time.
\begin{eBox}
    \textit{One-wayness of causation}
    \begin{align}
        \label{eq:oneway_causation}
        \text{For any $I_j = r,a$ with $j \in [1,n-1]$,} \quad
        G_{a I_1 \cdots I_{n-1}} (x, y_1, \cdots, y_{n-1}) = 0
        \quad \text{if} \quad
        {}^\forall i \in [1, n-1],~ x^0 > y^0_i\,.
    \end{align}
\end{eBox}
\paragraph*{\textit{Proof.---}}
Suppose that the largest time of $G_{a I_1 \cdots I_{n-1}}$ is $z_0$.
Then, the turning point of the Schwinger--Keldysh contour must be $z_0$ because the back-and-forth time evolution without an operator insertion is trivial, \textit{i.e.,} $\mathcal{U}(t,z_0) \mathcal{U}^\dag(t,z_0) = 1$.
The turning point can be regarded as either the end point of the contour $1$ or the starting point of the contour $2$.
Hence, the operator inserted at the largest time $z_0$ must be the $r$-type operator.
Otherwise the correlator $G_{a I_1 \cdots I_{n-1}}$ vanishes.
This completes the proof of Eq.~\eqref{eq:oneway_causation}.
Note that this proof relies on the fact that the back-and-forth time evolution is unitary since otherwise we do not have the cancellation of the contour by $\mathcal{U}(t,z_0) \mathcal{U}^\dag(t,z_0) = 1$. This property is also used in the proof of the cutting rule given in Sec.~\ref{subsec:cutting_proof}.

\vspace{5mm}

\noindent
Another one also seems intuitively obvious, indicating that the causation cannot be circular.
\begin{eBox}
    \textit{No closed loop of causation}
    \begin{align} 
        \text{Any Feynman diagram involving a closed loop of causal flow vanishes identically.}
    \end{align}
\end{eBox}
\paragraph*{\textit{Proof.---}}
Consider a diagram in a real spacetime.
Suppose that $n$ vertices located at $x_j$ with $j = 1, \cdots, n$ is connected in numbered order by a causal flow from $x_1$ to $x_n$.
In order for this diagram to be non-vanishing, we must have $x_n^0 > x_{n-1}^0 > \cdots > x_1^0$.
A closed causal flow of $x_1 \to x_2 \to \cdots \to x_n \to x_1$ further enforces $x_1^0 > x_n^0$, which contradicts with the earlier requirement.

\vspace{5mm}

Summarizing the properties we have described so far, in the $r/a$ basis, 
one may write down Feynman diagrams in the following steps:
\begin{enumerate}

\item Write down diagrams without any causal flow assignment first.

\item Assign one of the three causal arrows in Eq.~\eqref{eq:prop_ra} for each propagator.

\item Make sure each vertex has an odd number of outgoing arrows and diagrams do not have a closed loop of causal flow.

\item Put a factor $(1/2)^{2n}$ at each vertex if it contains $2n+1$ outgoing arrows.

\end{enumerate}
It is often convenient to define $r$-type and $a$-type operators, $O_r$ and $O_a$, in terms of the number of $a$-fields;
$O_r$/$O_a$ is the operator with even/odd numbers of $a$-type fields, respectively.
Here we sum over all possible assignments of the $r/a$ indices 
as those are most relevant to the amputated diagrams of our interest below. 
For instance, if $O = \varphi^4$, we define
$O_r = \varphi_r^4 + 3 \varphi_r^2 \varphi_a^2/2 + \varphi_a^4/16$ and
$O_a = 4\varphi_r^3 \varphi_a + \varphi_r \varphi_a^3$.
One can show that they satisfy
\begin{align}
	O_r = \frac{1}{2}\left(O_1 + O_2\right),
	\quad
	O_a = O_1 - O_2.
\end{align}

\subsubsection*{Fully retarded function}
In the following, a generalization of the retarded self-energy to the $n$-point vertex function plays an important role.
Let us first recall the definition of the retarded self-energy, which is defined by the amputated diagram of $G_{ra}$ propagator, \textit{i.e.,}
\begin{align}
    \begin{tikzpicture}[baseline=(c)]
        \begin{feynman}[inline = (base.c),every blob={/tikz/fill=gray!30,/tikz/inner sep=2pt}]
            \vertex [label=\({\scriptstyle \varphi_r}\)](f1);
            \vertex [right = 1 of f1, blob, shape=ellipse ,minimum height=0.75cm,minimum width=1.5cm] (v1){$\longleftarrow$};
            \vertex [left = 0.75 cm of v1, square dot] (vl){};
            \vertex [right = 0.75 cm of v1, square dot] (vr){};
            \vertex [right = 1 of vr,label=\({\scriptstyle \varphi_a}\)] (f2);
            \vertex [below = 0.1 of f1] (c);
            \diagram*{
            (vl) -- [fermion] (f1),
            (vr) -- [anti fermion] (f2),
            };
        \end{feynman}
    \end{tikzpicture}
    \,=\,
    \begin{tikzpicture}[baseline=(c)]
        \begin{feynman}[inline = (base.c),every blob={/tikz/fill=gray!30,/tikz/inner sep=2pt}]
            \vertex [label=\({\scriptstyle \varphi_r}\)](f1);
            \vertex [right = 1 of f1,label=\({\scriptstyle \varphi_a}\)] (vl);
            \diagram*{
            (f1) -- [anti fermion] (vl),
            };
        \end{feynman}
    \end{tikzpicture}
    \,\times\,
    \begin{tikzpicture}[baseline=(c)]
        \begin{feynman}[inline = (base.c),every blob={/tikz/fill=gray!30,/tikz/inner sep=2pt}]
            \vertex (f1);
            \vertex [right = 0.4 of f1, blob, shape=ellipse ,minimum height=0.75cm,minimum width=1.5cm] (v1){$\longleftarrow$};
            \vertex [left = 0.75 cm of v1, label={[left]\({\scriptstyle O_r}\)}, square dot] (vl){};
            \vertex [right = 0.75 cm of v1, label={[right]\({\scriptstyle O_a}\)}, square dot] (vr){};
            \vertex [right = 0.4 of vr] (f2);
            \vertex [below = 0.1 of f1] (c);
        \end{feynman}
    \end{tikzpicture}
    \, \times \,
    \begin{tikzpicture}[baseline=(c)]
        \begin{feynman}[inline = (base.c),every blob={/tikz/fill=gray!30,/tikz/inner sep=2pt}]
            \vertex [label=\({\scriptstyle \varphi_r}\)](f1);
            \vertex [right = 1 of f1,label=\({\scriptstyle \varphi_a}\)] (vl);
            \diagram*{
            (f1) -- [anti fermion] (vl),
            };
        \end{feynman}
    \end{tikzpicture}\,.
\end{align}
Here we denote an operator couples to $\varphi$ as $O$ collectively.
To avoid clutter, we use the same notation $O$ for all the operators,
but the discussion below equally applies to the case with distinct operators,
as long as the microcausality holds among them.
The retarded self-energy reads
\begin{align}
    \Pi_{ra} (x,y)
    =
    \begin{tikzpicture}[baseline=(c)]
        \begin{feynman}[inline = (base.c),every blob={/tikz/fill=gray!30,/tikz/inner sep=2pt}]
            \vertex (f1);
            \vertex [right = 0.4 of f1, blob, shape=ellipse ,minimum height=0.75cm,minimum width=1.5cm] (v1){$\longleftarrow$};
            \vertex [left = 0.75 cm of v1, label={[left]\({\scriptstyle O_r}\)}, square dot] (vl){};
            \vertex [right = 0.75 cm of v1, label={[right]\({\scriptstyle O_a}\)}, square dot] (vr){};
            \vertex [right = 0.4 of vr] (f2);
            \vertex [below = 0.1 of f1] (c);
        \end{feynman}
    \end{tikzpicture}
    = \left\langle O_r(x) O_a (y) \right\rangle
    = \theta (x_0 - y_0) \left\langle \qty[ O(x), O(y) ] \right\rangle\,.
    \label{eq:def_2pt_ret}
\end{align}
Recall Eq.~\eqref{eq:Gra_ret} for the last equality.
We expect the microcausality for $O(x)$, \textit{i.e.}, $\qty[O(x),O(y)] = 0$ for $(x - y)^2 < 0$.
Hence, the retarded self-energy fulfills the causal requirements:
\begin{eBox}
    \textit{Causality of the retarded self-energy}
    \begin{align}
        \Pi_{ra} (x,y) = 0
        \quad \text{if} \quad
        x_0 < y_0 \lor (x-y)^2 < 0.
        \label{eq:ret_2pt}
    \end{align}
\end{eBox}

One may extend the notion of the retarded function to the $n$-point vertex, defined by the amputated diagram of $G_{ra\cdots a}$, \textit{i.e.,}
\begin{align}
    \begin{tikzpicture}[baseline=(c)]
        \begin{feynman}[inline = (base.c),every blob={/tikz/fill=gray!30,/tikz/inner sep=2pt}]
            \vertex [label=\({\scriptstyle \varphi_r}\)](f1);
            \vertex [right = 1 of f1, blob ,minimum height=1cm] (v1){$\longleftarrow$};
            \vertex [right = 0.5 cm of v1] (vr);
            \vertex [right = 0.8 of vr] (f2);
            \vertex [right = 0.05 of vr] (fm){};
            \vertex [above = 0.11 of fm] (vdots){$\vdots$};
            \vertex [left = 0.5 cm of v1, square dot] (vl){};
            \vertex [left = 0.1 cm of vr] (vrux);
            \vertex [above = 0.25 cm of vrux, square dot] (vru){};
            \vertex [below = 0.25 cm of vrux, square dot] (vrd){};
            \vertex [above = 0.75 cm of f2,label=\({\scriptstyle \varphi_a}\)] (f2u);
            \vertex [below = 1.5 cm of f2u,label=\({\scriptstyle \varphi_a}\)] (f2d);
            \vertex [below = 0.1 of f1] (c);
            \diagram*{
            (vl) -- [fermion] (f1),
            (vru) -- [anti fermion] (f2u),
            (vrd) -- [anti fermion] (f2d),
            };
        \end{feynman}
    \end{tikzpicture}
    =
    \begin{tikzpicture}[baseline=(c)]
        \begin{feynman}[inline = (base.c),every blob={/tikz/fill=gray!30,/tikz/inner sep=2pt}]
            \vertex [label=\({\scriptstyle \varphi_r}\)](f1);
            \vertex [right = 1 of f1,label=\({\scriptstyle \varphi_a}\)] (vl);
            \diagram*{
            (f1) -- [anti fermion] (vl),
            };
        \end{feynman}
    \end{tikzpicture}
    \,\times\,
    \begin{tikzpicture}[baseline=(c)]
        \begin{feynman}[inline = (base.c),every blob={/tikz/fill=gray!30,/tikz/inner sep=2pt}]
            \vertex [blob ,minimum height=1cm] (v1){$\longleftarrow$};
            \vertex [right = 0.5 cm of v1] (vr);
            \vertex [right = 0.8 of vr] (f2);
            \vertex [right = 0.05 of vr] (fm){};
            \vertex [above = 0.11 of fm] (vdots){$\vdots$};
            \vertex [left = 0.5 cm of v1, label={[left]\({\scriptstyle O_r}\)}, square dot] (vl){};
            \vertex [left = 0.1 cm of vr] (vrux);
            \vertex [above = 0.25 cm of vrux, square dot] (vru){};
            \vertex [above = 0.05 cm of vru, label={[right]\({\scriptstyle O_a}\)}] (vrun){};
            \vertex [below = 0.25 cm of vrux, label={[below right]\({\scriptstyle O_a}\)}, square dot] (vrd){};
            \vertex [below = 0.1 of v1] (c);
        \end{feynman}
    \end{tikzpicture}
    \,\times\,
    \begin{tikzpicture}[baseline=(c)]
        \begin{feynman}[inline = (base.c),every blob={/tikz/fill=gray!30,/tikz/inner sep=2pt}]
            \vertex (vr);
            \vertex [right = 0.8 of vr] (f2);
            \vertex [right = 0.1 of vr] (fm){};
            \vertex [above = 0.11 of fm] (vdots){$\vdots$};
            \vertex [left = 0.1 cm of vr] (vrux);
            \vertex [above = 0.4 cm of vrux, label=\({\scriptstyle \varphi_r}\)] (vru);
            \vertex [below = 0.4 cm of vrux, label=\({\scriptstyle \varphi_r}\)] (vrd);
            \vertex [above = 0.75 cm of f2,label=\({\scriptstyle \varphi_a}\)] (f2u);
            \vertex [below = 1.5 cm of f2u,label=\({\scriptstyle \varphi_a}\)] (f2d);
            \vertex [below = 0.1 of vr] (c);
            \diagram*{
            (vru) -- [anti fermion] (f2u),
            (vrd) -- [anti fermion] (f2d),
            };
        \end{feynman}
    \end{tikzpicture}\,.
\end{align}
The fully retarded $n$-point function is defined by
\begin{align}
    \mathcal{V}_{ra \cdots a} (x,y_1, \cdots , y_{n-1}) \coloneqq
    \begin{tikzpicture}[baseline=(c)]
        \begin{feynman}[inline = (base.c),every blob={/tikz/fill=gray!30,/tikz/inner sep=2pt}]
            \vertex [blob ,minimum height=1cm] (v1){$\longleftarrow$};
            \vertex [right = 0.5 cm of v1] (vr);
            \vertex [right = 0.8 of vr] (f2);
            \vertex [right = 0.05 of vr] (fm){};
            \vertex [above = 0.11 of fm] (vdots){$\vdots$};
            \vertex [left = 0.5 cm of v1, label={[left]\({\scriptstyle O_r (x)}\)}, square dot] (vl){};
            \vertex [left = 0.1 cm of vr] (vrux);
            \vertex [above = 0.25 cm of vrux, square dot] (vru){};
            \vertex [above = 0.05 cm of vru, label={[right]\({\scriptstyle O_a (y_1)}\)}] (vrun){};
            \vertex [below = 0.25 cm of vrux, label={[below right]\({\scriptstyle O_a (y_{n-1})}\)}, square dot] (vrd){};
            \vertex [below = 0.1 of v1] (c);
        \end{feynman}
    \end{tikzpicture}
    =
    \left\langle
        O_r(x) O_a (y_1) \cdots O_a(y_{n-1})
    \right\rangle\,.
\end{align}
Under the assumption of microcausality, $\qty[O(x), O(y)] = 0$ for $(x - y)^2 < 0$, one may show the following causal property for the fully retarded function:
\begin{eBox}
    \textit{Causality of the fully retarded function}
    \begin{align}
        \mathcal{V}_{ra \cdots a} (x,y_1, \cdots , y_{n-1}) = 0
        \quad \text{if} \quad
        {}^\exists i \in [1,n-1] ~\text{s.t.}~ x^0 < y^0_i \lor \qty(x - y_i)^2 < 0\,.
        \label{eq:causality_full_ret}
    \end{align}
\end{eBox}
\paragraph{\textit{Proof.---}}
The first condition is readily obtained from Eq.~\eqref{eq:oneway_causation}, which implies that the largest time must be $x_0$ for a nonzero $\mathcal{V}_{ra \cdots a}$, in other words, ${}^\exists i \in [1,n-1]$ s.t.~$x^0 < y_i^0$, $\mathcal{V}_{ra\cdots a} (x,y_1, \cdots, y_{n-1}) = 0$.

The second condition can be shown by induction.
Let $\hat{\mathcal{V}}$ being an operator product without taking an expectation value: $\hat{\mathcal{V}}_{ra \cdots a} (x, y_1, \cdots, y_{n-1}) \coloneqq O_r(x)O_a (y_1) \cdots O_a (y_{n-1})$.
The statement $P(n)$ that we will show is the following: $\hat{\mathcal{V}}_{ra \cdots a} (x, y_1, \cdots, y_{n-1}) = 0$ if ${}^\exists i \in [1,n-1]$ s.t.~$(x - y_i)^2 < 0$.
For $n=2$, we have already shown $P(2)$ in Eq.~\eqref{eq:def_2pt_ret}.
Suppose that $P(n)$ holds for $n = k$.
Let us prove $P(n)$ in the case of $n = k + 1$. 
For a given $\hat{\mathcal{V}}_{ra \cdots a} (x, y_1, \cdots, y_{k})$, one may relabel $y_1, \cdots, y_{k}$ such that $y_1^0 \geq y_2^0 \geq \cdots \geq y_{k}^0$ without loss of generality.
This implies that $O_1(y_k)$ appears in the right end while $O_2(y_k)$ appears in the left end, which indicates
\begin{align}
    \hat{\mathcal{V}}_{ra \cdots a} (x, y_1, \cdots, y_{k})
    &= \hat{\mathcal{V}}_{ra \cdots a} (x, y_1, \cdots, y_{k-1}) O_a (y_{k}) \nonumber \\
    &= \hat{\mathcal{V}}_{ra \cdots a} (x, y_1, \cdots, y_{k-1}) O (y_{k}) - O (y_{k}) \hat{\mathcal{V}}_{ra \cdots a} (x, y_1, \cdots, y_{k-1})\,.
    \label{eq:pr_causality}
\end{align}
In the second line, we have used the definition of the $a$-type operator: $O_a = O_1 - O_2$.
Owing to the assumption of $P(k)$, the non-trivial case for $P(k + 1)$ is just $(x - y_k)^2 < 0$.
For $(x - y_k)^2 < 0$, if there exists a certain $y_j$ with $j \in [1, k-1]$ s.t.~$(y_k - y_j)^2 \geq 0$, namely, $y_j$ resides within the future light-cone of $y_k$, $x$ is outside the future light-cone of $y_j$, \textit{i.e.,} $(x - y_j)^2 < 0$.\footnote{
	One can show this in Minkowski spacetime or the comoving coordinates in the FLRW background
	purely algebraically. Suppose
	that $(y_k-y_j)^2 \geq 0$ and $(x-y_j)^2 \geq 0$. It then follows that $(x-y_k)^2 \geq (x-y_j)^2 + (y_j - y_k)^2
	+ 2(x^0 - y_j^0)(y_j^0 - y_k^0) - 2\vert \vec{x}-\vec{y}_j\vert \vert \vec{y}_j -\vec{y}_k\vert
	\geq (x-y_j)^2 + (y_j - y_k)^2 \geq 0$, in contradiction with the original assumption.
	Therefore, $(y_k-y_j)^2 < 0$ and/or $(x-y_j)^2 < 0$ if $(x-y_k)^2 < 0$.
}
Then, $\hat{\mathcal{V}}_{ra \cdots a} (x, y_1, \cdots, y_{k-1}) = 0$ owing to the assumption on $P(k)$, and thereby $\hat{\mathcal{V}}_{ra \cdots a} (x, y_1, \cdots, y_{k}) = 0$ as can be seen from Eq.~\eqref{eq:pr_causality}.
On the other hand, if $(y_k - y_j)^2 < 0$ for all $j \in [1, k-1]$, $O(y_k)$ commutes with all the operators in $\hat{\mathcal{V}}_{ra \cdots a} (x, y_1, \cdots, y_{k-1})$ because of the microcausality.
Therefore, the right-hand-side of Eq.~\eqref{eq:pr_causality} vanishes identically as one can move all $O(y_k)$ in the second line to, \textit{e.g.,} the right end.
Since we have shown the condition for the operator product, the same condition for its expectation value holds trivially, which completes the proof.

\subsection{Cutting rule: general proof}
\label{subsec:cutting_proof}

Equipped with the $r/a$ basis, we now state and prove our cutting rule.
We start from the case that the bulk part of the diagram consists of two-point correlators,
and then generalize it to arbitrary $n$-point correlators.

\subsubsection*{Two-point bulk correlator}

Suppose that the bulk and boundary are connected by a single operator such as $\varphi^n {O}_\chi / n !$ with ${O}_\chi$ being a composite operator of bulk fields $\chi$.\footnote{
	This is just for notational simplicity. The discussion equally applies to the case with
	multiple distinct operators connecting the bulk and boundary, and can be generalized to
	the case with derivatives acting on fields, as should be clear from our proof.
}
Let us start with the simplest example where the bulk correlator is solely given by the two-point function of ${O}_\chi$.
The boundary correlator of our interest is the following:
\begin{align}
	\includegraphics[valign=c]{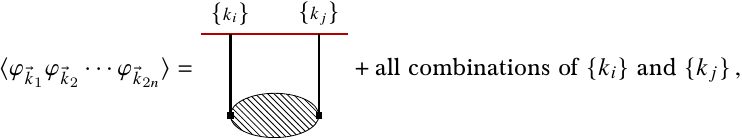}
\end{align}
where the left-hand side is the $(2n)$-point boundary correlator at equal time, the elliptic blob represents the bulk two-point correlator with the square dot being ${O}_\chi$, and a set of $n$ momentum is denoted by $\{k_i\} = \{k_{i_1}, k_{i_2}, \cdots, k_{i_n} \}$.
For notational simplicity, we use the same notation $\varphi$ for all the boundary fields, 
with an application to cosmological correlators of adiabatic modes in mind. 
However, this is not necessary and the discussion below equally applies to the case with multiple fields, 
such as cross-correlators of adiabatic and isocurvature modes.
The black thick line represents the bulk-to-boundary propagators collectively.
For example, in the case of $n = 2$, we have
\begin{align}
	\includegraphics[valign=c]{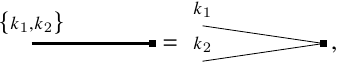}
\end{align}
where the black thin line represents the single bulk-to-boundary propagator.
Also, for $n=2$, we have the following relations of propagators in Keldysh $r/a$ basis:
\begin{align}
	&\includegraphics[valign=c]{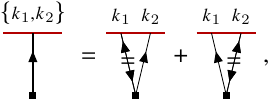}
	\label{eq:a-type_propagator}
	\\
	&\includegraphics[valign=c]{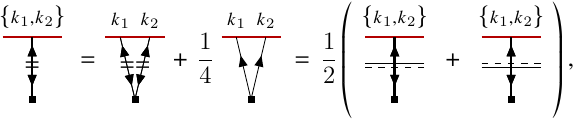}
	\label{eq:r-type_propagator}
\end{align}
where we introduce the following diagrammatic notations for cutting
\begin{align}
	\includegraphics[valign=c]{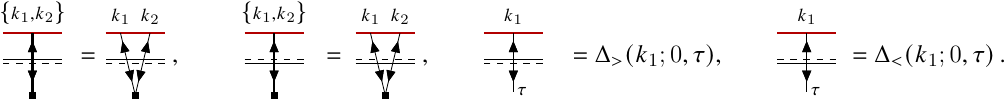}
\end{align}
The solid-dashed double line indicates the cutting where it is replaced with the positive or negative Wightman function depending on the ordering of solid and dashed lines.
We call Eqs.~\eqref{eq:a-type_propagator} and~\eqref{eq:r-type_propagator} 
as the $a$-type and $r$-type bulk-to-boundary propagators, respectively.
Its generalization to $n \geq 3$ is straightforward.

The cutting rule of the two-point bulk correlator can be expressed as
\begin{eBox}
\begin{align}
	\includegraphics[valign=c]{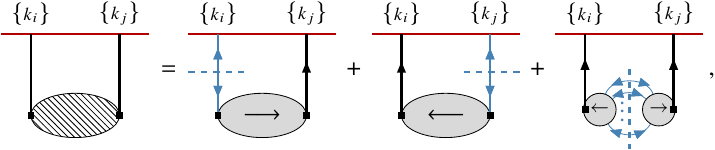}
	\label{eq:cutting_2pt}
\end{align}
\end{eBox}
where the causal flow in the vertex is imprinted as an arrow, representing the fully retarded function.
The cyan cut-diagram is defined by the average of all the possible orderings of the Wightman functions, namely
\begin{align}
	\includegraphics[valign=c]{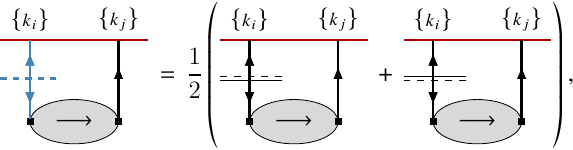}
\end{align}
\begin{align}
	\includegraphics[valign=c]{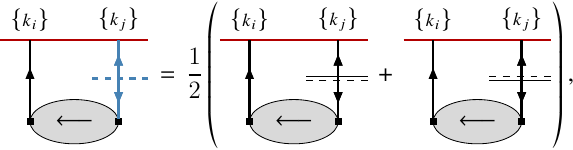}
\end{align}
and
\begin{align}
	\includegraphics[valign=c]{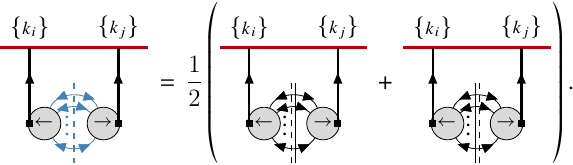}
\end{align}

\begin{figure}[t]
	\centering
	\includegraphics[width=\linewidth]{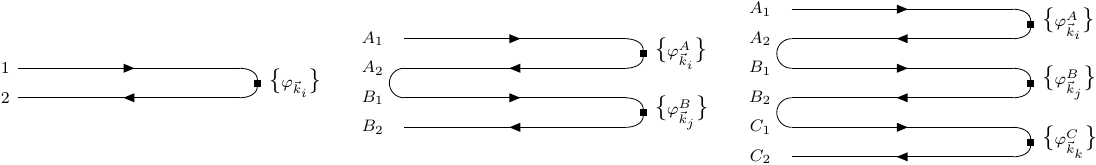}
	\caption{\small \emph{Left:} the standard $1/2$ contour of the in-in formalism.
	\emph{Middle:} the $A/B$ contour for the cutting rule of the two-point bulk correlators.
	\emph{Right:} the $A/B/C$ contour for the cutting rule of the three-point bulk correlators.
	One can duplicate the contour further for higher-point bulk correlators.
	}
	\label{fig:contours}
\end{figure}

\paragraph{\textit{Proof.---}}
To prove the cutting rule, we utilize the same trick as~\cite{Caron-Huot:2007zhp}.
Let us introduce an additional back-and-forth contour on top of the Schwinger--Keldysh contour as shown in the middle in Fig.~\ref{fig:contours}.
Since the equal-time correlator does not depend on the orderings, one may freely assign either $A$ or $B$ to each boundary field.
The result agrees with the original correlator since the time evolution of the $A_2$ and $B_1$ contours 
in Fig.~\ref{fig:contours} cancels with each other, assuming unitary evolution, 
which reduces the contour to the original $1/2$ one.
For our purpose, we consider the following assignment: $\{ \varphi_{\vec{k}_{i_1}}, \cdots, \varphi_{\vec{k}_{i_n}} \}$ on $A$, $\{ \varphi_{\vec{k}_{j_1}}, \cdots, \varphi_{\vec{k}_{j_n}} \}$ on $B$, and visa versa.
Owing to the equal-time nature, the result should be independent of whether each boundary field lies on the back- or forth-contour, which implies that the boundary fields can be regarded as the $r$-type fields for each $A$- and $B$-contour, \textit{e.g.}, $\{ \varphi_{\vec{k}_{i_1}}^A, \cdots, \varphi_{\vec{k}_{i_n}}^A \} = \{ \varphi_{\vec{k}_{i_1}}^{A,r}, \cdots, \varphi_{\vec{k}_{i_n}}^{A,r} \}$.

Suppose that we compute some correlator perturbatively on the $A$ plus $B$ contour.
Since the interactions are local, any interaction vertex can be classified as those living on the contour $A$ denoted by $\mathcal{V}_A$ or on the contour $B$ denoted by $\mathcal{V}_B$.
Consider a set of connected vertices which solely consist of $\mathcal{V}_A$  or $\mathcal{V}_B$.
We refer to it as an island defined by $\mathcal{I}_A \coloneqq \{ \mathcal{V}_A^1, \mathcal{V}_A^2, \cdots \}$, $\mathcal{I}_B \coloneqq \{ \mathcal{V}_B^1, \mathcal{V}_B^2, \cdots \}$.
Among the elements of $\mathcal{I}_{A/B}$, one of them has the largest time argument, which we denote as $\mathcal{V}_{A/B}^\text{max}$.
The resultant contribution of $\mathcal{I}_{A/B}$ does not depend on whether $\mathcal{V}_{A/B,1}^\text{max}$ or $\mathcal{V}_{A/B,2}^\text{max}$, \textit{i.e.,}
\begin{align}
	\includegraphics[valign=c]{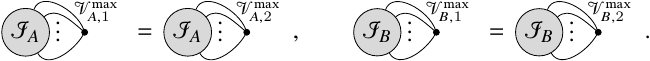}
	\label{eq:tmax_equality}
\end{align}
Owing to the back-and-forth nature, the interaction vertex on $A_1 (B_1)$ has an opposite sign with respect to $A_2 (B_2)$.
Hence, if $\mathcal{V}_{A/B}^\text{max}$ is not connected to the external line, \textit{i.e.,} the bulk-to-boundary propagator, the path-integral contribution of $\mathcal{V}_{A/B,1}^\text{max}$ is cancelled out with $\mathcal{V}_{A/B,2}^\text{max}$ because of the relation \eqref{eq:tmax_equality}.
An immediate consequence of this observation is the absence of isolated island $\mathcal{I}_{A/B}$ that is not connected to the bulk-to-boundary propagator.
Note here again that this fact is originated from the unitary evolution on the contour $A/B$ since the unitarity forces the largest-time operator on each contour to be $r$-type. See also Eq.~\eqref{eq:oneway_causation} and its proof.

Now we are ready to prove the cutting rule.
Let us consider the diagram in the left-hand side of Eq.~\eqref{eq:cutting_2pt} under the following assignment of $\{ \varphi_{\vec{k}_{i_1}}, \cdots, \varphi_{\vec{k}_{i_n}} \}$ on $A$, $\{ \varphi_{\vec{k}_{j_1}}, \cdots, \varphi_{\vec{k}_{j_n}} \}$ on $B$, and visa versa,
\begin{align}
	\label{eq:bulk-2pt_av}
	\includegraphics[valign=c]{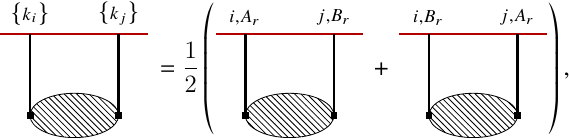}
\end{align}
where we have introduced a shorthand notation $(i, A_r)$ representing $\{ \varphi_{\vec{k}_{i_1}}^{A,r}, \cdots  \varphi_{\vec{k}_{i_n}}^{A,r}\}$.
Since the $B$-fields always appear later than the $A$-fields on the contour, we take the average of two assignments so as to reflect the property of the original correlator which does not have a particular ordering of $\{ \varphi_{\vec{k}_{i_1}}, \cdots, \varphi_{\vec{k}_{i_n}} \}$ and $\{ \varphi_{\vec{k}_{j_1}}, \cdots, \varphi_{\vec{k}_{j_n}} \}$.\footnote{
	This is not mandatory, and we simply obtain another (equally valid) cutting rule
	if we do not symmetrize the assignments.
	We find the symmetrized one more convenient for our purpose, and therefore focus on it in the following.
}

The first diagram in the bracket can be expressed as follows
\begin{align}
	\label{eq:bulk-2pt_AB}
	\includegraphics[valign=c]{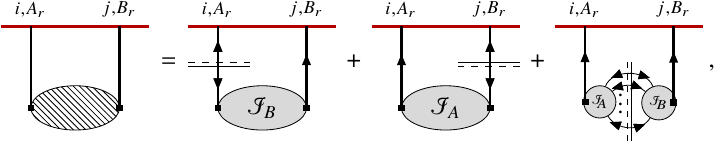}
\end{align}
depending on the location of the boundary between the $A$- and $B$-islands.
In the first term of the right-hand side, there exists a blob of $\mathcal{I}_B$, whose vertex with the largest time has to be connected to $(j, {B_r})$.
The bulk-to-boundary propagator in this case is the retarded one.
The left end of the $\mathcal{I}_B$ blob has to be incoming because of the causal flow, and hence the $\mathcal{I}_B$ blob is the retarded function.
The cut-propagator connecting the bulk (the left end of the $\mathcal{I}_B$ blob) to the boundary ($i,{A_r}$) must be the negative Wightman because the $B$-fields appear after the $A$-fields on the $A$ plus $B$ contour.
The same logic applies to the second diagram, where the cut-propagator connecting the bulk (the right end of the $\mathcal{I}_A$ blob) to the boundary ($j,B_r$) is the positive Wightman, and the $\mathcal{I}_A$ blob is the retarded function.
In the third diagram, the bulk-to-boundary propagators are both retarded.
All the bulk propagators connecting the $\mathcal{I}_A$ blob to the $\mathcal{I}_B$ blob are the positive Wightman functions.
This also implies that all the vertices from the bulk propagators are incoming causal flow, and hence the $\mathcal{I}_{A/B}$ blobs are the fully retarded functions
as the requirement that $\mathcal{V}_{A/B}^\text{max}$ should be connected to the external line indicates.
As an alternative way of understanding this, one may argue the following. 
The cut is at the boundary between the $A/B$-fields.
Since the ordering of the $A/B$-fields is independent of the $1/2$ indices, 
the fields consumed as the Wightman functions are the sum of the $1/2$-fields, or equivalently the $r$-fields.
It is then appropriate to assign the causal arrows incoming to the $\mathcal{I}_{A/B}$ blobs, indicating that these are fully retarded.
In the same way, one can show
\begin{align}
	\label{eq:bulk-2pt_BA}
	\includegraphics[valign=c]{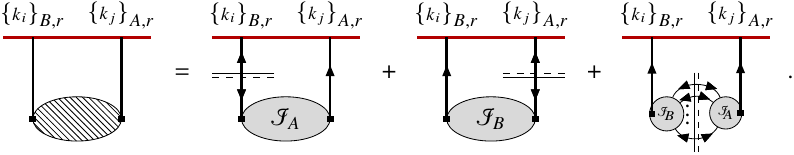}
\end{align}
Note that, in this case \eqref{eq:bulk-2pt_BA}, the cut-diagrams of the positive (negative) Wightman functions in Eq.~\eqref{eq:bulk-2pt_AB} are replaced with the negative (positive) Wightman functions.
Plugging Eqs.~\eqref{eq:bulk-2pt_AB} and \eqref{eq:bulk-2pt_BA} into Eq.~\eqref{eq:bulk-2pt_av}, we complete the proof of Eq.~\eqref{eq:cutting_2pt}.

\subsubsection*{$n$-point bulk correlator}
It is straightforward to generalize the cutting rule to the $n$-point bulk correlator. As an illustration, here we provide the cutting rule for the three-point bulk correlator:
\begin{eBox}
\begin{align}
	\includegraphics[valign=c]{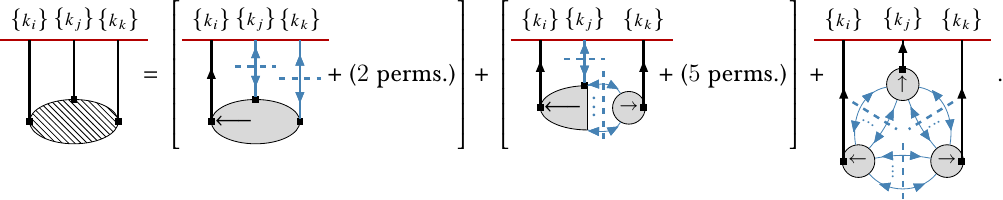}
	\label{eq:cutting_3pt}
\end{align}
\end{eBox}
Again the cyan cut-diagrams are given by the average of all possible cuts.
The first term in the left bracket can be expressed as
\begin{align}
	\label{eq:3pt_i_avcut}
	\includegraphics[valign=c]{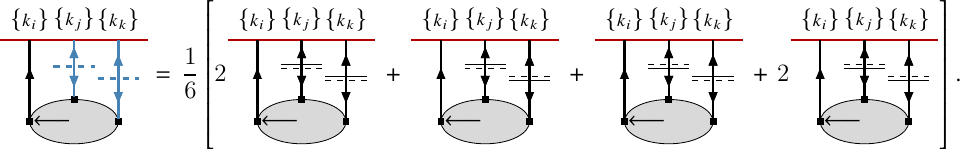}
\end{align}
The first term in the second bracket is given by
\begin{align}
	\includegraphics[valign=c]{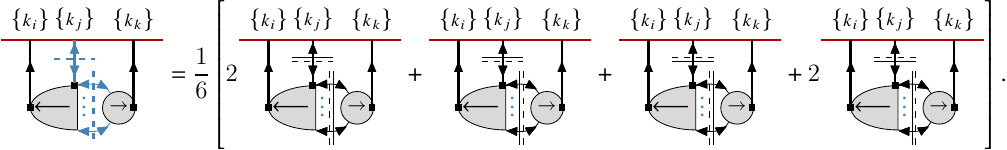}
	\label{eq:3pt_ii_avcut}
\end{align}
The last term is obtained from
\begin{align}
	\includegraphics[valign=c]{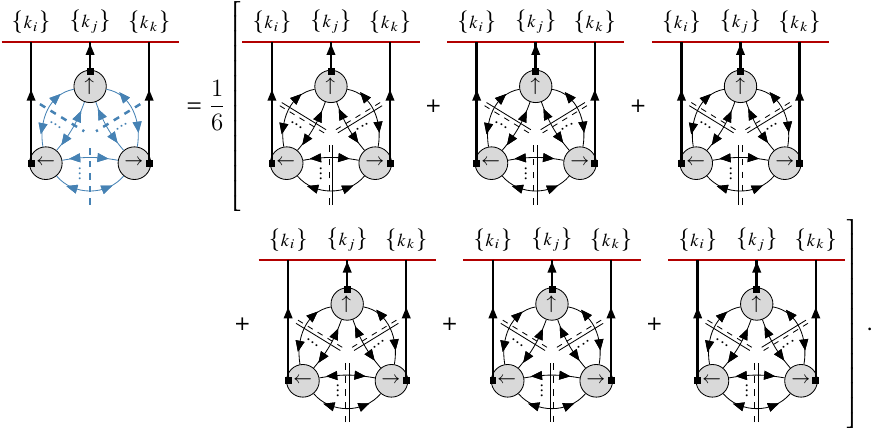}
	\label{eq:3pt_iii_avcut}
\end{align}
All the vertices with an arrow represent the fully retarded functions as can also be confirmed from the causal flow.

\paragraph{\textit{Proof.---}}
One may take a similar strategy as the two-point correlator.
Let us introduce another back-and-fourth contour $C$ on top of $A$ and $B$ (see Fig.~\ref{fig:contours}), and assign $A$, $B$, and $C$ to three sets of boundary fields respectively.\footnote{
	One can derive another form of the cutting rule without introducing $C$, 
	in the same way as the two-point bulk correlators.
	We nevertheless focus on the cutting rule with the additional contour $C$ 
	since it is more suitable for our purpose.
}
As the result should not depend on the ordering of the three sets of boundary fields owing to the equal-time and commuting nature, we take the average of all the possible assignments of $A$, $B$, and $C$ as follows
\begin{align}
	\includegraphics[valign=c]{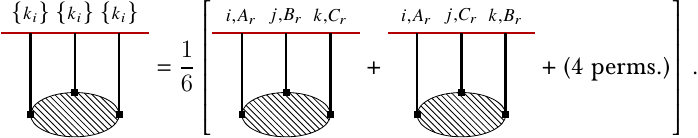}
	\label{eq:pr_3pt_1}
\end{align}
The bulk correlators are classified into the following three cases:
(i) all three vertices lie on the same contour,
(ii) two of them lie on the same contour,
(iii) each vertex lies on different contours.
In case~(i), we have three possibilities; all the vertices live on either the contour $A$, $B$, or $C$.

The first diagram in the bracket of Eq.~\eqref{eq:pr_3pt_1} can be expressed as follows if the three vertices lie on the same contour
\begin{align}
	\includegraphics[valign=c]{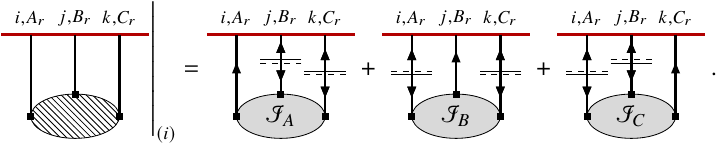}
	\label{eq:3pt_i}
\end{align}
By taking the average over all the permutations of $A$, $B$, $C$, one may express the first diagram in Eq.~\eqref{eq:3pt_i} as
\begin{align}
	\frac{\text{(6 perms. of the $1$st term in \eqref{eq:3pt_i})}}{6}
	&=
	\frac{1}{6} \left[
	\begin{tikzpicture}[baseline=(c)]
		\begin{feynman}[inline = (base.c),every blob={/tikz/fill=gray!30,/tikz/inner sep=2pt}]
			\vertex (f1);
			\vertex [right = 0.5 of f1,label=\({\scriptstyle {i},{A_r}}\)] (c1);
			\vertex [right = 0.75 of c1,label=\({\scriptstyle {j},{B_r}}\)] (m1);
			\vertex [right = 1.5 of c1,label=\({\scriptstyle {k},{C_r}}\)] (c2);
			\vertex [right = 0.5 of c2] (f2);
			\vertex [below = of m1, blob, shape=ellipse ,minimum height=0.75cm,minimum width=1.5cm] (v1){$\mathcal{I}_A$};
			\vertex [below = 0.45 of m1] (ecutt);
			\vertex [left = 0.35 of ecutt] (ecut1);
			\vertex [right = 0.35 of ecutt] (ecut2);
			\vertex [right = 1.5 of c1] (c2);
			\vertex [below = 0.65 of c2] (ecutt2);
			\vertex [left = 0.35 of ecutt2] (ecut3);
			\vertex [right = 0.35 of ecutt2] (ecut4);
			\vertex [above = 0.375 cm of v1, square dot](v3p){};
			\vertex [left = 0.75 cm of v1, square dot](v1p){};
			\vertex [right = 0.75 cm of v1, square dot](v2p){};
			\vertex [below = 0.7 of c1] (c);
			\draw[decoration={dashsoliddouble}, decorate, very thick](ecut2) -- (ecut1);
			\draw[decoration={dashsoliddouble}, decorate, very thick](ecut4) -- (ecut3);
			\diagram*{
			(f1) -- [very thick, darkred] (f2),
			(c1) -- [anti fermion, very thick](v1p),
			(c2) -- [anti majorana,  very thick](v2p),
			(m1) -- [anti majorana,  very thick](v3p),
			};
		\end{feynman}
	\end{tikzpicture}
	+
	\begin{tikzpicture}[baseline=(c)]
		\begin{feynman}[inline = (base.c),every blob={/tikz/fill=gray!30,/tikz/inner sep=2pt}]
			\vertex (f1);
			\vertex [right = 0.5 of f1,label=\({\scriptstyle {i},{A_r}}\)] (c1);
			\vertex [right = 0.75 of c1,label=\({\scriptstyle {j},{C_r}}\)] (m1);
			\vertex [right = 1.5 of c1,label=\({\scriptstyle {k},{B_r}}\)] (c2);
			\vertex [right = 0.5 of c2] (f2);
			\vertex [below = of m1, blob, shape=ellipse ,minimum height=0.75cm,minimum width=1.5cm] (v1){$\mathcal{I}_A$};
			\vertex [below = 0.45 of m1] (ecutt);
			\vertex [left = 0.35 of ecutt] (ecut1);
			\vertex [right = 0.35 of ecutt] (ecut2);
			\vertex [right = 1.5 of c1] (c2);
			\vertex [below = 0.65 of c2] (ecutt2);
			\vertex [left = 0.35 of ecutt2] (ecut3);
			\vertex [right = 0.35 of ecutt2] (ecut4);
			\vertex [above = 0.375 cm of v1, square dot](v3p){};
			\vertex [left = 0.75 cm of v1, square dot](v1p){};
			\vertex [right = 0.75 cm of v1, square dot](v2p){};
			\vertex [below = 0.7 of c1] (c);
			\draw[decoration={dashsoliddouble}, decorate, very thick](ecut2) -- (ecut1);
			\draw[decoration={dashsoliddouble}, decorate, very thick](ecut4) -- (ecut3);
			\diagram*{
			(f1) -- [very thick, darkred] (f2),
			(c1) -- [anti fermion, very thick](v1p),
			(c2) -- [anti majorana,  very thick](v2p),
			(m1) -- [anti majorana,  very thick](v3p),
			};
		\end{feynman}
	\end{tikzpicture}
	+
	\begin{tikzpicture}[baseline=(c)]
		\begin{feynman}[inline = (base.c),every blob={/tikz/fill=gray!30,/tikz/inner sep=2pt}]
			\vertex (f1);
			\vertex [right = 0.5 of f1,label=\({\scriptstyle {i},{B_r}}\)] (c1);
			\vertex [right = 0.75 of c1,label=\({\scriptstyle {j},{C_r}}\)] (m1);
			\vertex [right = 1.5 of c1,label=\({\scriptstyle {k},{A_r}}\)] (c2);
			\vertex [right = 0.5 of c2] (f2);
			\vertex [below = of m1, blob, shape=ellipse ,minimum height=0.75cm,minimum width=1.5cm] (v1){$\mathcal{I}_B$};
			\vertex [below = 0.45 of m1] (ecutt);
			\vertex [left = 0.35 of ecutt] (ecut1);
			\vertex [right = 0.35 of ecutt] (ecut2);
			\vertex [right = 1.5 of c1] (c2);
			\vertex [below = 0.65 of c2] (ecutt2);
			\vertex [left = 0.35 of ecutt2] (ecut3);
			\vertex [right = 0.35 of ecutt2] (ecut4);
			\vertex [above = 0.375 cm of v1, square dot](v3p){};
			\vertex [left = 0.75 cm of v1, square dot](v1p){};
			\vertex [right = 0.75 cm of v1, square dot](v2p){};
			\vertex [below = 0.7 of c1] (c);
			\draw[decoration={dashsoliddouble}, decorate, very thick](ecut2) -- (ecut1);
			\draw[decoration={dashsoliddouble}, decorate, very thick](ecut3) -- (ecut4);
			\diagram*{
			(f1) -- [very thick, darkred] (f2),
			(c1) -- [anti fermion, very thick](v1p),
			(c2) -- [anti majorana,  very thick](v2p),
			(m1) -- [anti majorana,  very thick](v3p),
			};
		\end{feynman}
	\end{tikzpicture}
	\right. \nonumber \\
	&\qquad \left. +
	\begin{tikzpicture}[baseline=(c)]
		\begin{feynman}[inline = (base.c),every blob={/tikz/fill=gray!30,/tikz/inner sep=2pt}]
			\vertex (f1);
			\vertex [right = 0.5 of f1,label=\({\scriptstyle {i},{B_r}}\)] (c1);
			\vertex [right = 0.75 of c1,label=\({\scriptstyle {j},{A_r}}\)] (m1);
			\vertex [right = 1.5 of c1,label=\({\scriptstyle {k},{C_r}}\)] (c2);
			\vertex [right = 0.5 of c2] (f2);
			\vertex [below = of m1, blob, shape=ellipse ,minimum height=0.75cm,minimum width=1.5cm] (v1){$\mathcal{I}_B$};
			\vertex [below = 0.45 of m1] (ecutt);
			\vertex [left = 0.35 of ecutt] (ecut1);
			\vertex [right = 0.35 of ecutt] (ecut2);
			\vertex [right = 1.5 of c1] (c2);
			\vertex [below = 0.65 of c2] (ecutt2);
			\vertex [left = 0.35 of ecutt2] (ecut3);
			\vertex [right = 0.35 of ecutt2] (ecut4);
			\vertex [above = 0.375 cm of v1, square dot](v3p){};
			\vertex [left = 0.75 cm of v1, square dot](v1p){};
			\vertex [right = 0.75 cm of v1, square dot](v2p){};
			\vertex [below = 0.7 of c1] (c);
			\draw[decoration={dashsoliddouble}, decorate, very thick](ecut1) -- (ecut2);
			\draw[decoration={dashsoliddouble}, decorate, very thick](ecut4) -- (ecut3);
			\diagram*{
			(f1) -- [very thick, darkred] (f2),
			(c1) -- [anti fermion, very thick](v1p),
			(c2) -- [anti majorana,  very thick](v2p),
			(m1) -- [anti majorana,  very thick](v3p),
			};
		\end{feynman}
	\end{tikzpicture}
	+
	\begin{tikzpicture}[baseline=(c)]
		\begin{feynman}[inline = (base.c),every blob={/tikz/fill=gray!30,/tikz/inner sep=2pt}]
			\vertex (f1);
			\vertex [right = 0.5 of f1,label=\({\scriptstyle {i},{C_r}}\)] (c1);
			\vertex [right = 0.75 of c1,label=\({\scriptstyle {j},{A_r}}\)] (m1);
			\vertex [right = 1.5 of c1,label=\({\scriptstyle {k},{B_r}}\)] (c2);
			\vertex [right = 0.5 of c2] (f2);
			\vertex [below = of m1, blob, shape=ellipse ,minimum height=0.75cm,minimum width=1.5cm] (v1){$\mathcal{I}_C$};
			\vertex [below = 0.45 of m1] (ecutt);
			\vertex [left = 0.35 of ecutt] (ecut1);
			\vertex [right = 0.35 of ecutt] (ecut2);
			\vertex [right = 1.5 of c1] (c2);
			\vertex [below = 0.65 of c2] (ecutt2);
			\vertex [left = 0.35 of ecutt2] (ecut3);
			\vertex [right = 0.35 of ecutt2] (ecut4);
			\vertex [above = 0.375 cm of v1, square dot](v3p){};
			\vertex [left = 0.75 cm of v1, square dot](v1p){};
			\vertex [right = 0.75 cm of v1, square dot](v2p){};
			\vertex [below = 0.7 of c1] (c);
			\draw[decoration={dashsoliddouble}, decorate, very thick](ecut1) -- (ecut2);
			\draw[decoration={dashsoliddouble}, decorate, very thick](ecut3) -- (ecut4);
			\diagram*{
			(f1) -- [very thick, darkred] (f2),
			(c1) -- [anti fermion, very thick](v1p),
			(c2) -- [anti majorana,  very thick](v2p),
			(m1) -- [anti majorana,  very thick](v3p),
			};
		\end{feynman}
	\end{tikzpicture}
	+
	\begin{tikzpicture}[baseline=(c)]
		\begin{feynman}[inline = (base.c),every blob={/tikz/fill=gray!30,/tikz/inner sep=2pt}]
			\vertex (f1);
			\vertex [right = 0.5 of f1,label=\({\scriptstyle {i},{C_r}}\)] (c1);
			\vertex [right = 0.75 of c1,label=\({\scriptstyle {j},{B_r}}\)] (m1);
			\vertex [right = 1.5 of c1,label=\({\scriptstyle {k},{A_r}}\)] (c2);
			\vertex [right = 0.5 of c2] (f2);
			\vertex [below = of m1, blob, shape=ellipse ,minimum height=0.75cm,minimum width=1.5cm] (v1){$\mathcal{I}_C$};
			\vertex [below = 0.45 of m1] (ecutt);
			\vertex [left = 0.35 of ecutt] (ecut1);
			\vertex [right = 0.35 of ecutt] (ecut2);
			\vertex [right = 1.5 of c1] (c2);
			\vertex [below = 0.65 of c2] (ecutt2);
			\vertex [left = 0.35 of ecutt2] (ecut3);
			\vertex [right = 0.35 of ecutt2] (ecut4);
			\vertex [above = 0.375 cm of v1, square dot](v3p){};
			\vertex [left = 0.75 cm of v1, square dot](v1p){};
			\vertex [right = 0.75 cm of v1, square dot](v2p){};
			\vertex [below = 0.7 of c1] (c);
			\draw[decoration={dashsoliddouble}, decorate, very thick](ecut1) -- (ecut2);
			\draw[decoration={dashsoliddouble}, decorate, very thick](ecut3) -- (ecut4);
			\diagram*{
			(f1) -- [very thick, darkred] (f2),
			(c1) -- [anti fermion, very thick](v1p),
			(c2) -- [anti majorana,  very thick](v2p),
			(m1) -- [anti majorana,  very thick](v3p),
			};
		\end{feynman}
	\end{tikzpicture}
	\right]\,.
	\label{eq:pr_3pt_i_avcut}
\end{align}
One may readily see from the causal flows that all the blobs are the fully retarded functions, and hence, Eq.~\eqref{eq:3pt_i_avcut} coincides with Eq.~\eqref{eq:pr_3pt_i_avcut}.
The second and third terms in Eq.~\eqref{eq:3pt_i} can be obtained by replacing the first and the second (the first and the third) bulk-to-boundary propagators, which is responsible for the ($2$ perms.) in the first bracket of Eq.~\eqref{eq:cutting_3pt}.

The first diagram of Eq.~\eqref{eq:pr_3pt_1} for the case (ii) can be expressed as
\begin{align}
	\includegraphics[valign=c]{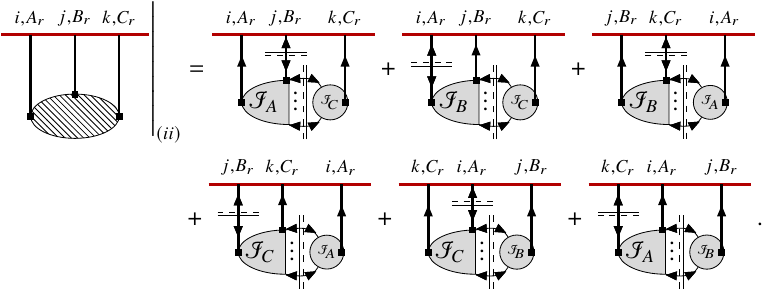}
	\label{eq:3pt_ii}
\end{align}
By taking the summation over all the permutations of $A,B,C$, we obtain the following expression for the first term of Eq.~\eqref{eq:3pt_ii}
\begin{align}
	\frac{\text{(6 perms. of the $1$st term in \eqref{eq:3pt_ii})}}{6}
	&=
	\frac{1}{6} \left[
	\begin{tikzpicture}[baseline=(c)]
		\begin{feynman}[inline = (base.c),every blob={/tikz/fill=gray!30,/tikz/inner sep=2pt}]
			\vertex (f1);
			\vertex [right = 0.5 of f1, label=\({\scriptstyle i,A_r}\)] (c1);
			\vertex [right = 0.75 of c1,label=\({\scriptstyle j,B_r}\)] (m1);
			\vertex [right = 1.75 of c1, label=\({\scriptstyle k,C_r}\)] (c2);
			\vertex [below = 0.325 of m1] (ecutt);
			\vertex [left = 0.35 of ecutt] (ecut1);
			\vertex [right = 0.35 of ecutt] (ecut2);
			\vertex [right = 0.5 of c2] (f2);
			\vertex [below = of m1] (v1){};
			\vertex [above = 0.375 cm of v1](v3p);
			\vertex [right = 0.05 cm of v3p](v3pp);
			\vertex [below = 0.75 cm of v3pp](v4p);
			\vertex [below = 0.7 of c1] (c);
			\draw[fill=gray!30] (v4p) -- (v3pp) arc (90:270:0.8cm and 0.375cm) -- cycle;
			\vertex [right = 0.75 of v1, blob, minimum height=0.5cm,minimum width=0.5cm] (v2){${\scriptscriptstyle \mathcal{I}_{\! \scriptscriptstyle C}}$};
			\vertex [above = 0.375 cm of v1, square dot](v3ppp){};
			\vertex [right = 0.32 cm of v3ppp](vcutt){};
			\vertex [right = 0.16 cm of v1](vdott){};
			\vertex [above = 0.12 cm of vdott](vdot){$\vdots$};
			\vertex [above = 0.25 cm of vcutt](vcut1);
			\vertex [below = 1.25 cm of vcut1](vcut2);
			\vertex [left = 0.75 cm of v1, square dot](v1p){};
			\vertex [right = 1 of v1, square dot](v2p){};
			\vertex [right = 0.375 cm of v1p](bname){$\mathcal{I}_A$};
			\draw[decoration={dashsoliddouble}, decorate, very thick](ecut2) -- (ecut1);
			\draw[decoration={dashsoliddouble}, decorate, very thick](vcut2) -- (vcut1);
			\diagram*{
			(f1) -- [very thick, darkred] (f2),
			(c1) -- [anti fermion, very thick](v1p),
			(c2) -- [anti fermion,  very thick](v2p),
			(m1) -- [anti majorana,  very thick](v3ppp),
			(v3pp) -- [anti majorana, out=10, in=120](v2),
			(v4p) -- [anti majorana, out=350, in=240](v2),
			};
		\end{feynman}
	\end{tikzpicture}
	+
	\begin{tikzpicture}[baseline=(c)]
		\begin{feynman}[inline = (base.c),every blob={/tikz/fill=gray!30,/tikz/inner sep=2pt}]
			\vertex (f1);
			\vertex [right = 0.5 of f1, label=\({\scriptstyle i,A_r}\)] (c1);
			\vertex [right = 0.75 of c1,label=\({\scriptstyle j,C_r}\)] (m1);
			\vertex [right = 1.75 of c1, label=\({\scriptstyle k,B_r}\)] (c2);
			\vertex [below = 0.325 of m1] (ecutt);
			\vertex [left = 0.35 of ecutt] (ecut1);
			\vertex [right = 0.35 of ecutt] (ecut2);
			\vertex [right = 0.5 of c2] (f2);
			\vertex [below = of m1] (v1){};
			\vertex [above = 0.375 cm of v1](v3p);
			\vertex [right = 0.05 cm of v3p](v3pp);
			\vertex [below = 0.75 cm of v3pp](v4p);
			\vertex [below = 0.7 of c1] (c);
			\draw[fill=gray!30] (v4p) -- (v3pp) arc (90:270:0.8cm and 0.375cm) -- cycle;
			\vertex [right = 0.75 of v1, blob, minimum height=0.5cm,minimum width=0.5cm] (v2){${\scriptscriptstyle \mathcal{I}_{\! \scriptscriptstyle B}}$};
			\vertex [above = 0.375 cm of v1, square dot](v3ppp){};
			\vertex [right = 0.32 cm of v3ppp](vcutt){};
			\vertex [right = 0.16 cm of v1](vdott){};
			\vertex [above = 0.12 cm of vdott](vdot){$\vdots$};
			\vertex [above = 0.25 cm of vcutt](vcut1);
			\vertex [below = 1.25 cm of vcut1](vcut2);
			\vertex [left = 0.75 cm of v1, square dot](v1p){};
			\vertex [right = 1 of v1, square dot](v2p){};
			\vertex [right = 0.375 cm of v1p](bname){$\mathcal{I}_A$};
			\draw[decoration={dashsoliddouble}, decorate, very thick](ecut2) -- (ecut1);
			\draw[decoration={dashsoliddouble}, decorate, very thick](vcut2) -- (vcut1);
			\diagram*{
			(f1) -- [very thick, darkred] (f2),
			(c1) -- [anti fermion, very thick](v1p),
			(c2) -- [anti fermion,  very thick](v2p),
			(m1) -- [anti majorana,  very thick](v3ppp),
			(v3pp) -- [anti majorana, out=10, in=120](v2),
			(v4p) -- [anti majorana, out=350, in=240](v2),
			};
		\end{feynman}
	\end{tikzpicture}
	+
	\begin{tikzpicture}[baseline=(c)]
		\begin{feynman}[inline = (base.c),every blob={/tikz/fill=gray!30,/tikz/inner sep=2pt}]
			\vertex (f1);
			\vertex [right = 0.5 of f1, label=\({\scriptstyle i,B_r}\)] (c1);
			\vertex [right = 0.75 of c1,label=\({\scriptstyle j,C_r}\)] (m1);
			\vertex [right = 1.75 of c1, label=\({\scriptstyle k,A_r}\)] (c2);
			\vertex [below = 0.325 of m1] (ecutt);
			\vertex [left = 0.35 of ecutt] (ecut1);
			\vertex [right = 0.35 of ecutt] (ecut2);
			\vertex [right = 0.5 of c2] (f2);
			\vertex [below = of m1] (v1){};
			\vertex [above = 0.375 cm of v1](v3p);
			\vertex [right = 0.05 cm of v3p](v3pp);
			\vertex [below = 0.75 cm of v3pp](v4p);
			\vertex [below = 0.7 of c1] (c);
			\draw[fill=gray!30] (v4p) -- (v3pp) arc (90:270:0.8cm and 0.375cm) -- cycle;
			\vertex [right = 0.75 of v1, blob, minimum height=0.5cm,minimum width=0.5cm] (v2){${\scriptscriptstyle \mathcal{I}_{\! \scriptscriptstyle A}}$};
			\vertex [above = 0.375 cm of v1, square dot](v3ppp){};
			\vertex [right = 0.32 cm of v3ppp](vcutt){};
			\vertex [right = 0.16 cm of v1](vdott){};
			\vertex [above = 0.12 cm of vdott](vdot){$\vdots$};
			\vertex [above = 0.25 cm of vcutt](vcut1);
			\vertex [below = 1.25 cm of vcut1](vcut2);
			\vertex [left = 0.75 cm of v1, square dot](v1p){};
			\vertex [right = 1 of v1, square dot](v2p){};
			\vertex [right = 0.375 cm of v1p](bname){$\mathcal{I}_B$};
			\draw[decoration={dashsoliddouble}, decorate, very thick](ecut2) -- (ecut1);
			\draw[decoration={dashsoliddouble}, decorate, very thick](vcut1) -- (vcut2);
			\diagram*{
			(f1) -- [very thick, darkred] (f2),
			(c1) -- [anti fermion, very thick](v1p),
			(c2) -- [anti fermion,  very thick](v2p),
			(m1) -- [anti majorana,  very thick](v3ppp),
			(v3pp) -- [anti majorana, out=10, in=120](v2),
			(v4p) -- [anti majorana, out=350, in=240](v2),
			};
		\end{feynman}
	\end{tikzpicture}
	\right. \nonumber \\
	& \left. \qquad +
	\begin{tikzpicture}[baseline=(c)]
		\begin{feynman}[inline = (base.c),every blob={/tikz/fill=gray!30,/tikz/inner sep=2pt}]
			\vertex (f1);
			\vertex [right = 0.5 of f1, label=\({\scriptstyle i,B_r}\)] (c1);
			\vertex [right = 0.75 of c1,label=\({\scriptstyle j,A_r}\)] (m1);
			\vertex [right = 1.75 of c1, label=\({\scriptstyle k,C_r}\)] (c2);
			\vertex [below = 0.325 of m1] (ecutt);
			\vertex [left = 0.35 of ecutt] (ecut1);
			\vertex [right = 0.35 of ecutt] (ecut2);
			\vertex [right = 0.5 of c2] (f2);
			\vertex [below = of m1] (v1){};
			\vertex [above = 0.375 cm of v1](v3p);
			\vertex [right = 0.05 cm of v3p](v3pp);
			\vertex [below = 0.75 cm of v3pp](v4p);
			\vertex [below = 0.7 of c1] (c);
			\draw[fill=gray!30] (v4p) -- (v3pp) arc (90:270:0.8cm and 0.375cm) -- cycle;
			\vertex [right = 0.75 of v1, blob, minimum height=0.5cm,minimum width=0.5cm] (v2){${\scriptscriptstyle \mathcal{I}_{\! \scriptscriptstyle C}}$};
			\vertex [above = 0.375 cm of v1, square dot](v3ppp){};
			\vertex [right = 0.32 cm of v3ppp](vcutt){};
			\vertex [right = 0.16 cm of v1](vdott){};
			\vertex [above = 0.12 cm of vdott](vdot){$\vdots$};
			\vertex [above = 0.25 cm of vcutt](vcut1);
			\vertex [below = 1.25 cm of vcut1](vcut2);
			\vertex [left = 0.75 cm of v1, square dot](v1p){};
			\vertex [right = 1 of v1, square dot](v2p){};
			\vertex [right = 0.375 cm of v1p](bname){$\mathcal{I}_B$};
			\draw[decoration={dashsoliddouble}, decorate, very thick](ecut1) -- (ecut2);
			\draw[decoration={dashsoliddouble}, decorate, very thick](vcut2) -- (vcut1);
			\diagram*{
			(f1) -- [very thick, darkred] (f2),
			(c1) -- [anti fermion, very thick](v1p),
			(c2) -- [anti fermion,  very thick](v2p),
			(m1) -- [anti majorana,  very thick](v3ppp),
			(v3pp) -- [anti majorana, out=10, in=120](v2),
			(v4p) -- [anti majorana, out=350, in=240](v2),
			};
		\end{feynman}
	\end{tikzpicture}
	+
	\begin{tikzpicture}[baseline=(c)]
		\begin{feynman}[inline = (base.c),every blob={/tikz/fill=gray!30,/tikz/inner sep=2pt}]
			\vertex (f1);
			\vertex [right = 0.5 of f1, label=\({\scriptstyle i,C_r}\)] (c1);
			\vertex [right = 0.75 of c1,label=\({\scriptstyle j,A_r}\)] (m1);
			\vertex [right = 1.75 of c1, label=\({\scriptstyle k,B_r}\)] (c2);
			\vertex [below = 0.325 of m1] (ecutt);
			\vertex [left = 0.35 of ecutt] (ecut1);
			\vertex [right = 0.35 of ecutt] (ecut2);
			\vertex [right = 0.5 of c2] (f2);
			\vertex [below = of m1] (v1){};
			\vertex [above = 0.375 cm of v1](v3p);
			\vertex [right = 0.05 cm of v3p](v3pp);
			\vertex [below = 0.75 cm of v3pp](v4p);
			\vertex [below = 0.7 of c1] (c);
			\draw[fill=gray!30] (v4p) -- (v3pp) arc (90:270:0.8cm and 0.375cm) -- cycle;
			\vertex [right = 0.75 of v1, blob, minimum height=0.5cm,minimum width=0.5cm] (v2){${\scriptscriptstyle \mathcal{I}_{\! \scriptscriptstyle B}}$};
			\vertex [above = 0.375 cm of v1, square dot](v3ppp){};
			\vertex [right = 0.32 cm of v3ppp](vcutt){};
			\vertex [right = 0.16 cm of v1](vdott){};
			\vertex [above = 0.12 cm of vdott](vdot){$\vdots$};
			\vertex [above = 0.25 cm of vcutt](vcut1);
			\vertex [below = 1.25 cm of vcut1](vcut2);
			\vertex [left = 0.75 cm of v1, square dot](v1p){};
			\vertex [right = 1 of v1, square dot](v2p){};
			\vertex [right = 0.375 cm of v1p](bname){$\mathcal{I}_C$};
			\draw[decoration={dashsoliddouble}, decorate, very thick](ecut1) -- (ecut2);
			\draw[decoration={dashsoliddouble}, decorate, very thick](vcut1) -- (vcut2);
			\diagram*{
			(f1) -- [very thick, darkred] (f2),
			(c1) -- [anti fermion, very thick](v1p),
			(c2) -- [anti fermion,  very thick](v2p),
			(m1) -- [anti majorana,  very thick](v3ppp),
			(v3pp) -- [anti majorana, out=10, in=120](v2),
			(v4p) -- [anti majorana, out=350, in=240](v2),
			};
		\end{feynman}
	\end{tikzpicture}
	+
	\begin{tikzpicture}[baseline=(c)]
		\begin{feynman}[inline = (base.c),every blob={/tikz/fill=gray!30,/tikz/inner sep=2pt}]
			\vertex (f1);
			\vertex [right = 0.5 of f1, label=\({\scriptstyle i,C_r}\)] (c1);
			\vertex [right = 0.75 of c1,label=\({\scriptstyle j,B_r}\)] (m1);
			\vertex [right = 1.75 of c1, label=\({\scriptstyle k,A_r}\)] (c2);
			\vertex [below = 0.325 of m1] (ecutt);
			\vertex [left = 0.35 of ecutt] (ecut1);
			\vertex [right = 0.35 of ecutt] (ecut2);
			\vertex [right = 0.5 of c2] (f2);
			\vertex [below = of m1] (v1){};
			\vertex [above = 0.375 cm of v1](v3p);
			\vertex [right = 0.05 cm of v3p](v3pp);
			\vertex [below = 0.75 cm of v3pp](v4p);
			\vertex [below = 0.7 of c1] (c);
			\draw[fill=gray!30] (v4p) -- (v3pp) arc (90:270:0.8cm and 0.375cm) -- cycle;
			\vertex [right = 0.75 of v1, blob, minimum height=0.5cm,minimum width=0.5cm] (v2){${\scriptscriptstyle \mathcal{I}_{\! \scriptscriptstyle B}}$};
			\vertex [above = 0.375 cm of v1, square dot](v3ppp){};
			\vertex [right = 0.32 cm of v3ppp](vcutt){};
			\vertex [right = 0.16 cm of v1](vdott){};
			\vertex [above = 0.12 cm of vdott](vdot){$\vdots$};
			\vertex [above = 0.25 cm of vcutt](vcut1);
			\vertex [below = 1.25 cm of vcut1](vcut2);
			\vertex [left = 0.75 cm of v1, square dot](v1p){};
			\vertex [right = 1 of v1, square dot](v2p){};
			\vertex [right = 0.375 cm of v1p](bname){$\mathcal{I}_C$};
			\draw[decoration={dashsoliddouble}, decorate, very thick](ecut1) -- (ecut2);
			\draw[decoration={dashsoliddouble}, decorate, very thick](vcut1) -- (vcut2);
			\diagram*{
			(f1) -- [very thick, darkred] (f2),
			(c1) -- [anti fermion, very thick](v1p),
			(c2) -- [anti fermion,  very thick](v2p),
			(m1) -- [anti majorana,  very thick](v3ppp),
			(v3pp) -- [anti majorana, out=10, in=120](v2),
			(v4p) -- [anti majorana, out=350, in=240](v2),
			};
		\end{feynman}
	\end{tikzpicture}
	\right]\,.
	\label{eq:pr_3pt_ii_avcut}
\end{align}
Again, all the vertices are fully retarded as can be seen from the causal flow, and hence Eq.~\eqref{eq:pr_3pt_ii_avcut} coincides with Eq.~\eqref{eq:3pt_ii_avcut}.
The average over permutations of $A,B,C$ for the remaining each term in Eqs.~\eqref{eq:3pt_ii} can be obtained from an appropriate permutations of Eq.~\eqref{eq:pr_3pt_ii_avcut} with respect to $i,j,k$.

For the case (iii), the first diagram of Eq.~\eqref{eq:pr_3pt_1} is
\begin{align}
		\includegraphics[valign=c]{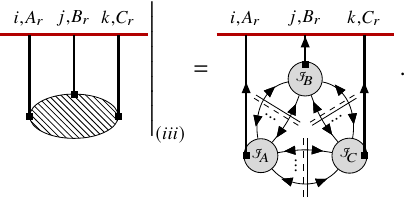}
		\label{eq:3pt_iii}
\end{align}
By taking the average over the permutations of $A,B,C$ for Eq.~\eqref{eq:3pt_iii}, we obtain Eq.~\eqref{eq:3pt_iii_avcut}.
One can generalize our proof to arbitrary $n$-point bulk correlators by further duplicating the contours.

\subsubsection*{Comparison with previous literature}

Cutting rules on cosmological correlators are discussed in previous literature.
In particular, here we compare our cutting rule with the one based on mapping the in-in formalism
to the in-out formalism~\cite{Donath:2024utn} since both directly apply to the correlators.
For a brief discussion on the comparison with other cutting rules in literature, 
see the end of Sec.~\ref{subsec:microcausality}.

In~\cite{Donath:2024utn}, 
it was argued that the in-in correlators can be equally calculated by the in-out formalism
as long as there is no dissipation. The authors then derived a cutting rule for the correlators
based on the largest time equation widely used in the in-out formalism.
Focusing on the examples discussed in this reference, our cutting rule essentially agrees with theirs
under the assumption of the equivalence between the in-in and in-out formalisms.
In fact, one may map our $A$-contour to the in-out contour, 
and $B$-contour to the ``out-in'' contour, respectively.
It then follows that the boundary between our $A$ and $B$ regions corresponds
to the boundary between the $T/\bar{T}$-product regions in~\cite{Donath:2024utn},
which identifies our cut with theirs.
Aside from this equivalence in the simple examples,
we have clarified several points not discussed in~\cite{Donath:2024utn}, \emph{i.e.},
(1) our cutting rule clarifies the causal structure of the diagram, in particular that the sub-diagrams separated by the cuts are the fully retarded functions,
(2) our derivation applies to diagrams with general topologies while~\cite{Donath:2024utn} considered only diagrams with simple topologies and had technical difficulties 
in generalizing it,
and (3) we have discussed generalization to $n$-point bulk correlators 
by further duplicating the in-in contours 
(in this sense, the largest time equation with only the $T/\bar{T}$-products duplicates the contour just once).
Therefore, one may view our cutting rule as a generalization of theirs,
with a diagrammatic derivation that allows immediate extention to 
arbitrary topologies and $n$-point bulk correlators.


\subsection{Cutting rule: explicit checks with examples}
\label{subsec:cutting_examples}

In Sec.~\ref{subsec:cutting_proof}, we have derived the cutting rule of equal-time in-in correlators.
Our proof is quite general and the cutting rule holds for general theories with any particle contents
and (local) interactions, 
as we rely only on minimal assumptions such as the unitarity and causal structure of the in-in formalism.
At the same time, this may leave an impression that our proof is somewhat abstract
and perhaps hard to believe for skeptical readers.
Therefore, in this subsection, we consider several examples.
Our goal here is to convince (possible) skeptical readers of the validity of our cutting rule,
and hence we start from the in-in Feynman rules in the $r/a$ basis
and confirm that the cutting rule indeed holds for these diagrams.

Before starting our discussion, we emphasize that the following calculation
crucially depends on unitarity as it relies on the Feynman rules discussed
in~\ref{subsec:Keldysh_basis}, which in turn relies on unitarity.
This is expected since our general proof in Sec.~\ref{subsec:cutting_proof}
relies on the cancellation of the $A_2$- and $B_1$-contours under the assumption of unitary evolution.
For example, we assume that an odd number of arrows is outgoing from each vertex in the following,
and this is true only for a unitary theory.
To illustrate this point, let us consider a quartic interaction with a complex coupling $\lambda$.
The in-in action is given by
\begin{align}
	\frac{\lambda}{4!} \phi_1^4 - \frac{\lambda^*}{4!} \phi_2^4.
\end{align}
If the theory is unitary, $\lambda = \lambda^*$,
this interaction is odd under $\phi_1 \leftrightarrow \phi_2$,
or in other words, it contains only odd numbers of $\phi_a$.
However, if $\lambda \neq \lambda^*$, this interaction is given by
\begin{align}
	\frac{\lambda + \lambda^*}{12}\left(\phi_r^3 \phi_a + \frac{1}{4}\phi_r \phi_a^3\right)
	+ \frac{\lambda - \lambda^*}{24}\left(\phi_r^4 + \frac{3}{2}\phi_r^2 \phi_a^2 + \frac{1}{16}\phi_a^4\right),
\end{align}
which contains both even and odd numbers of $\phi_a$.\footnote{
	One may instead define the $2$-contour as the inverse of the $1$-contour, 
	not the hermitian conjugate, to extend our cutting rule to a non-hermitian case,
	although this deviates from the in-in formalism as the bra and ket states are no longer
	hermitian conjugate of each other.
	We do not explore this possibility any further in the current study.
} In the following, we focus on the unitary theory so that the rules in Sec.~\ref{subsec:Keldysh_basis} apply.

\subsubsection*{Melon diagrams}

Our first example is the so-called melon diagrams, given by
\begin{align}
	\langle \varphi_{\vec{k}_1}\varphi_{\vec{k}_2} \cdots \varphi_{\vec{k}_{n}}
	\rangle
	=
	\begin{tikzpicture}[baseline=(v1)]
	\begin{feynman}[inline = (base.v1)]
		\vertex (f1);
		\vertex [right = 0.5 of f1] (c1);
		\vertex [right = 0.75 of c1] (m1);
		\vertex [right = 1.5 of c1] (c2);
		\vertex [right = 0.5 of c2] (f2);
		\vertex [below = 1.25 of c1, square dot] (v1){};
		\vertex [right = 1.5 of v1, square dot] (v2){};
		\vertex [right = 0.75 of v1] (c);
		\vertex [right = 0.75 of v1] (d2);
		\node [above = -0.3055 of d2] {\vdots};
		\diagram*{
		(f1) -- [very thick, darkred] (f2),
		(c1) -- [very thick] (v1),
		(c2) -- [very thick] (v2),
		(v1) -- [half right] (v2) -- [half right] (v1),
		(v1) -- [out=290, in=250] (v2),
		(v1) -- [out=70, in=110] (v2),
		};
	\end{feynman}
	\end{tikzpicture},
\end{align}
where the dots indicate an arbitrary number of propagators that connect the two vertices.
This diagram vanishes if the bulk-to-boundary propagators are both $r$-type 
since it necessarily has a closed loop of causal flow.
The contributions with one $r$-type bulk-to-boundary propagator correspond to cutting that propagator,
and the remaining contribution is with two $a$-type bulk-to-boundary propagators.
For this contribution, we have an even number of the internal retarded or advanced propagators connecting the two vertices,
as we need an odd number of the outgoing arrow at each vertex (see Sec.~\ref{subsec:Keldysh_basis}).
Moreover, we can use only either retarded or advanced propagators to connect the two vertices 
to avoid causal arrow loops.
For instance, the two-loop contribution is expressed as
\begin{align}
	\includegraphics[valign=c]{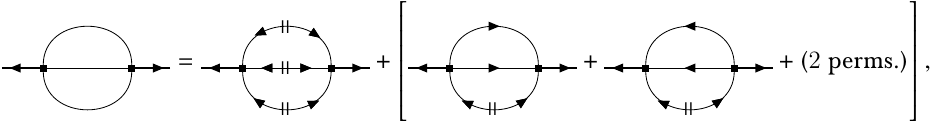}
\end{align}
where ``($2$ perms.)'' indicates the permutations of the position of the retarded/advanced propagators.
The diagrams with the opposite direction of the causal arrows can be combined to cancel out the Heaviside theta functions.
By noting that adding a pair of the $a$-fields at each vertex contains a factor of $1/4$, 
we obtain the desired result:
\begin{align}
	\begin{tikzpicture}[baseline=(v1)]
	\begin{feynman}[inline = (base.v1)] 
		(c1);
		\vertex [right = 0.65 of c1, square dot, label=100:\({\scriptstyle x_1}\)] (v1){};
		\vertex [right = 1.5 of v1, square dot, label=80:\({\scriptstyle x_2}\)] (v2){};
		\vertex [right = 0.65 of v2] (c2);
		\vertex [right = 0.75 of v1] (c);
		\vertex [right = 0.75 of v1] (d2);
		\node [above = -0.3055 of d2] {\vdots};
		\diagram*{
		(c1) -- [very thick, anti fermion] (v1),
		(c2) -- [very thick, anti fermion] (v2),
		(v1) -- [half right] (v2) -- [half right] (v1),
		(v1) -- [out=290, in=250] (v2),
		(v1) -- [out=70, in=110] (v2),
		};
	\end{feynman}
	\end{tikzpicture}
	&=
	\left[G_{rr}^{n+1}
	+ \begin{pmatrix} n \\ 2 \end{pmatrix}
	G_{rr}^{n+1-2}\left(\frac{G_\rho}{2}\right)^{2}
	+ \cdots
	\right]
	=
	\frac{1}{2}
	\left[\left(G_{rr}+\frac{G_\rho}{2}\right)^{n+1}
	+
	\left(G_{rr} - \frac{G_\rho}{2}\right)^{n+1}
	\right]
	\nonumber \\
	&= \frac{1}{2}
	\left[G_>^{n+1}(x_1,x_2) + G_<^{n+1}(x_1,x_2)\right]
	= 	
	\begin{tikzpicture}[baseline=(v1)]
	\begin{feynman}[inline = (base.v1)] 
		(c1);
		\vertex [right = 0.65 of c1, square dot, label=100:\({\scriptstyle x_1}\)] (v1){};
		\vertex [right = 1.5 of v1, square dot, label=80:\({\scriptstyle x_2}\)] (v2){};
		\vertex [right = 0.65 of v2] (c2);
		\vertex [right = 0.75 of v1] (c);
		\vertex [right = 0.75 of v1] (d2);
		\node [above = -0.025 of c] (vdots);
		\node [left = 0.325 of vdots, steelblue] {\vdots};
		\node [right = 0.325 of vdots, steelblue] {\vdots};
		\vertex [above = 0.9 of c] (m1);
		\vertex [below = 0.9 of c] (m2);
		\diagram*{
		(c1) -- [very thick, anti fermion] (v1),
		(c2) -- [very thick, anti fermion] (v2),
		(v1) -- [half right, anti majorana, steelblue] (v2),
		(v1) -- [half left, anti majorana, steelblue] (v2),
		(v1) -- [out=290, in=250, anti majorana, steelblue] (v2),
		(v1) -- [out=70, in=110, anti majorana, steelblue] (v2),
		(m1) -- [dashed, steelblue, very thick] (m2),
		};
	\end{feynman}
	\end{tikzpicture},
\end{align}
where $n$ is the number of the loops. Here we define the spectral function
\begin{align}
	G_\rho(x_1,x_2) = G_> (x_1,x_2) - G_<(x_1,x_2),
\end{align}
and we focus on the structure of the internal propagators and hence ignore coupling 
constants and integration over the position of the vertices, $x_1$ and $x_2$.
Therefore, starting from the in-in Feynman rules, we conclude that
\begin{align}
	\includegraphics[valign=c]{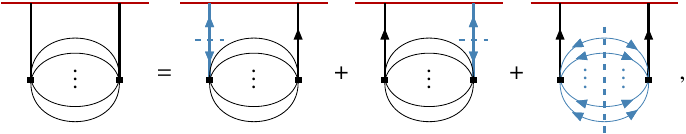}
\end{align}
which is nothing but the cutting rule~\eqref{eq:cutting_2pt}.
For notational simplicity, we have assumed that all the interactions do not involve derivatives here,
but the generalization to the case with derivatives is straightforward.

\subsubsection*{Two-loop diagram}

We next consider a more non-trivial example, a two-loop diagram of the following topology:
\begin{align}
	\langle \varphi_{\vec{k}_1}\varphi_{\vec{k}_2} \cdots \varphi_{\vec{k}_{n}}
	\rangle
	=
	\begin{tikzpicture}[baseline=(v1)]
	\begin{feynman}[inline = (base.v1)]
		\vertex (f1);
		\vertex [right = 0.5 of f1] (c1);
		\vertex [right = 0.75 of c1] (m1);
		\vertex [right = 1.5 of c1] (c2);
		\vertex [right = 0.5 of c2] (f2);
		\vertex [below = of c1, square dot, label=180:\({\scriptstyle x_1}\)] (v1){};
		\vertex [right = 1.5 of v1, square dot, label=0:\({\scriptstyle x_2}\)] (v2){};
		\vertex [right = 0.75 of v1] (c);
		\vertex [above = 0.55 of c, label=180:\({\scriptstyle y_1}\)] (v3);
		\vertex [below = 0.55 of c, label=0:\({\scriptstyle y_2}\)] (v4);
		\diagram*{
		(f1) -- [very thick, darkred] (f2),
		(c1) -- [very thick] (v1),
		(c2) -- [very thick] (v2),
		(v1) -- (v3),
		(v2) -- (v3),
		(v1) -- (v4),
		(v2) -- (v4),
		(v3) -- (v4),
		};
	\end{feynman}
	\end{tikzpicture},
\end{align}
where we assign the spacetime coordinates $x_1, x_2, y_1, y_2$ to the vertices as shown.
Our goal is, again, to confirm that the cutting rule holds for this diagram by starting from the standard in-in Feynman rule.
The contributions with one $r$-type bulk-to-boundary propagator are the ones with cuts on that propagator and are trivial.
Therefore we focus on the contribution with two $a$-type bulk-to-boundary propagators.
By exhausting all possible non-vanishing arrow assignments, it is given by
\begin{align}
	\includegraphics[valign=c]{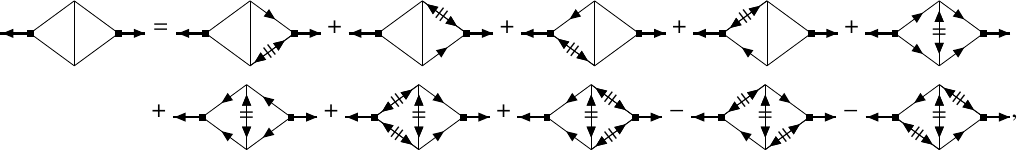}
\end{align}
where the lines without the arrow represent the contributions with all possible arrow assignments
consistent with the rules discussed in Sec.~\ref{subsec:Keldysh_basis},
and the last two terms originate from the double-counting of the first four terms.
We note that
\begin{align}
	\begin{tikzpicture}[baseline=(v1)]
	\begin{feynman}[inline = (base.v1)]
		(c1);
		\vertex [right = 0.45 of c1, square dot] (v1){};
		\vertex [right = 0.75 of v1] (c);
		\vertex [above = 0.55 of c] (v3);
		\vertex [below = 0.55 of c] (v4);
		\vertex [right = 0.45 of v3] (c2);
		\vertex [right = 0.45 of v4] (c3);
		\diagram*{
		(c1) -- [very thick, anti fermion] (v1),
		(v1) -- (v3),
		(v1) -- (v4),
		(v3) -- (v4),
		(v3) -- [fermion] (c2),
		(c3) -- [fermion] (v4),
		};
	\end{feynman}
	\end{tikzpicture}
	&= 
	\begin{tikzpicture}[baseline=(v1)]
	\begin{feynman}[inline = (base.v1)]
		(c1);
		\vertex [right = 0.45 of c1, square dot] (v1){};
		\vertex [right = 0.75 of v1] (c);
		\vertex [above = 0.55 of c] (v3);
		\vertex [below = 0.55 of c] (v4);
		\vertex [right = 0.45 of v3] (c2);
		\vertex [right = 0.45 of v4] (c3);		
		\draw (v1) --node[midway,rotate=127.5,anchor=center]{$=$} (v3);
		\draw (v3) --node[midway]{$=$} (v4);
		\diagram*{
		(c1) -- [very thick, anti fermion] (v1),
		(v1) -- [anti majorana] (v3),
		(v4) -- [fermion] (v1),
		(v3) -- [anti majorana] (v4),
		(v3) -- [fermion] (c2),
		(c3) -- [fermion] (v4),
		};
	\end{feynman}
	\end{tikzpicture}
	+
	\begin{tikzpicture}[baseline=(v1)]
	\begin{feynman}[inline = (base.v1)]
		(c1);
		\vertex [right = 0.45 of c1, square dot] (v1){};
		\vertex [right = 0.75 of v1] (c);
		\vertex [above = 0.55 of c] (v3);
		\vertex [below = 0.55 of c] (v4);
		\vertex [right = 0.45 of v3] (c2);
		\vertex [right = 0.45 of v4] (c3);		
		\draw (v1) --node[midway,rotate=127.5,anchor=center]{$=$} (v3);
		\draw (v1) --node[midway,rotate=52.5,anchor=center]{$=$} (v4);
		\diagram*{
		(c1) -- [very thick, anti fermion] (v1),
		(v1) -- [anti majorana] (v3),
		(v1) -- [anti majorana] (v4),
		(v4) -- [fermion] (v3),
		(v3) -- [fermion] (c2),
		(c3) -- [fermion] (v4),
		};
	\end{feynman}
	\end{tikzpicture}
	+
	\begin{tikzpicture}[baseline=(v1)]
	\begin{feynman}[inline = (base.v1)]
		(c1);
		\vertex [right = 0.45 of c1, square dot] (v1){};
		\vertex [right = 0.75 of v1] (c);
		\vertex [above = 0.55 of c] (v3);
		\vertex [below = 0.55 of c] (v4);
		\vertex [right = 0.45 of v3] (c2);
		\vertex [right = 0.45 of v4] (c3);		
		\diagram*{
		(c1) -- [very thick, anti fermion] (v1),
		(v3) -- [fermion] (v1),
		(v4) -- [fermion] (v1),
		(v3) -- [fermion] (v4),
		(v3) -- [fermion] (c2),
		(c3) -- [fermion] (v4),
		};
	\end{feynman}
	\end{tikzpicture}
	+
	\begin{tikzpicture}[baseline=(v1)]
	\begin{feynman}[inline = (base.v1)]
		(c1);
		\vertex [right = 0.45 of c1, square dot] (v1){};
		\vertex [right = 0.75 of v1] (c);
		\vertex [above = 0.55 of c] (v3);
		\vertex [below = 0.55 of c] (v4);
		\vertex [right = 0.45 of v3] (c2);
		\vertex [right = 0.45 of v4] (c3);		
		\diagram*{
		(c1) -- [very thick, anti fermion] (v1),
		(v1) -- [fermion] (v3),
		(v1) -- [fermion] (v4),
		(v4) -- [fermion] (v3),
		(v3) -- [fermion] (c2),
		(c3) -- [fermion] (v4),
		};
	\end{feynman}
	\end{tikzpicture}
	\nonumber \\
	&= \theta(x_1^0-y_2^0)G_{rr}(x_1,y_1)G_{rr}(y_1,y_2)G_\rho(x_1,y_2)
	+ \theta(y_1^0-y_2^0)G_{rr}(x_1,y_1)G_\rho(y_1,y_2)G_{rr}(x_1,y_2)
	\nonumber \\
	&+ \frac{1}{4}\left[\theta(y_1^0-y_2^0)\theta(y_2^0-x_1^0)-\theta(y_2^0-y_1^0)\theta(x_1^0-y_2^0)\right]
	G_\rho(x_1,y_1)G_\rho(y_1,y_2)G_\rho(x_1,y_2),
\end{align}
where we focus only on the structure of the internal propagators.
By using $\theta(y_2^0-x_1^0) = 1-\theta(x_1^0-y_2^0)$ and  $\theta(y_1^0-y_2^0) + \theta(y_2^0-y_1^0) = 1$,
we can show that
\begin{align}
	\includegraphics[valign=c]{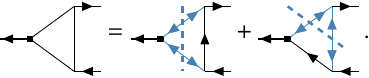}
	\label{eq:triangle_2a}
\end{align}
By noting that
\begin{align}
	\includegraphics[valign=c]{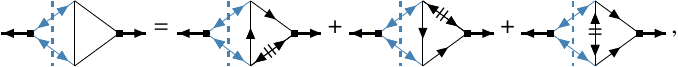}
\end{align}
we thus obtain
\begin{align}
	\includegraphics[valign=c]{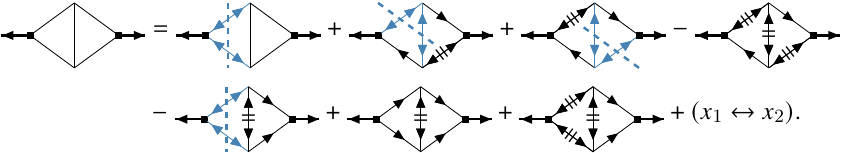}
\end{align}
To simplify the last three terms, we use that
\begin{align}
	\begin{tikzpicture}[baseline=(v1)]
	\begin{feynman}[inline = (base.v1)]
		(c1);
		\vertex [right = 0.45 of c1, square dot] (v1){};
		\vertex [right = 0.75 of v1] (c);
		\vertex [above = 0.55 of c] (v3);
		\vertex [below = 0.55 of c] (v4);
		\vertex [right = 0.45 of v2] (c2);
		\vertex [left = 0.375 of v3] (cut1);
		\vertex [left = 0.375 of v4] (cut2);
		\diagram*{
		(v1) -- [anti majorana, steelblue] (v3),
		(v1) -- [anti majorana, steelblue] (v4),
		(cut1) -- [dashed, steelblue, very thick] (cut2),
		};
	\end{feynman}
	\end{tikzpicture}
	\,-\,
	\begin{tikzpicture}[baseline=(v1)]
	\begin{feynman}[inline = (base.v1)]
		(c1);
		\vertex [right = 0.45 of c1, square dot] (v1){};
		\vertex [right = 0.75 of v1] (c);
		\vertex [above = 0.55 of c] (v3);
		\vertex [below = 0.55 of c] (v4);
		\vertex [right = 0.45 of v2] (c2);
		\draw (v1) --node[midway,rotate=127.5,anchor=center]{$=$} (v3);
		\draw (v1) --node[midway,rotate=52.5,anchor=center]{$=$} (v4);
		\vertex [left = 0.375 of v3] (cut1);
		\vertex [left = 0.375 of v4] (cut2);
		\diagram*{
		(v1) -- [anti majorana] (v3),
		(v1) -- [anti majorana] (v4),
		};
	\end{feynman}
	\end{tikzpicture}
	\,-\,
	\begin{tikzpicture}[baseline=(v1)]
	\begin{feynman}[inline = (base.v1)]
		(c1);
		\vertex [right = 0.45 of c1, square dot] (v1){};
		\vertex [right = 0.75 of v1] (c);
		\vertex [above = 0.55 of c] (v3);
		\vertex [below = 0.55 of c] (v4);
		\vertex [right = 0.45 of v2] (c2);
		\vertex [left = 0.375 of v3] (cut1);
		\vertex [left = 0.375 of v4] (cut2);
		\diagram*{
		(v1) -- [fermion] (v3),
		(v1) -- [fermion] (v4),
		};
	\end{feynman}
	\end{tikzpicture}
	&= \frac{1}{4}\left[1-\theta(y_1^0-x_1^0)\theta(y_2^0-x_1^0)\right]
	G_\rho(x_1,y_1)G_\rho(x_1,y_2),
\end{align}
and therefore
\begin{align}
	&
	\begin{tikzpicture}[baseline=(v1)]
	\begin{feynman}[inline = (base.v1)]
		(c1);
		\vertex [right = 0.45 of c1, square dot] (v1){};
		\vertex [right = 1.5 of v1, square dot] (v2){};
		\vertex [right = 0.75 of v1] (c);
		\vertex [above = 0.55 of c] (v3);
		\vertex [below = 0.55 of c] (v4);
		\vertex [right = 0.45 of v2] (c2);
		\draw (v3) --node[midway]{$=$} (v4);
		\vertex [left = 0.375 of v3] (cut1);
		\vertex [left = 0.375 of v4] (cut2);
		\diagram*{
		(c1) -- [very thick, anti fermion] (v1),
		(c2) -- [very thick, anti fermion] (v2),
		(v1) -- [anti majorana, steelblue] (v3),
		(v1) -- [anti majorana, steelblue] (v4),
		(v4) -- [fermion] (v2),
		(v3) -- [fermion] (v2),
		(v3) -- [anti majorana] (v4),
		(cut1) -- [dashed, steelblue, very thick] (cut2),
		};
	\end{feynman}
	\end{tikzpicture}
	-
	\begin{tikzpicture}[baseline=(v1)]
	\begin{feynman}[inline = (base.v1)]
		(c1);
		\vertex [right = 0.45 of c1, square dot] (v1){};
		\vertex [right = 1.5 of v1, square dot] (v2){};
		\vertex [right = 0.75 of v1] (c);
		\vertex [above = 0.55 of c] (v3);
		\vertex [below = 0.55 of c] (v4);
		\vertex [right = 0.45 of v2] (c2);
		\draw (v3) --node[midway]{$=$} (v4);
		\diagram*{
		(c1) -- [very thick, anti fermion] (v1),
		(c2) -- [very thick, anti fermion] (v2),
		(v1) -- [fermion] (v3),
		(v4) -- [fermion] (v2),
		(v1) -- [fermion] (v4),
		(v3) -- [fermion] (v2),
		(v3) -- [anti majorana] (v4),
		};
	\end{feynman}
	\end{tikzpicture}
	-
	\begin{tikzpicture}[baseline=(v1)]
	\begin{feynman}[inline = (base.v1)]
		(c1);
		\vertex [right = 0.45 of c1, square dot] (v1){};
		\vertex [right = 1.5 of v1, square dot] (v2){};
		\vertex [right = 0.75 of v1] (c);
		\vertex [above = 0.55 of c] (v3);
		\vertex [below = 0.55 of c] (v4);
		\vertex [right = 0.45 of v2] (c2);
		\draw (v1) --node[midway,rotate=127.5,anchor=center]{$=$} (v3);
		\draw (v1) --node[midway,rotate=52.5,anchor=center]{$=$} (v4);
		\draw (v3) --node[midway]{$=$} (v4);
		\diagram*{
		(c1) -- [very thick, anti fermion] (v1),
		(c2) -- [very thick, anti fermion] (v2),
		(v1) -- [anti majorana] (v3),
		(v4) -- [fermion] (v2),
		(v1) -- [anti majorana] (v4),
		(v3) -- [fermion] (v2),
		(v3) -- [anti majorana] (v4),
		};
	\end{feynman}
	\end{tikzpicture}
	\nonumber \\
	&= \frac{1}{4}\left[1-\theta(y_1^0-x_1^0)\theta(y_2^0-x_1^0)\right]\theta(x_2^0-y_1^0)\theta(x_2^0-y_2^0)
	G_{rr}(y_1,y_2)G_\rho(x_1,y_1)G_\rho(x_1,y_2)G_\rho(x_2,y_1)G_\rho(x_2,y_2).
\end{align}
By using the identities of the theta functions, one can show that
\begin{align}
	\left[1-\theta(y_1^0-x_1^0)\theta(y_2^0-x_1^0)\right]\theta(x_2^0-y_1^0)\theta(x_2^0-y_2^0)
	+ (x_1\leftrightarrow x_2)
	&= \theta(x_1^0-y_1^0)\theta(x_2^0-y_2^0)
	+ \theta(x_1^0-y_2^0)\theta(x_2^0-y_1^0),
\end{align}
and hence we obtain
\begin{align}
	\includegraphics[valign=c]{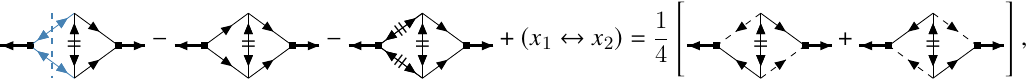}
\end{align}
where we define
\begin{align}
	\begin{tikzpicture}[baseline=(c1)]
	\begin{feynman}[inline = (base.c1)]
		\vertex [label=\({\scriptstyle x_1}\)] (c1);
		\vertex [right = of c1, label=\({\scriptstyle x_2}\)] (c2);
		\diagram*{
		(c1) -- [charged scalar] (c2),
		};
	\end{feynman}
	\end{tikzpicture}
	&= G_\rho(x_2,x_1).
\end{align}
Finally, by using
\begin{align}
	\includegraphics[valign=c]{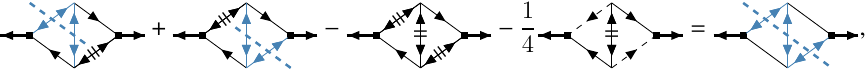}
\end{align}
we arrive at
\begin{align}
	\includegraphics[valign=c]{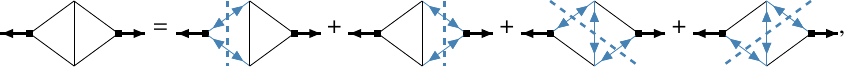}
\end{align}
which correctly reproduces the cutting rule~\eqref{eq:cutting_2pt}.
Again, for notational simplicity, we have assumed that all the interactions do not involve derivatives,
but the generalization to the case with derivatives is straightforward.
This example demonstrates that the cutting rule offers an efficient way of organizing the terms,
which is particularly useful for CC physics as we will see in Sec.~\ref{sec:application}.
Even though one can in principle find such a way brute-forcely from the standard Feynman rules as we have done here,
it would be tedious and impractical to do so, especially for diagrams with complicated topologies, 
without the knowledge of the cutting rule.

\subsubsection*{One-loop triangle diagram}

So far we have considered the examples with the two-point bulk correlators.
However, as discussed in Sec.~\ref{subsec:cutting_proof}, the cutting rule can be extended
to the case with arbitrary $n (\geq 2)$-point bulk correlators.
As the simplest example of such a higher-point bulk correlator, we now consider the triangle diagram given by
\begin{align}
	\langle \varphi_{\vec{k}_1}\varphi_{\vec{k}_2} \cdots \varphi_{\vec{k}_{n}}
	\rangle
	=
	\begin{tikzpicture}[baseline=(v2)]
	\begin{feynman}[inline = (base.v2)]
		\vertex (f1);
		\vertex [right = 0.5 of f1] (c1);
		\vertex [right = 0.75 of c1] (c2);
		\vertex [right = 1.5 of c1] (c3);
		\vertex [right = 0.5 of c3] (f2);
		\vertex [below = 1.5 of c1, square dot] (v1){};
		\vertex [right = 0.75 of v1] (c);
		\vertex [above = 0.75 of c, square dot] (v2){};
		\vertex [right = 1.5 of v1, square dot] (v3){};
		\diagram*{
		(f1) -- [very thick, darkred] (f2),
		(c1) -- [very thick] (v1),
		(c2) -- [very thick] (v2),
		(c3) -- [very thick] (v3),
		(v1) -- (v3),
		(v2) -- (v3),
		(v1) -- (v2),
		};
	\end{feynman}
	\end{tikzpicture}.
\end{align}
We take the opposite route in this example; we start from the cutting rule and show that 
it reduces to the diagrams with the standard in-in Feynman rules.

The cutting rule tells us that 
\begin{align}
	\includegraphics[valign=c]{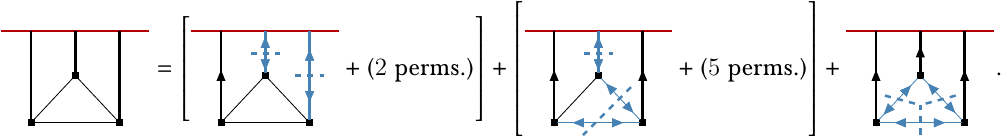}
\end{align}
The contribution with the cuttings on two bulk-to-boundary propagators can be organized as
\begin{align}
	\includegraphics[valign=c]{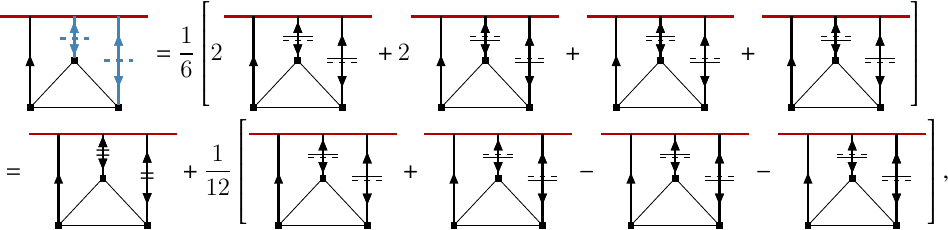}
	\label{eq:triangle_zero_bulkcut}
\end{align}
while the one with the cutting on one bulk-to-boundary propagator can be expressed as
\begin{align}
	\includegraphics[valign=c]{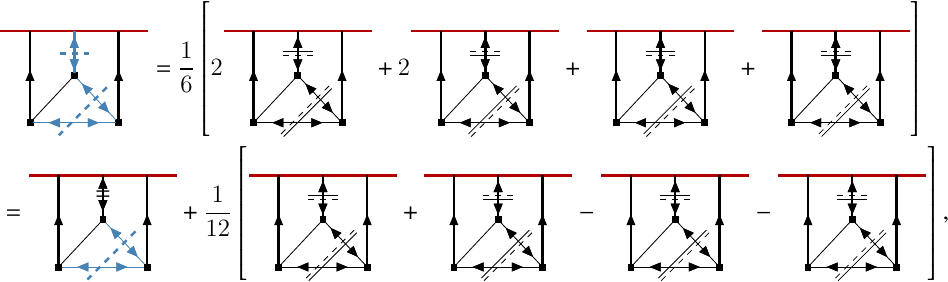}
\end{align}
and therefore
\begin{align}
	\includegraphics[valign=c]{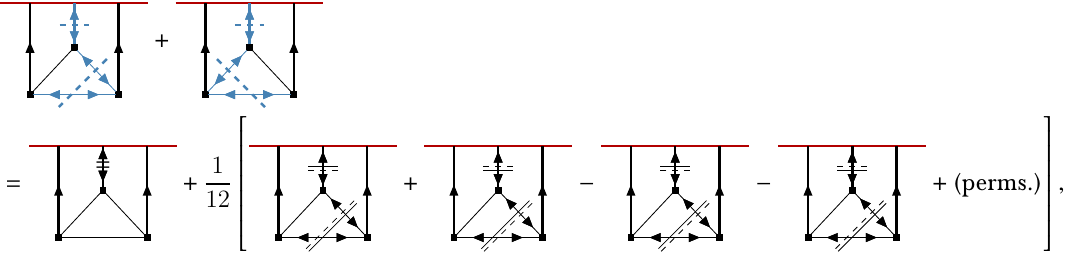}
	\label{eq:triangle_one_bulkcut}
\end{align}
where ``(perms.)'' in this expression indicates that the position of the cut on the bulk triangle is flipped,
and we have used Eq.~\eqref{eq:triangle_2a} for the first term.
We may denote the second terms in Eqs.~\eqref{eq:triangle_zero_bulkcut} and~\eqref{eq:triangle_one_bulkcut}
with the subscript ``$3a$'' since the bulk-to-boundary propagators can be expressed 
as a product of three retarded functions.
After some computation, we can show that
\begin{align}
	\includegraphics[valign=c]{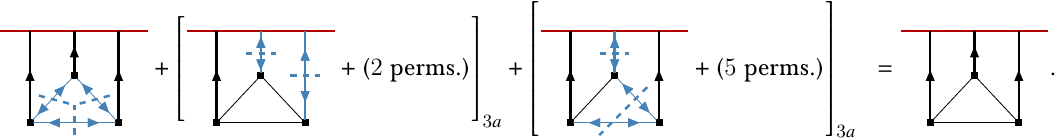}
\end{align}
Therefore we obtain
\begin{align}
	\includegraphics[valign=c]{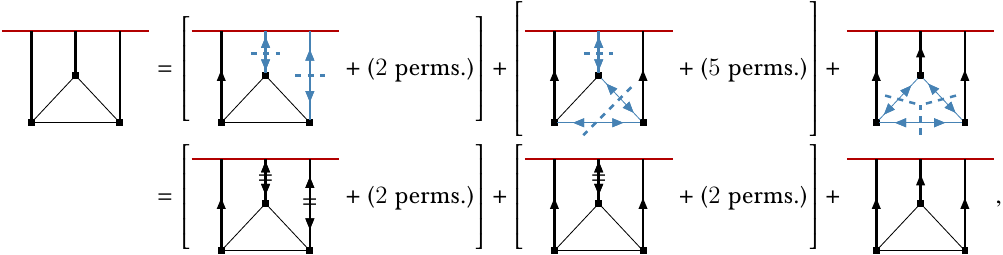}
\end{align}
which reproduces the in-in Feynman rule in the $r/a$ basis.

Here is a side remark. 
We may classify the diagrams by the number of the retarded bulk-to-boundary propagators as we have done above.
If we consider a two-point bulk correlator, there is a one-to-one correspondence between this classification 
and the position of the cut; the ``$1a$'' diagram corresponds to the cut on a bulk-to-boundary propagator, 
while the ``$2a$'' diagram corresponds to the cut on the bulk correlator.
However, this correspondence no longer holds for higher-point bulk correlators and a single topology of the cuts
in general contains different contributions as evident from 
Eqs.~\eqref{eq:triangle_zero_bulkcut} and~\eqref{eq:triangle_one_bulkcut}.

\section{Application to cosmological collider}
\label{sec:application}

In Sec.~\ref{sec:cutting}, we have proven the cutting rule for equal-time in-in correlators that decomposes
diagrams into fully retarded functions and cut-propagators.
Our proof relies only on unitarity, locality, and the causal structure of the in-in formalism,
and therefore our discussion so far applies to general theories with any particle contents and any local interactions, 
beyond the context of cosmological correlators.
Restricting ourselves to the cosmological context, we can use our cutting rule, \textit{e.g.}, 
not only for the de~Sitter background but for slow-roll inflationary or even non-inflationary (such as power-law) backgrounds.

In this section, we narrow down our focus and study CC signals during inflation as an application of our cutting rule.
In particular, assuming microcausality, we show that non-local CC signals arise solely from cut-propagators.
A corollary is that only diagrams with cuts on bulk correlators, which we call ``bulk-cut" diagrams, 
induce non-local CC signals.
This is of practical importance since the cut corresponds to the positive/negative Wightman functions without
the (anti-)time-ordering theta functions, and therefore the cutting rule allows us to factorize the conformal time integrals
if we focus on non-local CC signals.
The nested time integrals arising from (anti-)time-ordering 
are often the bottle-neck of evaluating CC signals,
and hence we believe that our cutting rule helps to resolve this issue.

In the following, we prove our claim on non-local CC signals in Sec.~\ref{subsec:microcausality}.
We then apply our cutting rule to several examples; tree-level scalar single exchange in Sec.~\ref{subsec:nlCC_tree},
one-loop scalar bubble diagram in Sec.~\ref{subsec:nlCC_bubble}, and one-loop scalar triangle diagram
in Sec.~\ref{subsec:nlCC_triangle}, respectively.

\subsection{Microcausality and cutting rule for non-local CC signals}
\label{subsec:microcausality}

Cosmological collider physics~\cite{Chen:2009we,Chen:2009zp,Baumann:2011nk,Assassi:2012zq,
Sefusatti:2012ye,Norena:2012yi,Chen:2012ge,Noumi:2012vr,Cespedes:2013rda,Gong:2013sma,Kehagias:2015jha,
Liu:2015tza,Arkani-Hamed:2015bza} 
aims at detecting signals of particle production during inflation imprinted in non-gaussianity of (iso-)curvature perturbations.\footnote{
	See, \emph{e.g.},~\cite{Pinol:2021aun,Chen:2023txq} for CC signals involving multiple degrees of freedom.
}
These signals behave non-analytically in terms of external momenta,
and are distinct from higher dimensional operators that generate only analytic dependence;
see~\cite{Cabass:2024wob,Sohn:2024xzd} for recent observational studies.
As an illustration, let us consider the diagram with a two-point bulk correlator:
\begin{align}
		\langle \varphi_{\vec{k}_1}\varphi_{\vec{k}_2} \cdots \varphi_{\vec{k}_{n}} \rangle
		=
		\begin{tikzpicture}[baseline=(c)]
		\begin{feynman}[inline = (base.c)]
			\vertex (f1);
			\vertex [right = 0.5 of f1, label=\({\left\{\scriptstyle k_{i}\right\}}\)] (c1);
			\vertex [right = 0.75 of c1] (m1);
			\vertex [right = 1.5 of c1, label=\({ \left\{ \scriptstyle k_j \right\}}\)] (c2);
			\vertex [right = 0.5 of c2] (f2);
			\vertex [below = of m1, blob, shape=ellipse,minimum height=0.75cm,minimum width=1.5cm,] (v1){};
			\vertex [left = 0.75 cm of v1, square dot](v1p){};
			\vertex [right = 0.75 cm of v1, square dot](v2p){};
			\vertex [below = 0.7 of c1] (c);
			\diagram*{
			(f1) -- [very thick, darkred] (f2),
			(c1) -- [plain, very thick](v1p),
			(c2) -- [plain,  very thick](v2p),
			};
		\end{feynman}
	\end{tikzpicture}.
\end{align}
We denote $k_L = \sum k_i$, $k_R = \sum k_j$, and $k_s = \vert \sum \vec{k}_i \vert = \vert \sum \vec{k}_j \vert$.
The CC signals then take the form of
\begin{align}
	\langle \varphi_{\vec{k}_1}\varphi_{\vec{k}_2} \cdots \varphi_{\vec{k}_{n}} \rangle
	\propto \left(\frac{k_s}{k_{L/R}}\right)^{\pm\nu},~\left(\frac{k_{L}}{k_{R}}\right)^{\pm\nu},
\end{align}
where $\nu$ depends on the masses of particles showing up in the bulk correlator and is non-integer in general,
which leads to non-analyticity.
The former contribution is called a ``non-local'' CC signal and arises from a pair production of particles
propagating to two vertices, while the latter contribution is called a ``local'' CC signal and 
is interpreted in~\cite{Wang:2020ioa,Tong:2021wai} as a particle production at one vertex propagating 
to the other vertex.
One should not confuse it with contributions from higher dimensional operators which we refer to as ``background''.
For general $n$-point bulk correlators, we call a CC signal non-local if it depends on a vector sum of the external momenta,
and local if it depends only on a scalar sum (or an ``energy'' sum).
In this paper, we focus on the non-local CC signals 
since the local ones are more subtle as we shall see.
This distinction is justified as long as we have more than one boundary field on each side 
which leads to $k_L, k_R \neq k_s$, and we focus on such a case in the following.
Even though the non-local CC signals are often
more important as only those are enhanced \textit{e.g.,} by the chemical potential~\cite{Tong:2021wai}, 
it would be more satisfactory if we could extend our arguments to local CC signals,
which we leave for future work.

As we have seen in Sec.~\ref{sec:cutting}, the cutting rule decomposes diagrams into fully retarded functions
and cut-propagators. We now show that
\begin{eBox}
\centering
non-local CC signals are generated solely from cut-propagators.
\end{eBox}
Since the bulk-to-boundary propagators do not have non-analytic dependence on the momenta, 
as a corollary, it follows that
\begin{eBox}
\centering
only bulk-cut diagrams induce non-local CC signals.
\end{eBox}
For instance, for the two-point bulk correlator, our claim can be expressed as
\begin{eBox}
\begin{align}
	\left[
	\begin{tikzpicture}[baseline=(c)]
		\begin{feynman}[inline = (base.c)]
			\vertex (f1);
			\vertex [right = 0.5 of f1, label=\({\left\{\scriptstyle k_{i}\right\}}\)] (c1);
			\vertex [right = 0.75 of c1] (m1);
			\vertex [right = 1.5 of c1, label=\({ \left\{ \scriptstyle k_j \right\}}\)] (c2);
			\vertex [right = 0.5 of c2] (f2);
			\vertex [below = of m1, blob, shape=ellipse,minimum height=0.75cm,minimum width=1.5cm,] (v1){};
			\vertex [left = 0.75 cm of v1, square dot] (v1p){};
			\vertex [right = 0.75 cm of v1, square dot](v2p){};
			\vertex [below = 0.7 of c1] (c);
			\diagram*{
			(f1) -- [very thick, darkred] (f2),
			(c1) -- [plain, very thick](v1p),
			(c2) -- [plain,  very thick](v2p),
			};
		\end{feynman}
	\end{tikzpicture}
	\right]_{\text{nlCC}}
	&= 
	\begin{tikzpicture}[baseline=(c)]
		\begin{feynman}[inline = (base.c),every blob={/tikz/fill=gray!30,/tikz/inner sep=2pt}]
			\vertex (f1);
			\vertex [right = 0.5 of f1, label=\({\left\{\scriptstyle k_{i}\right\}}\)] (c1);
			\vertex [right = 0.75 of c1] (m1);
			\vertex [right = 0.2 of m1] (m3);
			\vertex [right = 1.5 of c1, label=\({ \left\{ \scriptstyle k_j \right\}}\)] (c2);
			\vertex [right = 0.5 of c2] (f2);
			\vertex [right = 0.25 cm of c1](mv1);
			\vertex [left = 0.25 cm of c2](mv2);
			\vertex [below = of mv1, blob, minimum height=0.5cm,minimum width=0.5cm] (v1){$\shortleftarrow$};
			\vertex [below = of mv2, blob, minimum height=0.5cm,minimum width=0.5cm] (v2){$\shortrightarrow$};
			\vertex [left = 0.25 cm of v1, square dot](v1p){};
			\vertex [right = 0.25 cm of v2, square dot](v2p){};
			\vertex [below = 0.7 of c1] (c);
			\vertex [below = 0.6 of m1] (m0);
			\vertex [below = 1.35 of m0,label=270:\({\scriptstyle \color{steelblue}{\text{nlCC}}}\)] (m2);
			\vertex [left = 0.12 of m2] (vdots);
			\node [above = 0.32 of vdots, steelblue] {\vdots};
			\diagram*{
			(f1) -- [very thick, darkred] (f2),
			(c1) -- [anti fermion, very thick](v1p),
			(c2) -- [anti fermion,  very thick](v2p),
			(v1) -- [out=300, in=240, anti majorana, steelblue] (v2) 
			-- [out=110, in=70, anti majorana, steelblue] (v1),(v1) -- [out=40, in=140, anti majorana, steelblue] (v2),
			(m0) -- [dashed, steelblue, very thick] (m2),
			};
		\end{feynman}
	\end{tikzpicture},
	\label{eq:cutting_2pt_nlCC}
\end{align}
\end{eBox}
where the subscript ``nlCC'' on the left-hand side indicates that we focus only on non-local CC signals 
and the one on the right-hand side indicates that we extract the non-local part of the cut-propagators.\footnote{
	We \emph{do not} claim that the diagram on the right contains only non-local CC.
	As we will see below, it contains both non-local CC and (a part of) local CC signals even at the tree level.
}
For three-point bulk correlators, it is expressed as
\begin{eBox}
\begin{align}
	\left[
	\begin{tikzpicture}[baseline=(c)]
		\begin{feynman}[inline = (base.c)]
			\vertex (f1);
			\vertex [right = 0.5 of f1,label=\(\{{\scriptstyle k_{i}}\}\)] (c1);
			\vertex [right = 0.75 of c1,label=\(\{{\scriptstyle k_{j}}\}\)] (m1);
			\vertex [right = 1.5 of c1,label=\(\{{\scriptstyle k_{k}}\}\)] (c2);
			\vertex [right = 0.5 of c2] (f2);
			\vertex [below = of m1, blob, shape=ellipse ,minimum height=0.75cm,minimum width=1.5cm] (v1){};
			\vertex [above = 0.375 cm of v1, square dot](v3p){};
			\vertex [left = 0.75 cm of v1, square dot](v1p){};
			\vertex [right = 0.75 cm of v1, square dot](v2p){};
			\vertex [below = 0.7 of c1] (c);
			\diagram*{
			(f1) -- [very thick, darkred] (f2),
			(c1) -- [plain, very thick](v1p),
			(c2) -- [plain,  very thick](v2p),
			(m1) -- [plain,  very thick](v3p),
			};
		\end{feynman}
	\end{tikzpicture}
	\right]_{\text{nlCC}}
	&\, = \,
	\qty[
	\begin{tikzpicture}[baseline=(c)]
		\begin{feynman}[inline = (base.c),every blob={/tikz/fill=gray!30,/tikz/inner sep=2pt}]
			\vertex (f1);
			\vertex [right = 0.5 of f1,label=\(\{{\scriptstyle k_{i}}\}\)] (c1);
			\vertex [right = 0.75 of c1,label=\(\{{\scriptstyle k_{j}}\}\)] (m1);
			\vertex [below = 0.325 of m1] (ecutt);
			\vertex [left = 0.35 of ecutt] (ecut1);
			\vertex [right = 0.35 of ecutt] (ecut2);
			\vertex [right = 1.75 of c1,label=\(\{{\scriptstyle k_{k}}\}\)] (c2);
			\vertex [right = 0.5 of c2] (f2);
			\vertex [below = of m1] (v1){};
			\vertex [above = 0.375 cm of v1](v3p);
			\vertex [right = 0.05 cm of v3p](v3pp);
			\vertex [below = 0.75 cm of v3pp](v4p);
			\vertex [below = 0.7 of c1] (c);
			\draw[fill=gray!30] (v4p) -- (v3pp) arc (90:270:0.8cm and 0.375cm) -- cycle;
			\vertex [right = 0.75 of v1, blob, minimum height=0.5cm,minimum width=0.5cm] (v2){$\shortrightarrow$};
			\vertex [above = 0.375 cm of v1, square dot](v3ppp){};
			\vertex [right = 0.32 cm of v3ppp](vcutt){};
			\vertex [right = 0.16 cm of v1](vdott){};
			\vertex [above = 0.12 cm of vdott, steelblue](vdot){$\vdots$};
			\vertex [above = 0.25 cm of vcutt](vcut1);
			\vertex [below = 1.25 cm of vcut1, label=270:\({\scriptstyle \color{steelblue}{\text{nlCC}}}\)](vcut2);
			\vertex [left = 0.75 cm of v1, square dot](v1p){};
			\vertex [right = 1 of v1, square dot](v2p){};
			\vertex [right = 0.375 cm of v1p](bname){$\longleftarrow$};
			\diagram*{
			(f1) -- [very thick, darkred] (f2),
			(c1) -- [anti fermion, very thick](v1p),
			(c2) -- [anti fermion,  very thick](v2p),
			(m1) -- [anti majorana,  very thick, steelblue](v3ppp),
			(v3pp) -- [anti majorana, steelblue, out=10, in=120](v2),
			(v4p) -- [anti majorana, steelblue, out=350, in=240](v2),
			(vcut1) -- [dashed, steelblue, very thick](vcut2),
			(ecut1) -- [dashed, steelblue, very thick](ecut2),
			};
		\end{feynman}
	\end{tikzpicture}
	\, +
	\text{($5$ perms.)} ]
	\, + \,
	\begin{tikzpicture}[baseline=(center)]
		\begin{feynman}[inline = (base.center),every blob={/tikz/fill=gray!30,/tikz/inner sep=2pt}]
			\vertex (f1);
			\vertex [right = 0.5 of f1,label=\(\{{\scriptstyle k_{i}}\}\)] (c1);
			\vertex [below = 0.9 of c1] (bl);
			\vertex [right = 1 of c1,label=\(\{{\scriptstyle k_{j}}\}\)] (c2);
			\vertex [right = 2 of c1,label=\(\{{\scriptstyle k_{k}}\}\)] (c3);
			\vertex [right = 0.5 of c3] (f2);
			\vertex [below = 2 of c1, square dot] (v1){};
			\vertex [right = 0.25 of v1, blob, minimum height=0.5cm,minimum width=0.5cm] (v1p){$\shortleftarrow$};
			\vertex [below = 2 of c1, square dot] (v1){};
			\vertex [right = 1 of v1] (c);
			\vertex [above = 1.5 of c, square dot] (v2){};
			\vertex [below = 0.25 of v2, blob, minimum height=0.5cm,minimum width=0.5cm] (v2p){$\shortuparrow$};
			\vertex [above = 1.5 of c, square dot] (v2){};
			\vertex [right = 2 of v1, square dot] (v3){};
			\vertex [right = 0.85 of v2p](cutt1);
			\vertex [above = 0.54 of c](cuttc);
			\vertex [above = 0.54 of c](cuttc2);
			\vertex [right = 0.1 of cuttc](cut12);
			\vertex [left = 0.1 of cuttc2](cut22);
			\vertex [below = 0.3 of cutt1](cut11);
			\vertex [left = 1.7 of cut11](cut21);
			\vertex [above = 0.3 of c](cut31);
			\vertex [below = 1 of cut31, label=270:\({\scriptstyle \color{steelblue}{\text{nlCC}}}\)](cut32);
			\vertex [left = 0.25 of v3, blob, minimum height=0.5cm,minimum width=0.5cm] (v3p){$\shortrightarrow$};
			\vertex [right = 2 of v1, square dot] (v3){};
			\path (cut11) -- (cut12) node [pos=0.5,sloped, anchor=center, below, steelblue] {$\scriptstyle \cdots$};
			\path (cut21) -- (cut22) node [pos=0.5,sloped, anchor=center, below, steelblue] {$\scriptstyle \cdots$};
			\path (cut31) -- (cut32) node [pos=0.5,sloped, anchor=center, below, steelblue] {$\scriptstyle \cdots$};
			\vertex [below = 0.4 of bl] (center);
			\diagram*{
			(f1) -- [very thick, darkred] (f2),
			(c1) -- [anti fermion, very thick] (v1),
			(c2) -- [anti fermion, very thick] (v2),
			(c3) -- [anti fermion, very thick] (v3),
			(v1p) --[anti majorana, steelblue, out=10, in=170] (v3p),
			(v1p) --[anti majorana, steelblue, out=310, in=230] (v3p),
			(v2p) --[anti majorana, steelblue, out=350, in=80] (v3p),
			(v2p) --[anti majorana, steelblue, out=290, in=130] (v3p),
			(v1p) --[anti majorana, steelblue, out=100, in=190] (v2p),
			(v1p) --[anti majorana, steelblue, out=50, in=250] (v2p),
			(cut11) --[dashed, very thick, steelblue] (cut12),
			(cut21) --[dashed, very thick, steelblue] (cut22),
			(cut31) --[dashed, very thick, steelblue] (cut32),
			};
		\end{feynman}
		\end{tikzpicture}\,,
		\label{eq:cutting_3pt_nlCC}
\end{align}
\end{eBox}
where we need to extract the non-local signals from the cut-propagators 
in the bulk in the first diagrams on the right-hand side.
It is straightforward to generalize our claim to arbitrary $n$-point bulk functions.

\paragraph{\textit{Proof.---}}
Since the cutting rule decomposes diagrams into cut-propagators and fully retarded vertex functions, 
our claim can be rephrased that the fully retarded functions do not induce non-analytic dependence on spatial momenta.
To show this, let us remember that, as we have shown below Eq.~\eqref{eq:causality_full_ret}, 
the fully retarded function vanishes outside the light-cone as long as the microcausality holds, \textit{i.e.},
\begin{align}
	{}^{\exists}i \in [1,n]~\mathrm{s.t.}~(x-y_i)^2 < 0
	~~\Rightarrow~~
	\mathcal{V}_{ra\cdots a}(x,y_1,\cdots,y_n)
	= \langle {O}_r(x) {O}_a(y_1)\cdots {O}_a(y_n)\rangle
	&= 0.
\end{align}
Note again that, even though we use the same symbol ${O}$ for all the operators for notational
simplicity, they can be distinct operators for our discussion.
For the bulk-cut diagrams, one can think of ${O}$ as the vertex with the fields corresponding 
to the cut-propagators stripped off.
With the conformal times fixed, this indicates that the fully retarded function has finite supports 
with respect to its spatial arguments.\footnote{
	The conformal time is eventually integrated, but we assume that this does not modify our discussion here.
	This assumption is correct for the examples we consider below.
}
The Fourier transformation of the vertex function is given by
\begin{align}
	&\mathcal{V}_{ra\cdots a}(x^0, y_1^0, \cdots y_n^0; \vec{q}, \vec{k}_1,\cdots \vec{k}_n) \nonumber \\
	&\qquad = \int \dd^{d-1}x\,e^{-i\vec{x}\cdot (\vec{q}+\sum_i \vec{k}_i)}
	\int \left[\prod_{i=1}^{n} \dd^{d-1}y_i\,e^{-i\vec{k}_i\cdot (\vec{y}_i - \vec{x})}\right]
	\mathcal{V}_{ra\cdots a}(x,y_1,\cdots,y_n).
\end{align}
The integral over $\vec{x}$ gives us the delta function representing the overall spatial momentum conservation,
and the non-trivial dependence on $\vec{k}_i$ arises from the integrals over $\vec{y}_i$, 
or equivalently $\vec{y}_i - \vec{x}$.
Since the Fourier transform of a function with a finite support is an entire function due to
Paley--Wiener--Schwartz theorem, 
$\mathcal{V}_{ra\cdots a}(x^0, y_1^0, \cdots y_n^0; \vec{k}_1,\cdots \vec{k}_n)$ is analytic in $\vec{k}_1, \cdots, \vec{k}_n$,
which completes our proof.

\vspace{5mm}

Note that we have defined the microcausality based on the comoving coordinates in this proof.
This is appropriate in cosmological contexts since the conformal time corresponds to the comoving horizon.
In particular, in de~Sitter spacetime, the embedding distance $Z_{ij}$ is given by
\begin{align}
	Z_{ij} = 1 + \frac{(\tau_i - \tau_j)^2 - \vert \vec{x}_i - \vec{x}_j\vert^2}{2\tau_i \tau_j},
\end{align}
where $\tau_i$ is the conformal time,
and $Z_{ij} = 1$ separates the time-like and space-like distances (see \textit{e.g.,} \cite{Spradlin:2001pw}),
which in turn is determined by $(\tau_i - \tau_j)^2 - \vert \vec{x}_i - \vec{x}_j\vert^2 \gtrless 0$.

\subsubsection*{Comparison with previous literature}

Before closing this subsection, we compare our cutting rule with those discussed in previous literature.
In~\cite{Tong:2021wai}, a cutting rule on tree-level diagrams is discussed based on physical intuition.
Concerning the non-local CC signals,
our cutting rule essentially coincides with theirs at the tree level 
(apart from replacing irrelevant propagators by higher dimensional operators),
and therefore our cutting rule can be viewed as a generalization to loops,
with a rigorous proof based solely on the structure of the in-in formalism and microcausality.\footnote{
	The microcausality and its relation to the absence of non-local CC signals are also mentioned
	at the tree level in~\cite{Tong:2021wai}.
}
This reference also discusses a cutting rule for the local CC signals which is beyond the scope of our current discussion.

A cutting rule for non-local CC signals is extended to the loop level in~\cite{Qin:2023bjk,Qin:2023nhv}.
They rely on explicit forms of propagators, 
such as the analytic structure after performing the partial Mellin--Barnes transformation.
The main point of our paper is that the cutting rule follows 
solely from the structure of the in-in formalism and microcausality, independently of theory details.

Finally, a cutting rule on a wavefunction of the universe is discussed in~\cite{Goodhew:2020hob,Jazayeri:2021fvk,Melville:2021lst,Goodhew:2021oqg,AguiSalcedo:2023nds,
Stefanyszyn:2023qov,Ghosh:2024aqd},
relying on the basic requirements of the quantum field theory such as locality and unitary.
Furthermore, recently~\cite{Donath:2024utn} studies a cutting rule on cosmological correlators
based on the in-out formalism (see the end of Sec.~\ref{subsec:cutting_proof} for a detailed comparison),
and~\cite{Werth:2024mjg} discusses relations among cosmological correlators 
derived from the cosmological largest time equation.
Since those methods in principle describe the same physics, 
it would be interesting to see if and how their cutting rules are related to our cutting rule.
We leave this comparison as a future work.

\subsection{Tree-level scalar single exchange}
\label{subsec:nlCC_tree}

We now apply our cutting rule to extract non-local CC signals in several examples.
We begin with the simplest example: tree-level scalar single exchange diagram.
We take the interaction Lagrangian as
\begin{align}
	\mathcal{L} =  \lambda_3 a^2 \varphi'^2 \sigma,
\end{align}
where $\varphi$ is the adiabatic perturbation (which we take to be massless),
related to the curvature perturbation $\zeta$ as $\zeta = -H \varphi/\dot{{\phi}}$,\footnote{
	Strictly speaking, this relation holds only at the leading order~\cite{Maldacena:2002vr},
	but it does not affect our discussion in the following.
}
$\sigma$ is the massive particle whose CC signal is of our interest,
$a$ is the scale factor, and the prime denotes the derivative with respect to the conformal time $\tau$.
Including the contributions from the two contours, we obtain
\begin{align}
	\mathcal{L}_{1} - \mathcal{L}_{2} = \lambda_3  a^2 \left[\varphi_1'^2 \sigma_1 - \varphi_2'^2 \sigma_2\right]
	&= \lambda_3  a^2 \left[\left(\varphi_r'^2 + \frac{1}{4}\varphi_a'^2 \right) \sigma_a + 2\varphi_r'\varphi_a' \sigma_r
	\right].
\end{align}

The positive Wightman function for the massless mode $\varphi$ in de~Sitter background, $a(\tau) = -1/H\tau$, is given by
\begin{align}
	\Delta_>(k;\tau_1,\tau_2) = \frac{H^2}{2k^3}(1+ik\tau_1)(1-ik\tau_2)e^{-ik(\tau_1-\tau_2)},
\end{align}
and the bulk-to-boundary propagators in the Keldysh $r/a$ basis are given by
\begin{align}
	\Delta_{rr}(k;0,\tau) &= \frac{H^2}{2k^3}\left[\cos(k\tau) + k\tau \sin(k\tau)\right],
	\quad
	\Delta_{ra}(k;0,\tau) = \frac{iH^2}{k^3}\left[\sin(k\tau)-k\tau\cos(k\tau)\right],
\end{align}
where we do not have the Heaviside theta function for the retarded function since
$\tau = 0$ is the future boundary.
The positive Wightman function for the massive mode $\sigma$ is given by
\begin{align}
	G_>(k;\tau_1,\tau_2) = \frac{\pi (a_1a_2)^{\frac{1-d}{2}}}{4H}
	H_\nu^{(1)}(-k\tau_1) H_\nu^{(2)}(-k\tau_2),
	\quad
	\nu = \sqrt{\frac{(d-1)^2}{4}+d(d-1)\xi - \frac{m^2}{H^2}},
\end{align}
where $d$ is the spacetime dimension,
$H$ is the Hubble parameter, $m$ is the mass,
$\xi$ is the non-minimal coupling, and $H_\nu^{(1/2)}$ is the Hankel function of the first/second kind.
Here we take the Bunch--Davies vacuum as the initial condition.
For our purpose, it is more instructive to express it in terms of the Bessel function $J_{\pm\nu}(z)$ as
\begin{align}
	G_>(k;\tau_1,\tau_2) = \frac{\pi (a_1a_2)^{\frac{1-d}{2}}}{4H\sin^2 (\pi \nu)}
	&\left[J_{\nu}(z_1)J_{\nu}(z_2) + J_{-\nu}(z_1)J_{-\nu}(z_2)
	-e^{-i\pi\nu}J_{\nu}(z_1)J_{-\nu}(z_2) - e^{i\pi\nu}J_{-\nu}(z_1)J_{\nu}(z_2)
	\right],
	\label{eq:G_Bessel}
\end{align}
where $z_i = -k\tau_i$.
By noting that the Bessel function satisfies
\begin{align}
	J_{\pm\nu}(e^{2\pi i}z) = e^{\pm 2\pi i \nu} J_{\pm\nu}(z),
\end{align}
we see that the first two terms in Eq.~\eqref{eq:G_Bessel} obtain a non-trivial phase $e^{\pm 4\pi i \nu}$ 
under $k \to k e^{2\pi i}$ (which is equivalent to $z_1, z_2 \to z_1e^{2\pi i}, z_2 e^{2\pi i}$),
while the last two terms in Eq.~\eqref{eq:G_Bessel} are invariant.
This allows us to identify the first two terms as the non-local contribution and the second two terms as the local contribution,
and the late time expansions indeed agree with those given \textit{e.g.,}  in~\cite{Baumann:2018muz,Tong:2021wai}.
The propagators in the $r/a$ basis are given by
\begin{align}
	&G_{rr}(k;\tau_1,\tau_2) = \frac{\pi (a_1a_2)^{\frac{1-d}{2}}}{4H\sin^2(\pi \nu)}
	\left[J_\nu(z_1)J_\nu(z_2) - \cos(\pi\nu)J_\nu(z_1)J_{-\nu}(z_2) + (\nu \to -\nu)\right],
	\\
	&G_{ra}(k;\tau_1,\tau_2) = \theta(\tau_1-\tau_2)\frac{i\pi (a_1a_2)^{\frac{1-d}{2}}}{2H \sin(\pi \nu)}
	\left[J_\nu(z_1)J_{-\nu}(z_2) - J_{-\nu}(z_1)J_\nu(z_2)\right].
\end{align}
In particular, the retarded/advanced function does not contain the non-local contribution
since the non-local part is symmetric under $\tau_1 \leftrightarrow \tau_2$ in the Wightman function.
This is the simplest example of our discussion in Sec.~\ref{subsec:microcausality};
the fully retarded function does not contain the non-local CC signal due to the microcausality.

We now compute the four-point correlation function.
We focus on the ``$s$-channel'' process, 
\textit{i.e.}, those with $\varphi_{\vec{k}_1}$ and $\varphi_{\vec{k}_2}$ combined;
the other channels are obtained by an analogous calculation.
By applying the cutting rule for the two-point bulk correlator~\eqref{eq:cutting_2pt},
we obtain
\begin{align}
	\left.\langle \varphi_{\vec{k}_1}\varphi_{\vec{k}_2} \varphi_{\vec{k}_3} \varphi_{\vec{k}_{4}}
	\rangle
	\right\vert_{s}
	=
	\begin{tikzpicture}[baseline=(b)]
	\begin{feynman}[inline = (base.b)]
		\vertex (f1);
		\vertex [right = 0.5 of f1, label=\({\scriptstyle \{{k}_1, {k}_2\}}\)] (c1);
		\vertex [right = 1.5 of c1, label=\({\scriptstyle \{{k}_3, {k}_4\}}\)] (c2);
		\vertex [right = 0.5 of c2] (f2);
		\vertex [below = of c1, square dot] (v1){};
		\vertex [below = 0.65 of c1] (b);
		\vertex [right = 0.75 of v1] (c);
		\vertex [right = 1.5 of v1, square dot] (v2){};
		\vertex [below = 0.5 of c1] (cut0);
		\vertex [left = 0.25 of cut0] (cut1);
		\vertex [right = 0.25 of cut0] (cut2);
		\diagram*{
		(f1) -- [very thick, darkred] (f2),
		(c1) -- [very thick, anti majorana, steelblue] (v1),
		(v2) -- [very thick, fermion] (c2),
		(v1) -- [fermion] (v2),
		(cut1) -- [dashed, very thick, steelblue] (cut2),
		};
	\end{feynman}
	\end{tikzpicture}
	+
	\begin{tikzpicture}[baseline=(b)]
	\begin{feynman}[inline = (base.b)]
		\vertex (f1);
		\vertex [right = 0.5 of f1] (c1);
		\vertex [right = 1.5 of c1] (c2);
		\vertex [right = 0.5 of c2] (f2);
		\vertex [below = of c1, square dot] (v1){};
		\vertex [below = 0.65 of c1] (b);
		\vertex [right = 0.75 of v1] (c);
		\vertex [right = 1.5 of v1, square dot] (v2){};
		\vertex [below = 0.5 of c2] (cut0);
		\vertex [left = 0.25 of cut0] (cut1);
		\vertex [right = 0.25 of cut0] (cut2);
		\diagram*{
		(f1) -- [very thick, darkred] (f2),
		(v1) -- [very thick, fermion] (c1),
		(v2) -- [very thick, ,anti majorana, steelblue] (c2),
		(v2) -- [fermion] (v1),
		(cut1) -- [dashed, very thick, steelblue] (cut2),
		};
	\end{feynman}
	\end{tikzpicture}
	+
	\begin{tikzpicture}[baseline=(b)]
	\begin{feynman}[inline = (base.b)]
		\vertex (f1);
		\vertex [right = 0.5 of f1] (c1);
		\vertex [right = 1.5 of c1] (c2);
		\vertex [right = 0.5 of c2] (f2);
		\vertex [below = of c1, square dot] (v1){};
		\vertex [below = 0.65 of c1] (b);
		\vertex [right = 0.75 of v1] (c);
		\vertex [right = 1.5 of v1, square dot] (v2){};
		\vertex [above = 0.25 of c] (cut1);
		\vertex [below = 0.25 of c] (cut2);
		\diagram*{
		(f1) -- [very thick, darkred] (f2),
		(v1) -- [very thick, fermion] (c1),
		(v2) -- [very thick, fermion] (c2),
		(v1) -- [anti majorana, steelblue] (v2),
		(cut1) -- [dashed, very thick, steelblue] (cut2),
		};
	\end{feynman}
	\end{tikzpicture},
	\label{eq:cut_tree_all}
\end{align}
where
\begin{align}
	\begin{tikzpicture}[baseline=(c)]
		\begin{feynman}[inline = (base.c)]
			\vertex (f1);
			\vertex [right = 0.5 of f1, label=\({\left\{\scriptstyle k_1,k_2\right\}}\)] (c1);
			\vertex [right = 0.5 of c1] (f2);
			\vertex [below =  of c1, square dot] (v1){};
			\vertex [below = 0.5 of c1] (c);
			\diagram*{
			(f1) -- [very thick, darkred] (f2),
			(c1) -- [anti fermion, very thick] (v1),
			};
		\end{feynman}
	\end{tikzpicture}
	&=
	\Delta_{rr}'(k_1;0,\tau)\Delta_{ra}'(k_2;0,\tau)
	+ \Delta_{ra}'(k_1;0,\tau)\Delta_{rr}'(k_2;0,\tau) = \frac{iH^4\tau^2}{2k_1 k_2}\sin(k_{12}\tau),
	\\
	\begin{tikzpicture}[baseline=(c)]
		\begin{feynman}[inline = (base.c)]
			\vertex (f1);
			\vertex [right = 0.5 of f1, label=\({\left\{\scriptstyle k_1,k_2\right\}}\)] (c1);
			\vertex [right = 0.5 of c1] (f2);
			\vertex [below =  of c1, square dot] (v1){};
			\vertex [below = 0.5 of c1] (c);
			\vertex [below = 0.47 of f1] (m0);
			\vertex [below = 0.47 of f2] (m2);
			\diagram*{
			(f1) -- [very thick, darkred] (f2),
			(c1) -- [anti majorana, very thick, steelblue] (v1),
			(m0) -- [dashed, steelblue, thick] (m2),
			};
		\end{feynman}
	\end{tikzpicture}
	&=
	\Delta_{rr}'(k_1;0,\tau)\Delta_{rr}'(k_2;0,\tau)
	+ \frac{1}{4}\Delta_{ra}'(k_1;0,\tau)\Delta_{ra}'(k_2;0,\tau)
	= \frac{H^4\tau^2}{4k_1k_2}\cos(k_{12}\tau),
\end{align}
with the derivatives acting on $\tau$ originating from the interactions included here, 
and $k_{ij} = k_i + k_j$ (note that this is a scalar sum
and not a vector sum).
Since only the last term contains the non-local CC signals,
let us calculate it first. It is given by
\begin{align}
	&\begin{tikzpicture}[baseline=(b)]
	\begin{feynman}[inline = (base.b)]
		\vertex (f1);
		\vertex [right = 0.5 of f1, label=\({\left\{\scriptstyle k_1,k_2\right\}}\)] (c1);
		\vertex [right = 1.5 of c1, label=\({\left\{\scriptstyle k_3,k_4\right\}}\)] (c2);
		\vertex [right = 0.5 of c2] (f2);
		\vertex [below = of c1, square dot] (v1){};
		\vertex [below = 0.65 of c1] (b);
		\vertex [right = 0.75 of v1] (c);
		\vertex [right = 1.5 of v1, square dot] (v2){};
		\vertex [above = 0.25 of c] (cut1);
		\vertex [below = 0.25 of c] (cut2);
		\diagram*{
		(f1) -- [very thick, darkred] (f2),
		(v1) -- [very thick, fermion] (c1),
		(v2) -- [very thick, fermion] (c2),
		(v1) -- [anti majorana, steelblue] (v2),
		(cut1) -- [dashed, very thick, steelblue] (cut2),
		};
	\end{feynman}
	\end{tikzpicture}
	= \frac{\lambda_3^2 H^4}{k_1k_2k_3k_4}
	\int_{-\infty}^0 \dd\tau_L \int_{-\infty}^0 \dd\tau_R \sin(k_{12}\tau_L) \sin(k_{34}\tau_R) G_{rr}(k_s;\tau_L,\tau_R)
	\nonumber \\
	&= \frac{\pi \lambda_3^2 H^6}{4\sin^2(\pi\nu) k_1k_2k_3k_4k_{12}^{5/2}k_{34}^{5/2}}
	\left[S_{\nu;5/2}\left(\frac{k_s}{k_{12}}\right) \left(S_{\nu;5/2}\left(\frac{k_s}{k_{34}}\right)
	-\cos(\pi\nu)S_{-\nu;5/2}\left(\frac{k_s}{k_{34}}\right)\right) + (\nu \to -\nu)
	\right],
\end{align}
where $k_s = \vert \vec{k}_1 + \vec{k}_2\vert = \vert \vec{k}_3 + \vec{k}_4\vert $,
and we define
\begin{align}
	&S_{\nu;n}(x) = \lim_{\epsilon \to +0}\left[\int_0^\infty \dd z\,z^{n-1} \times e^{-\epsilon z} \sin z \times J_\nu(xz)\right],
\end{align}
with $\epsilon$ originating from the $\pm i\epsilon$ prescription of the in-in formalism~\cite{Chen:2017ryl}.
Due to the cutting rule, the conformal time integral is factorized
since $G_{rr}$ does not contain the theta function, simplifying the computation significantly.
Indeed, the conformal time integral is readily computed as
\begin{align}
	S_{\nu;n}(x) &= \left(\frac{x}{2}\right)^{\nu}\sin\left(\frac{(\nu+n)\pi}{2}\right)
	\frac{\Gamma(\nu+n)}{\Gamma(\nu+1)}{}_2F_1\left[\begin{matrix} \frac{\nu+n}{2}, \frac{\nu+n+1}{2} 
	\\ 1+\nu \end{matrix};x^2\right],
	\label{eq:Snun}
\end{align}
with ${}_2F_1$ being the Gauss's hypergeometric function, and we obtain
\begin{align}
	&\begin{tikzpicture}[baseline=(b)]
	\begin{feynman}[inline = (base.b)]
		\vertex (f1);
		\vertex [right = 0.5 of f1] (c1);
		\vertex [right = 1.5 of c1] (c2);
		\vertex [right = 0.5 of c2] (f2);
		\vertex [below = of c1, square dot] (v1){};
		\vertex [below = 0.65 of c1] (b);
		\vertex [right = 0.75 of v1] (c);
		\vertex [right = 1.5 of v1, square dot] (v2){};
		\vertex [above = 0.25 of c] (cut1);
		\vertex [below = 0.25 of c] (cut2);
		\diagram*{
		(f1) -- [very thick, darkred] (f2),
		(v1) -- [very thick, fermion] (c1),
		(v2) -- [very thick, fermion] (c2),
		(v1) -- [anti majorana, steelblue] (v2),
		(cut1) -- [dashed, very thick, steelblue] (cut2),
		};
	\end{feynman}
	\end{tikzpicture}
	\nonumber \\
	&= \frac{\lambda_3^2 H^6 [1+\sin(\pi\nu)]\Gamma^2(-\nu)\Gamma^2\left(\frac{5}{2}+\nu\right)}
	{8\pi k_1k_2k_3k_4 k_{12}^{5/2}k_{34}^{5/2}}
	\left(\frac{k_s^2}{4k_{12}k_{34}}\right)^{\nu}
	{}_2F_1\left[
	\begin{matrix}
	\frac{5+2\nu}{4},\frac{7+2\nu}{4} \\
	1+\nu
	\end{matrix};\frac{k_s^2}{k_{12}^2}\right]
	{}_2F_1\left[
	\begin{matrix}
	\frac{5+2\nu}{4},\frac{7+2\nu}{4} \\
	1+\nu
	\end{matrix};\frac{k_s^2}{k_{34}^2}\right]
	\nonumber \\
	&+\frac{\lambda_3^2 H^6 \cos(\pi \nu) \Gamma\left(\frac{5}{2}+\nu\right)\Gamma\left(\frac{5}{2}-\nu\right)}
	{8\nu\tan(\pi \nu)k_1k_2k_3k_4 k_{12}^{5/2}k_{34}^{5/2}}
	\left(\frac{k_{34}}{k_{12}}\right)^{\nu}
	{}_2F_1\left[
	\begin{matrix}
	\frac{5+2\nu}{4},\frac{7+2\nu}{4} \\
	1+\nu
	\end{matrix}; \frac{k_s^2}{k_{12}^2}\right]
	{}_2F_1\left[
	\begin{matrix}
	\frac{5-2\nu}{4},\frac{7-2\nu}{4} \\
	1-\nu
	\end{matrix};\frac{k_s^2}{k_{34}^2}\right]
	+ (\nu \to -\nu),
	\label{eq:cut_tree}
\end{align}
where the delta function for the total spatial momentum conservation is stripped out (here and hereafter).
The first term depends non-analytically on $k_s/k_{12}$ and $k_s/k_{34}$ and is the non-local CC signals.
As we have argued in Sec.~\ref{subsec:microcausality}, this is the full non-local CC contribution,
and this indeed reproduces the result in~\cite{Qin:2022lva}. 

Our main focus in this paper is the non-local CC signals, and the first term in Eq.~\eqref{eq:cut_tree} suffices for this purpose.
Nevertheless, it is interesting to investigate the local CC signals in this example.
Although analytic in $k_s/k_{12}$ and $k_s/k_{34}$,
the second term in Eq.~\eqref{eq:cut_tree} contains non-analytic dependence on $k_{34}/k_{12}$,
corresponding to the local CC signals discussed in~\cite{Wang:2020ioa,Tong:2021wai}.
Since higher dimensional operators can induce only analytic terms in $k_s/k_{12}$, $k_s/k_{34}$
and $k_{34}/k_{12}$, the second term should also arise from the effect of the particle production.
Therefore, the bulk-cut diagram contains only particle production contributions 
and no higher dimensional operators at the tree level.
Note that the inverse is \emph{false};
the boundary-cut diagrams, \emph{i.e.}, the first two diagrams in Eq.~\eqref{eq:cut_tree_all}, 
contain both background and local CC signals,
meaning that the second term in Eq.~\eqref{eq:cut_tree} is \emph{not} the full local CC signal.
To see this, we may take the folded limit $k_4 \to 0$. 
In this case, there is no distinction between the non-local and
local CC signals since $k_s/k_{12}, k_{34}/k_{12} \to k_3/k_{12}$.
The non-local CC signal in Eq.~\eqref{eq:cut_tree} diverges in this limit since the hypergeometric function becomes singular.
As the full result should be convergent in the folded limit in the Bunch--Davies vacuum~\cite{Flauger:2013hra,Arkani-Hamed:2015bza,Arkani-Hamed:2018kmz,Green:2020whw},
this divergence should be canceled with the local CC signals 
(the background contribution is analytic in $k_3/k_{12}$ and therefore cannot cancel it).
However, the local CC signal in Eq.~\eqref{eq:cut_tree} alone does not, 
indicating that there is another local CC contribution.

To take a closer look at the last point, we may implement a procedure similar to~\cite{Tong:2021wai}.
We first note that the divergence at $k_4 \to 0$ can be understood as follows.
The conformal time integral of the vertex on the right is expressed as
\begin{align}
	\int_0^\infty \dd z\,z^{n-1} \sin z 
	\left[J_{\nu}\left(\frac{k_s}{k_{34}}z\right) - \cos (\pi \nu) J_{-\nu}\left(\frac{k_s}{k_{34}}z\right)\right],
\end{align}
and the limit $k_4 \to 0$ corresponds to $k_s/k_{34} \to 1$.
In this limit, the oscillation frequencies of $\sin z$ and $J_\nu(k_s z/k_{34})$ at $z \to \infty$ coincide with each other
and the product contains a non-oscillatory constant term, resulting in divergence (even after the $i\epsilon$ prescription).
To get a convergent expression, we may organize the boundary-cut diagram as
\begin{align}
	\begin{tikzpicture}[baseline=(b)]
	\begin{feynman}[inline = (base.b)]
		\vertex (f1);
		\vertex [right = 0.5 of f1] (c1);
		\vertex [right = 1.5 of c1] (c2);
		\vertex [right = 0.5 of c2] (f2);
		\vertex [below = of c1, square dot] (v1){};
		\vertex [below = 0.65 of c1] (b);
		\vertex [right = 0.75 of v1] (c);
		\vertex [right = 1.5 of v1, square dot] (v2){};
		\vertex [below = 0.5 of c2] (cut0);
		\vertex [left = 0.25 of cut0] (cut1);
		\vertex [right = 0.25 of cut0] (cut2);
		\diagram*{
		(f1) -- [very thick, darkred] (f2),
		(v1) -- [very thick, fermion] (c1),
		(v2) -- [very thick, anti majorana, steelblue] (c2),
		(v2) -- [fermion] (v1),
		(cut1) -- [dashed, very thick, steelblue] (cut2),
		};
	\end{feynman}
	\end{tikzpicture}
	&=
	\begin{tikzpicture}[baseline=(b)]
	\begin{feynman}[inline = (base.b)]
		\vertex (f1);
		\vertex [right = 0.5 of f1] (c1);
		\vertex [right = 1.5 of c1] (c2);
		\vertex [right = 0.5 of c2] (f2);
		\vertex [below = of c1, square dot] (v1){};
		\vertex [below = 0.65 of c1] (b);
		\vertex [right = 0.75 of v1] (c);
		\vertex [right = 1.5 of v1, square dot] (v2){};
		\vertex [below = 0.5 of c2] (cut0);
		\vertex [left = 0.25 of cut0] (cut1);
		\vertex [right = 0.25 of cut0] (cut2);
		\diagram*{
		(f1) -- [very thick, darkred] (f2),
		(v1) -- [very thick, fermion] (c1),
		(v2) -- [very thick, anti majorana, steelblue] (c2),
		(v1) -- [fermion] (v2),
		(cut1) -- [dashed, very thick, steelblue] (cut2),
		};
	\end{feynman}
	\end{tikzpicture}
	+
	\begin{tikzpicture}[baseline=(b)]
	\begin{feynman}[inline = (base.b)]
		\vertex (f1);
		\vertex [right = 0.5 of f1] (c1);
		\vertex [right = 1.5 of c1] (c2);
		\vertex [right = 0.5 of c2] (f2);
		\vertex [below = of c1, square dot] (v1){};
		\vertex [below = 0.65 of c1] (b);
		\vertex [right = 0.75 of v1] (c);
		\vertex [right = 1.5 of v1, square dot] (v2){};
		\vertex [below = 0.5 of c2] (cut0);
		\vertex [left = 0.25 of cut0] (cut1);
		\vertex [right = 0.25 of cut0] (cut2);
		\diagram*{
		(f1) -- [very thick, darkred] (f2),
		(v1) -- [very thick, fermion] (c1),
		(v2) -- [very thick, ,anti majorana, steelblue] (c2),
		(v2) -- [charged scalar] (v1),
		(cut1) -- [dashed, very thick, steelblue] (cut2),
		};
	\end{feynman}
	\end{tikzpicture},
\end{align}
where the diagrams on the right-hand side should be understood as merely replacing the massive propagator
in the diagram on the left-hand side by using the identity of the theta function.
By adding the last diagram to Eq.~\eqref{eq:cut_tree}, we then obtain
\begin{align}
	&\begin{tikzpicture}[baseline=(b)]
	\begin{feynman}[inline = (base.b)]
		\vertex (f1);
		\vertex [right = 0.5 of f1] (c1);
		\vertex [right = 1.5 of c1] (c2);
		\vertex [right = 0.5 of c2] (f2);
		\vertex [below = of c1, square dot] (v1){};
		\vertex [below = 0.65 of c1] (b);
		\vertex [right = 0.75 of v1] (c);
		\vertex [right = 1.5 of v1, square dot] (v2){};
		\vertex [above = 0.25 of c] (cut1);
		\vertex [below = 0.25 of c] (cut2);
		\diagram*{
		(f1) -- [very thick, darkred] (f2),
		(v1) -- [very thick, fermion] (c1),
		(v2) -- [very thick, fermion] (c2),
		(v1) -- [anti majorana, steelblue] (v2),
		(cut1) -- [dashed, very thick, steelblue] (cut2),
		};
	\end{feynman}
	\end{tikzpicture}
	+
	\begin{tikzpicture}[baseline=(b)]
	\begin{feynman}[inline = (base.b)]
		\vertex (f1);
		\vertex [right = 0.5 of f1] (c1);
		\vertex [right = 1.5 of c1] (c2);
		\vertex [right = 0.5 of c2] (f2);
		\vertex [below = of c1, square dot] (v1){};
		\vertex [below = 0.65 of c1] (b);
		\vertex [right = 0.75 of v1] (c);
		\vertex [right = 1.5 of v1, square dot] (v2){};
		\vertex [below = 0.5 of c2] (cut0);
		\vertex [left = 0.25 of cut0] (cut1);
		\vertex [right = 0.25 of cut0] (cut2);
		\diagram*{
		(f1) -- [very thick, darkred] (f2),
		(v1) -- [very thick, fermion] (c1),
		(v2) -- [very thick, ,anti majorana, steelblue] (c2),
		(v2) -- [charged scalar] (v1),
		(cut1) -- [dashed, very thick, steelblue] (cut2),
		};
	\end{feynman}
	\end{tikzpicture}
	\nonumber \\
	&=
	\frac{\pi \lambda_3^2 H^6}{4\sin(\pi\nu)k_1k_2k_3k_4k_{12}^{5/2}k_{34}^{5/2}}
	S_{\nu;5/2}\left(\frac{k_s}{k_{12}}\right)
	\times
	\int_0^\infty \dd z\left[e^{iz} H_{-\nu}^{(1)}\left(\frac{k_s}{k_{34}}z\right)
	+ e^{-iz} H_{-\nu}^{(2)}\left(\frac{k_s}{k_{34}}z\right)\right]
	+ (\nu \to -\nu).
	\label{eq:folded_limit_tree}
\end{align}
The integrand now contains only the oscillatory terms even for $k_4 \to 0$
since $H_\nu^{(1)}(z) \sim e^{iz}/\sqrt{z}$ and $H_\nu^{(2)}(z) \sim e^{-iz}/\sqrt{z}$ at $z \to \infty$,
and hence the result is convergent after the $i\epsilon$ prescription.\footnote{
	This holds only for the Bunch--Davies vacuum and not for other choices of the initial condition for $\sigma$.
}
The second diagram in Eq.~\eqref{eq:folded_limit_tree} is essential to obtain this result.

Our findings at the tree-level single exchange are summarized as follows:
\begin{itemize}
\item The non-local CC signal arises solely from the bulk-cut diagram as expected.
\item Moreover, the bulk-cut diagram contains only non-local and local CC signals.
\item The boundary-cut diagram contains the local CC signal and backgrounds.
\end{itemize}
The cutting rule discussed in Sec.~\ref{sec:cutting} efficiently extracts the non-local CC signal,
but the last property indicates that the local CC signals are distributed in different cut diagrams and are more subtle.
It would be interesting to see if we can in general extend our cutting rule to incorporate the local CC signals
by, \textit{e.g.}, shuffling the diagrams further, as we have done above at the tree level.
We leave a detailed study on this point for future work and focus on the non-local CC signals in the rest.

\subsection{One-loop scalar bubble}
\label{subsec:nlCC_bubble}

We next consider a slightly more non-trivial example: a scalar one-loop bubble diagram.
We take the interaction as
\begin{align}
	\mathcal{L} = \frac{\lambda_4}{4}a^2 \varphi'^2 \sigma^2,
	\label{eq:Lint_loop}
\end{align}
where $\sigma$ is again the massive particle whose non-local CC is of our interest.
In the $r/a$ basis, this is translated to
\begin{align}
	\mathcal{L}_1 - \mathcal{L}_2 = \frac{\lambda_4}{2}a^2\left[
	\varphi_r' \varphi_a' \left(\chi_r^2 + \frac{\chi_a^2}{4}\right)
	+ \left(\varphi_r'^2 + \frac{\varphi_a'^2}{4}\right) \chi_r \chi_a
	\right].
\end{align}
Since our main interest is the non-local CC signals, we focus on the bulk-cut diagram.
It is given by
\begin{align}
	\begin{tikzpicture}[baseline=(v1)]
	\begin{feynman}[inline = (base.v1)]
		\vertex (f1);
		\vertex [right = 0.5 of f1, label=\({\scriptstyle \{{k}_1, {k}_2\}}\)] (c1);
		\vertex [right = 0.75 of c1] (m1);
		\vertex [right = 1.5 of c1, label=\({\scriptstyle \{{k}_3, {k}_4\}}\)] (c2);
		\vertex [right = 0.5 of c2] (f2);
		\vertex [below = of c1, square dot] (v1){};
		\vertex [right = 1.5 of v1, square dot] (v2){};
		\vertex [right = 0.75 of v1] (c);
		\vertex [right = 0.75 of v1] (d2);
		\node [above = -0.025 of c] (vdots);
		\vertex [above = 0.75 of c] (m1);
		\vertex [below = 0.75 of c] (m2);
		\vertex [above = 0.25 of c] (center);
		\diagram*{
		(f1) -- [very thick, darkred] (f2),
		(c1) -- [very thick, anti fermion] (v1),
		(c2) -- [very thick, anti fermion] (v2),
		(v1) -- [out=290, in=250, anti majorana, steelblue] (v2),
		(v1) -- [out=70, in=110, anti majorana, steelblue] (v2),
		(m1) -- [dashed, steelblue, very thick] (m2),
		};
	\end{feynman}
	\end{tikzpicture}
	&= \frac{\lambda_4^2 H^4}{16k_1k_2k_3k_4}
	\int_{-\infty}^0 \dd\tau_L \int_{-\infty}^0 \dd\tau_R
	\int\frac{\dd^3 l}{(2\pi)^3} \sin(k_{12}\tau_L) \sin(k_{34}\tau_R)
	\nonumber \\[-.5em]
	&\times
	\left[G_>(l;\tau_L,\tau_R)G_>(\vert \vec{l}+\vec{k}_s\vert;\tau_L,\tau_R)
	+ G_<(l;\tau_L,\tau_R)G_<(\vert \vec{l}+\vec{k}_s\vert;\tau_L,\tau_R)
	\right],
\end{align}
where $\vec{k}_s = \vec{k}_1 + \vec{k}_2$ 
and we have set $d = 3$ since the non-local CC signal does not diverge.
We again emphasize that this expression does not contain the Heaviside theta function due to the cutting rule,
and therefore the conformal time integrals are factorized.
This diagram contains not only the non-local CC signals but other contributions.
We extract the non-local CC signals by picking up the non-local part of the Wightman functions as
\begin{align}
	\begin{tikzpicture}[baseline=(center)]
	\begin{feynman}[inline = (base.center)]
		\vertex (f1);
		\vertex [right = 0.5 of f1] (c1);
		\vertex [right = 0.75 of c1] (m1);
		\vertex [right = 1.5 of c1] (c2);
		\vertex [right = 0.5 of c2] (f2);
		\vertex [below = of c1, square dot] (v1){};
		\vertex [right = 1.5 of v1, square dot] (v2){};
		\vertex [right = 0.75 of v1] (c);
		\vertex [right = 0.75 of v1] (d2);
		\node [above = -0.025 of c] (vdots);
		\vertex [above = 0.75 of c] (m1);
		\vertex [below = 0.75 of c, label=270:\({\scriptstyle \color{steelblue}{\text{nlCC}}}\)] (m2);
		\vertex [above = 0. of c] (center);
		\diagram*{
		(f1) -- [very thick, darkred] (f2),
		(c1) -- [very thick, anti fermion] (v1),
		(c2) -- [very thick, anti fermion] (v2),
		(v1) -- [out=290, in=250, anti majorana, steelblue] (v2),
		(v1) -- [out=70, in=110, anti majorana, steelblue] (v2),
		(m1) -- [dashed, steelblue, very thick] (m2),
		};
	\end{feynman}
	\end{tikzpicture}
	&= \frac{\pi^2 \lambda_4^2 H^8}{128 \sin^4(\pi \nu) k_1 k_2 k_3 k_4}
	\int_{-\infty}^{0} \dd\tau_L \int_{-\infty}^{0} \dd\tau_R
	\int\frac{\dd^3 l}{(2\pi)^3}
	\sin(k_{12}\tau_L)\sin(k_{34}\tau_R) \tau_L^3 \tau_R^3
	\nonumber \\[-1em]
	&\times
	\sum_{\nu_1,\nu_2 = \pm \nu}
	J_{\nu_1}(-l\tau_L)J_{\nu_1}(-l\tau_R) J_{\nu_2}(-\vert \vec{l}+\vec{k}_s\vert \tau_L) J_{\nu_2}
	(-\vert \vec{l}+\vec{k}_s\vert \tau_R).
	\label{eq:bubble_nlCC}
\end{align}
To evaluate the integrals, we may perform the polynomial expansion of the Bessel function.
By using
\begin{align}
	\int \frac{\dd^3 l}{(2\pi)^3} l^{2\nu_1} \vert \vec{l}+\vec{k}_s\vert^{2\nu_2}
	&= \frac{(k_s^2)^{3/2+\nu_1+\nu_2}}{(4\pi)^{3/2}}
	\Gamma\left[{\nu_1+3/2,\nu_2+3/2,-\nu_{12}-3/2}\atop{-\nu_1,-\nu_2,\nu_{12}+3}\right],
\end{align}
for non-integer $\nu_1, \nu_2$ with $\nu_{12} = \nu_1 + \nu_2$, and
\begin{align}
	&\lim_{\epsilon \to +0} \left[\int_0^\infty \dd z\,z^{n-1} e^{-\epsilon z}\sin z\right]
	= \Gamma(n) \sin\frac{n\pi}{2},
	\label{eq:time_int_sin}
\end{align}
we obtain
\begin{align}
	\begin{tikzpicture}[baseline=(center)]
	\begin{feynman}[inline = (base.center)]
		\vertex (f1);
		\vertex [right = 0.5 of f1] (c1);
		\vertex [right = 0.75 of c1] (m1);
		\vertex [right = 1.5 of c1] (c2);
		\vertex [right = 0.5 of c2] (f2);
		\vertex [below = of c1, square dot] (v1){};
		\vertex [right = 1.5 of v1, square dot] (v2){};
		\vertex [right = 0.75 of v1] (c);
		\vertex [right = 0.75 of v1] (d2);
		\node [above = -0.025 of c] (vdots);
		\vertex [above = 0.75 of c] (m1);
		\vertex [below = 0.75 of c, label=270:\({\scriptstyle \color{steelblue}{\text{nlCC}}}\)] (m2);
		\vertex [above = 0. of c] (center);
		\diagram*{
		(f1) -- [very thick, darkred] (f2),
		(c1) -- [very thick, anti fermion] (v1),
		(c2) -- [very thick, anti fermion] (v2),
		(v1) -- [out=290, in=250, anti majorana, steelblue] (v2),
		(v1) -- [out=70, in=110, anti majorana, steelblue] (v2),
		(m1) -- [dashed, steelblue, very thick] (m2),
		};
	\end{feynman}
	\end{tikzpicture}
	&= \frac{\lambda_4^2 H^8\sin^2(\pi \nu)}{128 \pi^{7/2} k_1 k_2 k_3 k_4 k_{12}^{5/2}k_{34}^{5/2}}
	\left(\frac{k_s^2}{4k_{12}k_{34}}\right)^{3/2+2\nu}
	\sum_{n_1,\cdots,n_4=0}^{\infty}
	(-)^{n_{1234}}
	\left(\frac{k_s}{2k_{12}}\right)^{2n_{12}}\left(\frac{k_s}{2k_{34}}\right)^{2n_{34}}
	\nonumber \\[-1em]
	&\times
	\Gamma\left[{2(\nu+n_{12}+2),2(\nu+n_{34}+2),-n_1-\nu,-n_2-\nu,-n_3-\nu,-n_4-\nu}
	\atop{n_1+1,n_2+1,n_3+1,n_4+1}\right]
	\nonumber \\
	&\times 
	\Gamma\left[{\nu+n_{13}+3/2,\nu+n_{24}+3/2,-2\nu-n_{1234}-3/2}
	\atop{-\nu-n_{13},-\nu-n_{24},2\nu+n_{1234}+3}\right]
	+ (\nu \leftrightarrow -\nu),
\end{align}
where we use the short-hand notation such as $n_{ij} = n_i + n_j$, $n_{1234} = n_1 + n_2 + n_3 + n_4$ and 
\begin{align}
	\Gamma\left[{a,b,c,\cdots}\atop{n,m,l,\cdots} \right]
	= \frac{\Gamma(a)\Gamma(b)\Gamma(c)\cdots}{\Gamma(n)\Gamma(m)\Gamma(l)\cdots}.
\end{align}
Note that only the terms with $\nu_1 = \nu_2$ in the sum of Eq.~\eqref{eq:bubble_nlCC} 
survive the conformal time integral~\eqref{eq:time_int_sin}.
If we have two propagators with different masses, the other terms also generate non-local CC signals
with the exponent given by the difference of the weights $\pm (\nu_1 - \nu_2)$~\cite{Aoki:2020zbj}.
This agrees with the result in~\cite{Qin:2022lva}.
One can obtain a more compact expression with the help of the spectral decomposition~\cite{Xianyu:2022jwk,Qin:2023bjk},
but this expression is enough for our purpose of demonstrating the cutting rule.

\subsection{One-loop scalar triangle}
\label{subsec:nlCC_triangle}

As the last example in this paper, we consider a scalar one-loop triangle diagram.
We take the interaction Lagrangian the same as the scalar one-loop bubble case, \textit{i.e.}, Eq.~\eqref{eq:Lint_loop}.
The cutting rule tells us that the non-local CC signals are given by
\begin{align}
	\left[
	\begin{tikzpicture}[baseline=(v2)]
	\begin{feynman}[inline = (base.v2)]
		\vertex (f1);
		\vertex [right = 0.5 of f1, label=\({\scriptstyle \{k_1,k_2\}}\)] (c1);
		\vertex [right = 1 of c1, label=\({\scriptstyle \{k_3,k_4\}}\)] (c2);
		\vertex [right = 2 of c1, label=\({\scriptstyle \{k_5,k_6\}}\)] (c3);
		\vertex [right = 0.5 of c3] (f2);
		\vertex [below = 1.5 of c1, square dot] (v1){};
		\vertex [right = of v1] (c);
		\vertex [above = 0.75 of c, square dot] (v2){};
		\vertex [right = 2 of v1, square dot] (v3){};
		\diagram*{
		(f1) -- [very thick, darkred] (f2),
		(c1) -- [very thick] (v1),
		(c2) -- [very thick] (v2),
		(c3) -- [very thick] (v3),
		(v1) -- (v3),
		(v2) -- (v3),
		(v1) -- (v2),
		};
	\end{feynman}
	\end{tikzpicture}
	\right]_{\text{nlCC}}
	&= 
	\left[
	\begin{tikzpicture}[baseline=(v2)]
	\begin{feynman}[inline = (base.v2)]
		\vertex (f1);
		\vertex [right = 0.5 of f1] (c1);
		\vertex [right = 0.75 of c1] (c2);
		\vertex [right = 1.5 of c1] (c3);
		\vertex [right = 0.5 of c3] (f2);
		\vertex [below = 1.5 of c1, square dot] (v1){};
		\vertex [right = 0.75 of v1] (c);
		\vertex [above = 0.75 of c, square dot] (v2){};
		\vertex [right = 1.5 of v1, square dot] (v3){};
		\vertex [below = 0.75 of c1] (m1);
		\vertex [below = 0.375 of c2] (m2);
		\vertex [below = 0.75 of c3] (m3);
		\vertex [below left = 0.53 of v2] (m4);
		\vertex [below right = 0.53 of v2] (m5);
		\vertex [right = 0.75 of v1] (m6);
		\vertex [left = 0.25 of m2] (cut1);
		\vertex [right = 0.25 of m2] (cut2);
		\vertex [above right = 0.25 of m5] (cut3);
		\vertex [below left = 0.9 of m5] (cut4);
		\vertex [below = 0.5 of v2] (center);
		\vertex [below = 0.95 of center, label=\({\scriptstyle \color{steelblue}{\text{nlCC}}}\)] (nlCC);
		\diagram*{
		(f1) -- [very thick, darkred] (f2),
		(v1) -- [very thick, fermion] (c1),
		(v2) -- [very thick, steelblue, anti majorana] (c2),
		(v3) -- [very thick, fermion] (c3),
		(v1) -- [steelblue, anti majorana] (v3),
		(v2) -- [steelblue, anti majorana] (v3),
		(v1) -- (v2),
		(cut1) -- [dashed, steelblue, very thick] (cut2),
		(cut3) -- [dashed, steelblue, very thick] (cut4),
		};
	\end{feynman}
	\end{tikzpicture}
	+ \text{($5$ perms.)}
	\right]
	+
	\begin{tikzpicture}[baseline=(v2)]
	\begin{feynman}[inline = (base.v2)]
		\vertex (f1);
		\vertex [right = 0.5 of f1] (c1);
		\vertex [right = 0.75 of c1] (c2);
		\vertex [right = 1.5 of c1] (c3);
		\vertex [right = 0.5 of c3] (f2);
		\vertex [below = 1.5 of c1, square dot] (v1){};
		\vertex [right = 0.75 of v1] (c);
		\vertex [above = 0.75 of c, square dot] (v2){};
		\vertex [right = 1.5 of v1, square dot] (v3){};
		\vertex [below = 0.75 of c1] (m1);
		\vertex [below = 0.375 of c2] (m2);
		\vertex [below = 0.75 of c3] (m3);
		\vertex [below left = 0.53 of v2] (m4);
		\vertex [below right = 0.53 of v2] (m5);
		\vertex [right = 0.75 of v1] (m6);
		\vertex [below = 0.5 of v2] (center);
		\vertex [left = 0.6 of center] (cut1m);
		\vertex [above = 0.16 of cut1m] (cut1);
		\vertex [right = 0.6 of center] (cut2m);
		\vertex [above = 0.16 of cut2m] (cut2);
		\vertex [below = 0.5 of center] (cut3);
		\vertex [below = 0.95 of center, label=\({\scriptstyle \color{steelblue}{\text{nlCC}}}\)] (nlCC);
		\diagram*{
		(f1) -- [very thick, darkred] (f2),
		(v1) -- [very thick, fermion] (c1),
		(v2) -- [very thick, fermion] (c2),
		(v3) -- [very thick, fermion] (c3),
		(v1) -- [steelblue, anti majorana] (v3),
		(v2) -- [steelblue, anti majorana] (v3),
		(v1) -- [steelblue, anti majorana] (v2),
		(cut1) -- [dashed, steelblue, very thick] (center),
		(cut2) -- [dashed, steelblue, very thick] (center),
		(cut3) -- [dashed, steelblue, very thick] (center),
		};
	\end{feynman}
	\end{tikzpicture},
\end{align}
where we consider a six-point correlator to separate the non-local CC signals from the local ones.
The first diagrams on the right-hand side contain only two cut-propagators in the bulk 
and are similar to the bulk-cut diagrams of the bubble topology.
Therefore we may focus on the last term in the following.
By taking the non-local parts from all the cut-propagators, we obtain
\begin{align}
	\begin{tikzpicture}[baseline=(center)]
	\begin{feynman}[inline = (base.center)]
		\vertex (f1);
		\vertex [right = 0.5 of f1] (c1);
		\vertex [right = 0.75 of c1] (c2);
		\vertex [right = 1.5 of c1] (c3);
		\vertex [right = 0.5 of c3] (f2);
		\vertex [below = 1.5 of c1, square dot] (v1){};
		\vertex [right = 0.75 of v1] (c);
		\vertex [above = 0.75 of c, square dot] (v2){};
		\vertex [right = 1.5 of v1, square dot] (v3){};
		\vertex [below = 0.75 of c1] (m1);
		\vertex [below = 0.375 of c2] (m2);
		\vertex [below = 0.75 of c3] (m3);
		\vertex [below left = 0.53 of v2] (m4);
		\vertex [below right = 0.53 of v2] (m5);
		\vertex [right = 0.75 of v1] (m6);
		\vertex [below = 0.5 of v2] (center);
		\vertex [left = 0.6 of center] (cut1m);
		\vertex [above = 0.16 of cut1m] (cut1);
		\vertex [right = 0.6 of center] (cut2m);
		\vertex [above = 0.16 of cut2m] (cut2);
		\vertex [below = 0.5 of center] (cut3);
		\vertex [below = 0.95 of center, label=\({\scriptstyle \color{steelblue}{\text{nlCC}}}\)] (nlCC);
		\diagram*{
		(f1) -- [very thick, darkred] (f2),
		(v1) -- [very thick, fermion] (c1),
		(v2) -- [very thick, fermion] (c2),
		(v3) -- [very thick, fermion] (c3),
		(v1) -- [steelblue, anti majorana] (v3),
		(v2) -- [steelblue, anti majorana] (v3),
		(v1) -- [steelblue, anti majorana] (v2),
		(cut1) -- [dashed, steelblue, very thick] (center),
		(cut2) -- [dashed, steelblue, very thick] (center),
		(cut3) -- [dashed, steelblue, very thick] (center),
		};
	\end{feynman}
	\end{tikzpicture}
	\!\!\!\!\!&= -\frac{\pi \lambda_4^3 H^{12}}{512 \sin^6(\pi \nu) k_1 k_2 k_3 k_4 k_5 k_6 k_{12}^{4}k_{34}^4 k_{56}^4} \nonumber \\[-1em]
	&\times \int_0^\infty \dd z_L \,z_L^3 \sin z_L 
	\int_0^\infty \dd z_M \,z_M^3 \sin z_M
	\int_0^\infty \dd z_R \,z_R^3 \sin z_R \,
	\int\frac{\dd^3 l}{(2\pi)^3}
	\sum_{\nu_1,\nu_2,\nu_3=\pm\nu}
	\nonumber \\
	&\times
	J_{\nu_1}\left(\frac{\vert \vec{l}-\vec{k}_{12}\vert}{k_{12}}z_L\right) 
	J_{\nu_1} \left(\frac{\vert \vec{l}-\vec{k}_{12}\vert}{k_{34}}z_M\right)
	J_{\nu_2}\left(\frac{lz_L}{k_{12}}\right) 
	J_{\nu_2} \left(\frac{lz_R}{k_{56}}\right)
	J_{\nu_3}\left(\frac{\vert \vec{l}+\vec{k}_{56}\vert}{k_{34}}z_M\right) 
	J_{\nu_3} \left(\frac{\vert \vec{l}+\vec{k}_{56}\vert}{k_{56}}z_R\right),
	\label{eq:triangle_nlCC}
\end{align}
where $\vec{k}_{ij} = \vec{k}_i + \vec{k}_j$ and we set $d = 4$ due to the absence of divergences in this contribution.
We may again perform the polynomial expansion of the Bessel functions,
and use
\begin{align}
	\int\frac{\dd^3 l}{(2\pi)^3}\vert \vec{l}-\vec{k}_{12}\vert^{2\nu_1} l^{2\nu_2} \vert \vec{l}+\vec{k}_{56}\vert^{2\nu_3}
	= \frac{\vert \vec{k}_{12}\vert^{\nu_{12} + 1}\vert \vec{k}_{34}\vert^{\nu_{13} + 1}
	\vert \vec{k}_{56}\vert^{\nu_{23} + 1}}{(4\pi)^{3/2}}
	\Gamma\left[\begin{matrix} -\nu_{123}-3/2 \\ -\nu_1, -\nu_2, -\nu_3 \end{matrix}\right]
	F(\nu_1,\nu_2,\nu_3),
\end{align}
for non-integer $\nu_1, \nu_2, \nu_3$ where
\begin{align}
	F(\nu_1,\nu_2,\nu_3) &=
	\int_0^1 \dd x\,\dd y\,\dd z\,\delta(x+y+z-1)
	x^{-\nu_1 - 1} y^{-\nu_2 - 1} z^{-\nu_3 -1}
	\frac{\left[xy \vert \vec{k}_{12} \vert^2 + zx \vert \vec{k}_{34} \vert^2 + yz 
	\vert \vec{k}_{56} \vert^2 \right]^{\nu_{123}+3/2}}
	{\vert \vec{k}_{12}\vert^{\nu_{12} + 1}\vert \vec{k}_{34}\vert^{\nu_{13} + 1}
	\vert \vec{k}_{56}\vert^{\nu_{23} + 1}},
\end{align}
with $\nu_{ij} = \nu_i + \nu_j$, $\nu_{123} = \nu_1 + \nu_2 + \nu_3$, and Eq.~\eqref{eq:time_int_sin}. 
We then obtain
\begin{align}
	\begin{tikzpicture}[baseline=(center)]
	\begin{feynman}[inline = (base.center)]
		\vertex (f1);
		\vertex [right = 0.5 of f1] (c1);
		\vertex [right = 0.75 of c1] (c2);
		\vertex [right = 1.5 of c1] (c3);
		\vertex [right = 0.5 of c3] (f2);
		\vertex [below = 1.5 of c1, square dot] (v1){};
		\vertex [right = 0.75 of v1] (c);
		\vertex [above = 0.75 of c, square dot] (v2){};
		\vertex [right = 1.5 of v1, square dot] (v3){};
		\vertex [below = 0.75 of c1] (m1);
		\vertex [below = 0.375 of c2] (m2);
		\vertex [below = 0.75 of c3] (m3);
		\vertex [below left = 0.53 of v2] (m4);
		\vertex [below right = 0.53 of v2] (m5);
		\vertex [right = 0.75 of v1] (m6);
		\vertex [below = 0.5 of v2] (center);
		\vertex [left = 0.6 of center] (cut1m);
		\vertex [above = 0.16 of cut1m] (cut1);
		\vertex [right = 0.6 of center] (cut2m);
		\vertex [above = 0.16 of cut2m] (cut2);
		\vertex [below = 0.5 of center] (cut3);
		\vertex [below = 0.95 of center, label=\({\scriptstyle \color{steelblue}{\text{nlCC}}}\)] (nlCC);
		\diagram*{
		(f1) -- [very thick, darkred] (f2),
		(v1) -- [very thick, fermion] (c1),
		(v2) -- [very thick, fermion] (c2),
		(v3) -- [very thick, fermion] (c3),
		(v1) -- [steelblue, anti majorana] (v3),
		(v2) -- [steelblue, anti majorana] (v3),
		(v1) -- [steelblue, anti majorana] (v2),
		(cut1) -- [dashed, steelblue, very thick] (center),
		(cut2) -- [dashed, steelblue, very thick] (center),
		(cut3) -- [dashed, steelblue, very thick] (center),
		};
	\end{feynman}
	\end{tikzpicture}
	&=
	-\frac{\lambda_4^3 H^{12}\sin^3 (\pi\nu) \Gamma^3\left(2\nu + 4\right) \Gamma^3(-\nu)}
	{4096 \pi^{13/2} k_1 k_2 k_3 k_4 k_5 k_6 k_{12}^3 k_{34}^3 k_{56}^3}
	\Gamma\left(-3\nu - \frac{3}{2}\right)
	\left(\frac{\vert \vec{k}_{12} \vert \vert \vec{k}_{34}\vert \vert \vec{k}_{56}\vert}{8k_{12}k_{34}k_{56}}\right)^{2\nu+1}
	F(\nu,\nu,\nu) 
	\nonumber \\[-.5em]
	&+ (\nu \to -\nu) + \cdots,
\end{align}
where we write down only the first terms in the polynomial expansion as the full expression is lengthy
(although straightforward to derive).
Note that only the contributions with $\nu_1 = \nu_2 = \nu_3$ in Eq.~\eqref{eq:triangle_nlCC}
survive the conformal time integrals, as in the case of the bubble diagram.
If the scalar fields inside the loop have different masses, those are given by the sums
of the three weights $\pm \nu_1, \pm\nu_2$, and $\pm\nu_3$.

\section{Summary and discussion}
\label{sec:summary}

In this paper, we have derived a cutting rule for equal-time in-in correlators; see Eqs.~\eqref{eq:cutting_2pt}
and~\eqref{eq:cutting_3pt} for two- and three-point bulk correlators, respectively.
Our proof relies only on basic properties of the in-in formalism such as unitarity, locality, 
and the causal structure, and therefore applies to arbitrary theories, even beyond the context
of cosmological perturbations, as long as they satisfy the basic assumptions of quantum field theory.
As an application, we have shown that with the assumption of microcausality, 
our cutting rule efficiently extracts non-local CC signals
as these arise solely from cut-propagators; see Eqs.~\eqref{eq:cutting_2pt_nlCC} and~\eqref{eq:cutting_3pt_nlCC}. 
This in particular tells us that the conformal time integrals
can be always factorized as far as non-local CC signals are concerned, which is of practical advantage
for evaluating the signals.

We believe that our cutting rule opens up many interesting applications, far beyond those discussed in this paper.
For instance, we have considered only scalar fields for explicit calculations,
but our cutting rule holds for fermions and gauge bosons independently of 
\textit{e.g.,}  the presence/absence of the chemical potentials.
Since the CC signals can be enhanced with the chemical potentials~\cite{Chen:2018xck,Hook:2019zxa,
Hook:2019vcn,Wang:2019gbi,Bodas:2020yho,Wang:2021qez,Niu:2022quw,Niu:2022fki,Stefanyszyn:2023qov,
Jazayeri:2023xcj,Jazayeri:2023kji}, 
it would be of practical importance to apply our cutting rule to these theories 
and evaluate the non-local CC signals analytically, beyond the soft limit.

Second, even though we have focused only on non-local CC signals in this paper,
it would be important to see if and how our results can be extended to local CC signals,
as it can mix with the non-local CC signals \textit{e.g.,}  for bispectrum.
Moreover, understanding background contributions, including UV divergences for loop processes,
is important for the actual evaluation of the signals.
We have seen that, at the tree level, the bulk-cut diagrams do not contain background contributions.
Therefore a natural question would be if, and if yes to what extent, this is extended to loop diagrams.
In general, the loop computations in this paper should be understood as primitive 
since our main focus was to derive the cutting rule,
and a more comprehensive study on loop corrections with our cutting rule would be desirable.

Partly related to the above point, it would be worth combining our cutting rule with modern techniques
of calculating the cosmological correlators.
In this paper, we have directly performed the conformal time integrals in our explicit computations.
However, it is sometimes more convenient to consider differential equations that the correlators satisfy,
called cosmological bootstrap~\cite{Arkani-Hamed:2018kmz,Sleight:2019hfp,Baumann:2019oyu,Baumann:2020dch,
Pajer:2020wxk,Baumann:2022jpr,Pimentel:2022fsc,Jazayeri:2022kjy,Wang:2022eop,
Qin:2023ejc,Arkani-Hamed:2023kig,Aoki:2023wdc,Aoki:2024uyi}.
We may see \textit{e.g.,}  if the differential equations that the bulk-cut diagrams obey contain inhomogeneous terms or not.
Combining our cutting rule with the (partial) Mellin--Barnes transformation and spectral representation~\cite{Bros:2009bz,
Marolf:2010zp,Sleight:2019mgd,Sleight:2020obc,Hogervorst:2021uvp,DiPietro:2021sjt,Qin:2022fbv,Xianyu:2022jwk,
Penedones:2023uqc,Loparco:2023rug,Xianyu:2023ytd,Werth:2024mjg}, 
or with the cosmological flow~\cite{Werth:2023pfl,Pinol:2023oux,Werth:2024aui}, could be another direction.

Finally, as we have noted in Sec.~\ref{subsec:microcausality}, alternative cutting rules exist in literature.
In particular, there exists a cutting rule acting not on the in-in correlators but on a wavefunction of the universe~\cite{Goodhew:2020hob,
Jazayeri:2021fvk,Melville:2021lst,Goodhew:2021oqg,AguiSalcedo:2023nds,Stefanyszyn:2023qov,Ghosh:2024aqd}, and another rule derived by mapping the in-in formalism to the in-out formalism~\cite{Donath:2024utn}
(see the end of Sec.~\ref{subsec:cutting_proof} for a detailed comparison).
Moreover, relations among cosmological correlators are derived from 
the cosmological largest time equations recently in~\cite{Werth:2024mjg}.
Since those cutting rules rely only on basic assumptions of the quantum field theory, it would be interesting to 
investigate possible relations between them and our cutting rule.

\paragraph{Acknowledgements}

We thank Soubhik Kumar, Lian-Tao Wang, Wei Xue, and Masahide Yamaguchi for helpful discussions. 
This work is supported in part by U.S.\ Department of Energy Grant No.~DESC0011842, and by JSPS KAKENHI Grant No.\ JP22K14044.
Y.E.\ and K.M.\ were supported in part by the Munich Institute for Astro-, Particle and BioPhysics (MIAPbP), 
which is funded by the Deutsche Forschungsgemeinschaft (DFG, German Research Foundation) 
under Germany's Excellence Strategy -- EXC-2094 -- 390783311.
Y.E.\ acknowledges the Aspen Center for Physics, where a part of this work was performed, 
supported by National Science Foundation grant PHY-2210452.
The diagrams in this paper were drawn with \texttt{TikZ-Feynman}~\cite{Ellis:2016jkw}.

\appendix

\small
\bibliographystyle{utphys}
\bibliography{ref}
  
\end{document}